\documentclass[prb,twocolumn,showpacs,preprintnumbers,amsmath,amssymb]{revtex4-1}

\usepackage{graphicx}
\usepackage{ifthen}
\usepackage{ifpdf}
\usepackage{color}
\definecolor{red}{rgb}{1.,0.,0.}
\definecolor{blue}{rgb}{0.,0.,1.}

\usepackage{latexsym}
\usepackage{amssymb}
\usepackage{amsmath}
\usepackage{bm}
\newcommand{\half}{\mbox{\small $\frac{1}{2}$}}

\newcommand{\im}{\mbox{Im}}
\newcommand{\re}{\mbox{Re}}
\newcommand{\eexp}{\mbox{e}^}

\newcommand{\h}{\mathcal{H}}
\newcommand{\tr}{\mbox{Tr}}
\newcommand{\xx}{{\bf x}}
\newcommand{\pp}{{\bf p}}

\newcommand{\beq}[1]{\begin{eqnarray}\ifthenelse{#1=-1}{\nonumber}
{\ifthenelse{#1=0}{}{\label{e#1}}}}
\newcommand{\eeq}{\end{eqnarray}}
\newcommand{\be}{\begin{equation}}
\newcommand{\ee}{\end{equation}}

\newcommand{\bea}{\begin{eqnarray}}
\newcommand{\eea}{\end{eqnarray}}

\newcommand{\hide}[1]{}

\begin{document}

\title{Double Quantum Dot scenario for spin resonance in current noise}
\author{Baruch Horovitz and Anatoly Golub }
\affiliation{ Department of Physics, Ben Gurion University,
Beer Sheva 84105 Israel}
\begin{abstract}
We show that interference between parallel currents through two quantum dots, in presence of spin orbit interactions and strong on-site Coulomb repulsion, leads to resonances in current noise at the corresponding Larmor frequencies. An additional resonance at the difference of Larmor frequencies is present even without spin-orbit interaction. The resonance lines have strength comparable to the background shot noise and therefore can account for the numerous observations of spin resonance in STM noise with non-polarized leads. We solve also several other models that show similar resonances.
\end{abstract}
\maketitle

Coherent control and detection of a single spin are fundamental challenges in nanoscience and nanotechnology, aiming to determine electronic structures as well as provide qubits for quantum information processing \cite{koppens,awschalom}. Of particular interest are studies that combine the high energy resolution of electron spin resonance (ESR) with the high spacial resolution of scanning tunneling microscope (STM). These ESR-STM studies are of two types, either monitoring the current power spectrum in a DC bias \cite{manassen1,balatsky1,manassen2}, or monitoring the DC current when an additional AC voltage is tuned to resonance conditions \cite{mulleger,baumann,willke}. In the latter case with a magnetic tip \cite{baumann,willke} the theory is well understood \cite{baumann,shavit}. In Ref. \onlinecite{mulleger} the tip is apparently nonmagnetic, hence it should be interpreted as the inverse phenomenon to that of the first type.

We focus here on the ESR-STM phenomenon of the first type, i.e. a DC bias alone.
The experimental technique is conceptually simple: an STM tip is placed above a localized spin center in presence of a DC magnetic field and the power spectrum, monitoring the current fluctuations, is measured; the data exhibits a sharp resonance at the expected Larmor frequency \cite{manassen1,balatsky1,manassen2} even at room temperature.  This phenomena has been further confirmed by an associated ENDOR effect \cite{manassen3}.
 The understanding of this ESR-STM phenomenon presents a theoretical challenge even at present \cite{balatsky1}. It was proposed early on that a spin-orbit coupling is essential for converting the spin fluctuations to current noise, assuming also that the tip and substrate are spin polarized \cite{bulaevskii,gurvitz,martinek}. However, the experimental data \cite{manassen1,balatsky1,manassen2} involves non-polarized tip and substrate. It was argued that an effective spin polarization is realized either as a fluctuation effect \cite{balatsky2,manassen2} or due to 1/f magnetic noise of the tunneling current \cite{manassen4}.
The first theoretical model that conclusively showed an ESR-STM phenomena in this case, i.e. non-polarized electrodes in a DC setup, was a nanoscopic interferometer model \cite{caso,golub}. In this model the current has an additional channel of direct tunneling from the tip to the substrate in parallel to the current via the spin states. The interference between the two channels leads to an ESR resonance, however, the signal is rather weak. Furthermore this model ignores on-site Coulomb interactions, that are expected to be significant at a localized spin site.

In the present work we propose a new mechanism for the ESR-STM phenomenon, a mechanism that provides a strong signal, comparable to that of the background shot noise, and allows for a strong Coulomb interactions at the spin site. The model assumes the presence of an additional spin such that the current passes in parallel via two spins, i.e. a double quantum dot (DQD). The additional spin is unintentional in the ESR-STM experiments so far, yet its presence can be tested by monitoring our predictions. In particular, in addition to the expected resonance at $\nu_1=g_1\mu_BH$  additional resonances are present at $g_2\mu_BH$ and at $|g_1-g_2|\mu_BH$; $g_1,\,g_2$ are the g-factors of the two spins, respectively, $\mu_B$ is the Bohr magneton and $H$ is the DC magnetic field. We solve also the single spin model \cite{caso,golub} with strong on-site Coulomb repulsion, as well as the non-interacting two spin model. We find that the DQD model provides a strong signal to noise ratio and is most likely to account for the ESR-STM data.
The properties of all the studied models are summarized in table I below.

\begin{figure}[t]  \centering
\includegraphics [width=.45\textwidth]{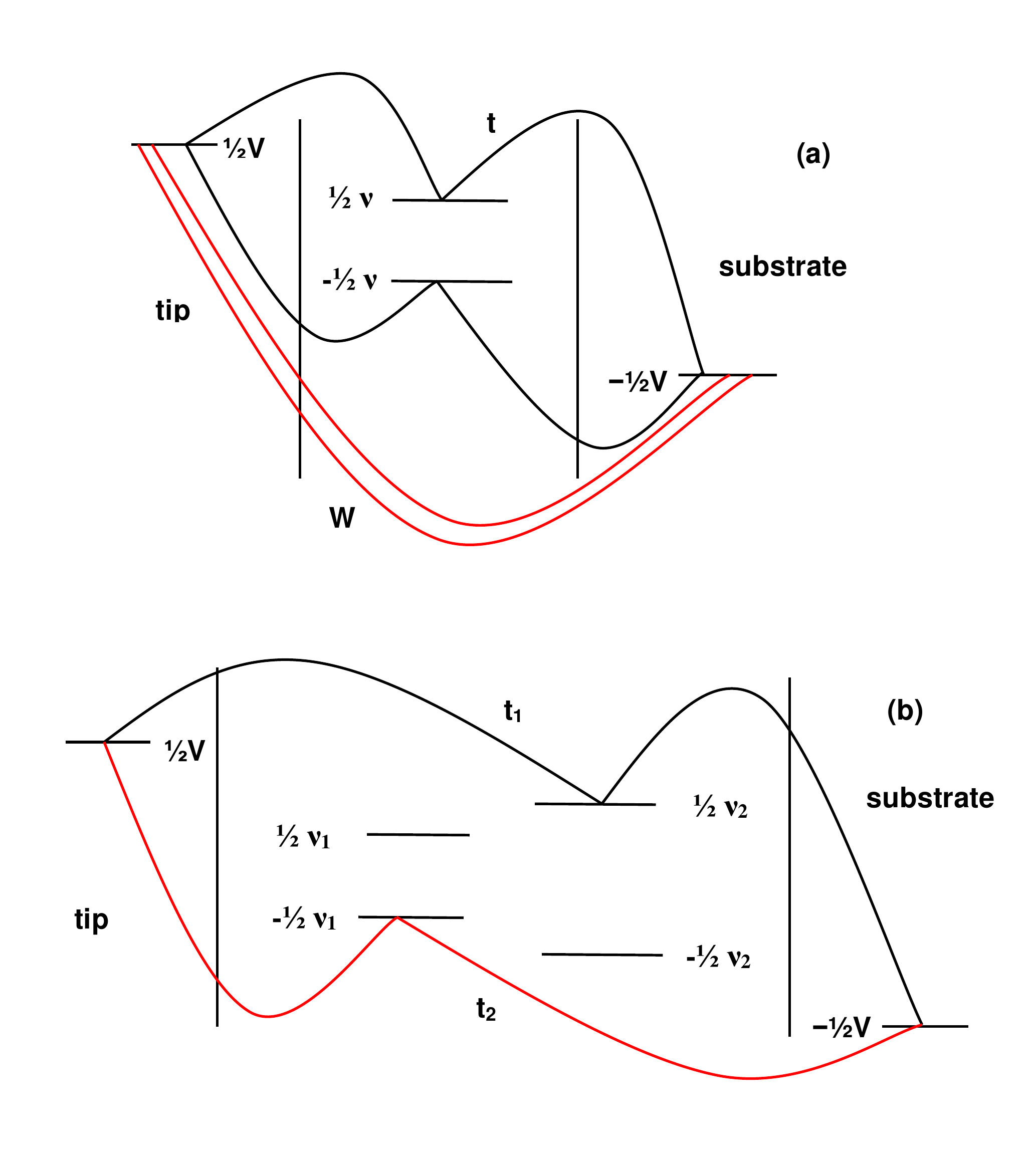}
\caption { Interference paths that lead to an ESR-STM effect: (a) paths via a single spin and a direct tunneling path (b) paths via two spins in parallel (for $\Delta=0$).}
\label{levels}
\end{figure}

We review first the previous model \cite{caso,golub} that involves interference between tunneling via  the spin and direct tunneling, as illustrated in Fig. \ref{levels}a.
Consider $l=L,R$ (left, right) fermion leads (i.e. tip and substrate) with the Hamiltonian ${\cal H}_0=\sum_{l,k,\sigma}\epsilon_{lk}c^\dagger_{lk\sigma}c_{lk\sigma}$ where $\sigma=\pm$ denotes the spin and $k$ are continuum states; $c_{lk\sigma}$ are the lead fermion operators whose dispersions $\epsilon_{lk}$ include the voltage and are spin independent, justified by the small ratio $10^{-5}$ of the Larmor frequency and a typical electron bandwidth. The spin site involves fermion operators $d_\sigma$ and a Hamiltonian ${\cal H}_d=\sum_\sigma (\epsilon_0+\half \nu\sigma) d^\dagger_\sigma d_\sigma$ where $\nu=g\mu_BH$ is the (single) Larmor frequency with g-factor $g$.
The reservoirs are connected by a direct tunneling as well as by tunneling via the spin, the latter allows for an SU(2) spin-orbit rotation \cite{caso,golub} $\hat u=\eexp{i\sigma_z\phi}\eexp{\half i\sigma_y\theta}$ where ${\bm \sigma}$ are the Pauli matrices. The total Hamiltonian is ${\cal H}_1={\cal H}_0+{\cal H}_d+{\cal H}^{(1)}_{tun}$ with the tunneling term,
\beq{01}
{\cal H}^{(1)}_{tun}=tc_L^\dagger d+t'c_R^\dagger \hat u d+ Wc^\dagger_Lc_R+h.c.
\eeq
where all operators are now spinors and $c_l^\dagger=\sum_k c_{lk}^\dagger$ is at the tunneling site.
The current noise for this model has been solved exactly \cite{caso,golub} with results summarized in the first column of table I, yet it is instructive to derive the main results heuristically. The resonance linewidth is seen from a Golden rule $\Gamma=2\pi t^2N(0)$, assuming now $t=t'$ and the density of states $N(0)$ per spin of both leads are taken equal, for simplicity. The resonance in the current correlation involves a closed loop with a given spin that passes at both spin levels, as illustrated in Fig. \ref{levels}a. Hence one needs at least 6 tunneling events, 4 via the spin and 2 direct tunnelings, as well as two spin flips of probability $\sin^2\half\theta$, hence an amplitude $\sim t^4W^2\sin^2\half\theta$  which multiplies a Lorenzian of width $\Gamma$, hence the peak amplitude is $\sim t^4W^2\sin^2\half\theta/\Gamma\sim t^2W^2\sin^2\half\theta$. This derivation is valid if the spin levels are within the voltage window \cite{caso,golub}, i.e. $|\epsilon_0|<\half eV[1+O(T/eV)]$ at temperature $T$.
The direct current $L\rightarrow R$ is also found by a Golden rule rate
  $2\pi W^2N(0)$ per spin times the final number of available states $eV$, i.e. $J_W=4\pi e^2 V W^2N^2(0)$. Assuming $W\gg t$ the background shot noise is $2eJ_W$, hence the Fano factor, i.e. the ratio of the resonance peak to that of the background, is $F\approx \Gamma\sin^2\half\theta/eV$. For \cite{balatsky1,manassen2} $\Gamma\approx 10$MHz, $V\approx 1$eV this ratio is $\approx 10^{-6}$, too small to account for ESR-STM data. (If $t\gg W$ the Fano factor would be even smaller, $F\sim W^2N^2(0)$).

  We consider next our new model, first its non-interacting variant. The model involves current transport via two spins, in parallel, i.e. the Hamiltonian is ${\cal H}_2={\cal H}_0+{\cal H}_d^{(2)}+{\cal H}^{(2)}_{tun}$ where $d_1,d_2$ are fermion spinor operators on the two spin sites,
  \beq{02}
  &&{\cal H}^{(2)}_d=\half \nu_1d^\dagger_1\sigma_z d_1+d^\dagger_2(\Delta+\half \nu_2\sigma_z)d_2\nonumber\\
  && {\cal H}_{tun}^{(2)}=c_L^\dagger[t_1d_1+t_2d_2]+c_R^\dagger[t'_1d_1+t'_2\hat ud_2]
  \eeq
We assume that only the R electrode (probably the tip) has significant spin-orbit interaction. In fact this is equivalent to two SU(2) spin rotation matrices $\hat u_1,\,\hat u_2$ for tunneling from R to sites 1 and 2, respectively, i.e. $t_1'c_R^\dagger \hat u_1d_1+t'_2c_R^\dagger\hat u_2d_2$. In general the wavefunctions of the two spins differ in their orbital part as well as in their locations, hence we expect $\hat u_1\neq\hat u_2$. In the supplementary material \cite{SM} we extend Bardeen's formula to include spin-orbit coupling and estimate the spin-flip angle $\theta$. We find that $0<|\tan \half\theta|\lesssim 1$, depending on the location of the spin site. Hence two spin locations can lead to fairly different $\hat u_1,\,\hat u_2$.
We then rotate $c_R^\dagger\rightarrow c_R^\dagger \hat u_1^\dagger$ so as to cancel $\hat u_1$ in the $d_1$ term while $\hat u=\hat u_1^\dagger \hat u_2$ for the $d_2$ term, resulting in Eq. (2).


  The current fluctuations correspond now to Fig. \ref{levels}b, i.e. a closed loop passing through both spins, at either level of each spin. A resonance appears then at the difference in energy levels, i.e. at $\half |\nu_1\pm \nu_2|$; the $+(-)$ sign is for trajectories through opposite (same) side levels. For a finite relative chemical potential $\Delta$ there are more resonance lines at
  $|\half \nu_1\pm\half \nu_2\pm \Delta|$. The significant virtue of the process in Fig. \ref{levels}b is that only 4 tunneling events are needed, hence a much stronger resonance.  We present \cite{SM} an exact solution of Eq. \eqref{e02} with numerical plots of typical results. The solution can be expanded for weak tunneling, with results summarized in the 3rd column of table I. While the Fano factor is strong the model is inadequate since it neglects on-site Coulomb interactions, expected to be strong for the experimental realizations. Furthermore, the resonance frequencies depend on the unknown $\Delta$ parameter. In fact, the addition of Coulomb interactions is essential for confining the dots as neutral so that the chemical potential $\Delta$ becomes irrelevant.

\begin{table*}[t]
  \caption{Summary of 4 models for weak tunneling, $\Gamma\ll \nu\ll eV$ and equal tunneling amplitudes $t$ (non-interacting models). The Aharononv-Bohm phase $\chi$ is finite only in 2nd column. The Fano factors in the 3rd column correspond to $|\half\nu_1+\half\nu_2\pm\Delta|$ and $|\half\nu_1-\half\nu_2\pm\Delta|$, respectively.
  In the 4th column the linewidth for $|\nu_1-\nu_2|$ differs (given by Eq. \eqref{e08}). The Fano factor in column 1 or 2 is shown for $W\gg t$ or $W\gg J$, respectively (the correction is shown in the DC current). The Fano factor in column 4 is shown for the $\nu_1,\,\nu_2$ resonances, other cases are in Eq. \eqref{e09}.}
\begin{tabular}{|c|c|c|c|c|}
&\multicolumn{2}{|c|}
  {single spin + direct tunneling} & \multicolumn{2}{|c|} {Two spins}  \\
\hline
& non-interacting &  strongly interacting & non-interacting  & strongly interacting (DQD)\\
\hline
Linewidth  $\Gamma$ & $ 2\pi t^2N(0)$ & $16\pi J^2N^2(0)eV$ & $4\pi t^2N(0)$ & $16\pi (J_1^2+J_2^2)N^2(0)eV $\\
DC current & $\,4\pi e^2VW^2N^2(0)+ 2e\Gamma\,$ & $\,2\pi e^2V(2W^2+3J^2)N^2(0)\,$ & $2e\Gamma$  &
$\frac{3}{8}e\Gamma$ \\
Resonance frequencies & $\nu$ & $\nu+\delta\nu\,\,\& \,\,0$ & $|\half\nu_1\pm\half\nu_2\pm\Delta|$ & $\nu_1,\,\nu_2,\,|\nu_1-\nu_2|, \, 0 $\\
Fano factors F &  $\frac{3\pi \Gamma}{16 eV}\sin^2\half\theta$ & $\sin^2\chi\sin^2\half\theta\,\,\,\,\& $ & $\frac{1}{8}\sin^2\half\theta$\,\,\, \&\,\, $\frac{1}{8}(1+\,\,$ &
\\
&&$2\sin^2\chi\sin^2\phi\cos^2\half\theta$& $\cos^2\half\theta-2\cos\phi\cos\half\theta)$&
{\large\(\frac{2\pi J_1^2J_2^2}{3(J_1^2+J_2^2)^2}\)}\(\sin^2\half\theta\)\\
\hline
\end{tabular}
  \label{tab:1}
\end{table*}

  We proceed now to solve both models Eqs. (\ref{e01},\ref{e02}) when strong on-site Coulomb interactions are present, as indeed is the case in atoms and small molecules. Considering first Eq. \eqref{e01}, we add a term $Un_\uparrow n_\downarrow$ where $n_\sigma=d^\dagger_\sigma d_\sigma$. The effective Hamiltonian for large $U, -\epsilon_0\gg t,\nu$ is well known from a Schrieffer-Wolff (SW) transformation \cite{hewson}
  \beq{03}
  {\cal H}_1^c={\cal H}_0+[2Jc_R^\dagger{\bm \sigma}c_L\cdot{\bf S}+
  W\eexp{-i\chi}c^\dagger_R\hat u^\dagger c_L+h.c.]+\nu S_z\nonumber\\
  \eeq
where ${\bf S}$ is the spin operator, $J=O(\frac{tt'}{U},\frac{tt'}{|\epsilon_0|})$ and an Aharonov-Bohm phase $\chi$ is introduced, useful in the following. $W$ may include potential scattering terms generated by the SW transformation. We note that $J$ is reduced by the strong Coulomb interaction, i.e. large $U$ and $-\epsilon_0$, an effect known as the Coulomb blockade.
We keep in \eqref{e03} only exchange terms that allow transport between the electrodes, other exchange terms that involve electrons only on one electrode are neglected since their contribution to transport would be of higher order.
We perform \cite{SM} a perturbation expansion to order $J^2W^2$ using the Keldysh method. The result shows, surprisingly, that the resonance term precisely vanishes when $\chi=0$. In ESR-STM experiments we expect $\chi=0$ since the nanometric dimensions of the setup allow only a negligible magnetic flux. To motivate this result, consider an interference along the loop $R\rightarrow \mbox{(via spin) }\rightarrow L\rightarrow R$ and an additional trajectory of going around the loop in the opposite direction $R\rightarrow L\rightarrow \mbox{(via spin) }\rightarrow R$. When $\chi=0$ these trajectories are related by time reversal, the single spin in the loop then yields a relative minus sign, i.e. cancellation. More specifically,
these two processes, when the localized spin is flipped up, sum up to
\beq{04}
&&\langle c_L^\dagger\sigma_- c_Rc_R^\dagger\hat u^\dagger c_L\rangle\eexp{-i\chi}
+\langle c_L^\dagger\hat u c_Rc_R^\dagger\sigma_- c_L\rangle\eexp{i\chi}\nonumber\\&&=
f_L(\epsilon_L)(1-f_R(\epsilon_R))\{\tr [\sigma_-\hat u^\dagger]\eexp{-i\chi}
+\tr [\sigma_-\hat u]\eexp{i\chi}\}\nonumber\\&&=f_L(\epsilon_L)(1-f_R(\epsilon_R)) 2i\sin\chi\sin\half\theta \eexp{i\phi}
\eeq
where $f_l(\epsilon)$ are Fermi functions and $\tr [\sigma_-\hat u^\dagger]=-\tr[\sigma_-\hat u]=2\sin\half\theta \eexp{i\phi}$. Hence the interference cancels at $\chi=0$. Energy conservation implies $\epsilon_L=\epsilon_R+O(\nu)$ and integration on $\epsilon_L$ yields then an $eV$ factor. Additional interference cycles that start at L involve $f_R(\epsilon)(1-f_L(\epsilon))$ are negligible for $V>0$ and $eV\gg \nu, T$. The result \eqref{e04} is confirmed by detailed perturbation expansion \cite{SM}, as summarized in the 2nd column of table I. Hence for the experimentally relevant case with $\chi=0$ this model may give a resonance only at orders higher than $J^2W^2$ and therefore does not account for ESR-STM data. We note also that replacing $\sigma_-\rightarrow\sigma_z$ in Eq. \eqref{e04} yields a resonance at $\omega=0$ with amplitude $\sim\tr[\sigma_z\hat u^\dagger]\eexp{-i\chi}+\tr[\sigma_z\hat u]\eexp{i\chi}=4\sin\chi\sin\phi\cos\half\theta$.

For completeness, we evaluate the resonance linewidth, relevant when $\chi\neq 0$. The simplest approach is  a Golden rule for the decay of a spin up by passing an electron from L to R,
\beq{05}
\Gamma_{\downarrow}&&=2\pi N^2(0)\int_{\epsilon_L,\epsilon_R}|4J\langle \uparrow|c_{L\downarrow}^\dagger c_{R\uparrow}S_+|\downarrow\rangle|^2\delta(\epsilon_L-\epsilon_R-\nu)
\nonumber\\&&=8\pi eVJ^2N^2(0)
\eeq
Similarly for $\Gamma_{\uparrow}$, so that $\frac{1}{T_1}=\Gamma_{\downarrow}+\Gamma_{\uparrow}=16\pi eVJ^2N^2(0)$, hence for the isotropic interaction in \eqref{e03} the linewidth is $1/T_1=1/T_2$. This result is confirmed by solving a Lindblad type equation \cite{SM} for the spin dynamics; it is also consistent with the linewidth as derived by higher orders in Keldysh diagrams \cite{paaske}, however, the framework of the Lindblad equation, being a proper 2nd order perturbation, is considerably more convenient.
The Lindblad equation also shows a shift in the resonance frequency $\delta\nu=-4\pi eVJWN^2(0)\sin\phi\cos\half\theta\cos\chi$, that may well be larger than the linewidth.

We proceed to our most interesting model, the DQD model with strong on-site Coulomb interactions. Proceeding with a SW type derivation \cite{SM} we find that \eqref{e02} is replaced by
\beq{06}
{\cal H}_2^c=&&{\cal H}_0+2J_1c^\dagger_R{\bm \sigma}c_L\cdot {\bf S}_1+
2J_2c^\dagger_R\hat u{\bm \sigma}c_L\cdot {\bf S}_2\nonumber\\&&
+\nu_1S_{1z}+\nu_2S_{2z}
\eeq
which is an obvious extension of the single spin case.
This Hamiltonian neglects potential scattering terms that may generate terms beyond those that we study of order $J_1^2J_2^2$; also $\chi=0$ here, for simplicity. Tunnelling between the two spin sites is neglected, leading to higher order terms for transport \cite{SM}; this tunneling yields also a direct exchange between the spins which shifts the Larmor frequencies, we neglect here this effect (e.g. if one spin is on the tip and the other on the surface this exchange is much weaker than either $J_1$ or $J_2$).

We note that the spin-orbit factor $\hat u$ is essential for observing a resonance at a Larmor frequency. If $\hat u=1$ then the tunneling elements conserve the total spin, while the $S_z$ terms in ${\cal H}_2^c$ allow conservation of the z component of the total spin. Thus a closed loop of a lead electron returning to its original spin cannot flip a single spin, i.e. no resonance at either $\nu_1$ or $\nu_2$. The loop can, however, flip both spins in opposite ways, hence a resonance at $|\nu_1-\nu_2|$ is possible even without spin-orbit effects. In fact, the same symmetry reasoning applies to all the models considered above. We further note that models with transport via a single spin, even if including spin orbit interaction, e.g. Eq. \eqref{e06} with $J_1=0$, do not show an ESR-STM phenomenon. This is seen by rotating $c_R^\dagger\rightarrow c_R^\dagger \hat u^\dagger$ so that $\hat u$ is canceled and then total $S_z$ conservation rules out a spin-flip resonance. This conclusion holds for models with other types of isotropic exchange interactions \cite{manassen2,balatsky2}, interactions that commute with the total $S_z$.

To appreciate the type of results, we consider the loops as in Eq. \eqref{e04} which for a single spin flip involve $\sigma_-$ on one spin while $\sigma_z$ on the other, hence
\beq{07}
&&\langle c_L^\dagger\sigma_- c_Rc_R^\dagger\hat u \sigma_z c_L\rangle
+\langle c_L^\dagger\sigma_z\hat u^\dagger c_Rc_R^\dagger\sigma_- c_L\rangle\nonumber\\&&=
f_L(\epsilon_L)(1-f_R(\epsilon_R))\{\tr [\sigma_-\hat u \sigma_z]
+\tr [\sigma_z\hat u^\dagger \sigma_-]\}\nonumber\\&&=-2f_L(\epsilon_L)(1-f_R(\epsilon_R))
\sin\half\theta \eexp{i\phi}
\eeq
Hence we expect resonances of the form $\sim J_1^2J_2^2\sin^2\half\theta \delta(\omega-\nu_i)$, i=1,2. An additional resonance at $|\nu_1-\nu_2|$ appears when $\sigma_z\rightarrow\sigma_+$ in Eq. \eqref{e07}, the matrix elements then lead to to $2\cos\half\theta\eexp{i\phi}$, hence a resonance
$\sim J_1^2J_2^2\cos^2\half\theta\delta(\omega-|\nu_1-\nu_2|)$. One further resonance is possible at $\omega=0$ when $\sigma_-\rightarrow \sigma_z$ in Eq. \eqref{e07}, i.e. no spin flips, leading to $4\cos\phi\cos\half\theta$, hence a resonance $\sim J_1^2J_2^2\cos^2\phi\cos^2\half\theta \delta(\omega)$. Finally, there is no resonance at $\nu_1+\nu_2$ since $\sigma_-^2=0$.

  \begin{figure}  \centering
\includegraphics [width=.52\textwidth]{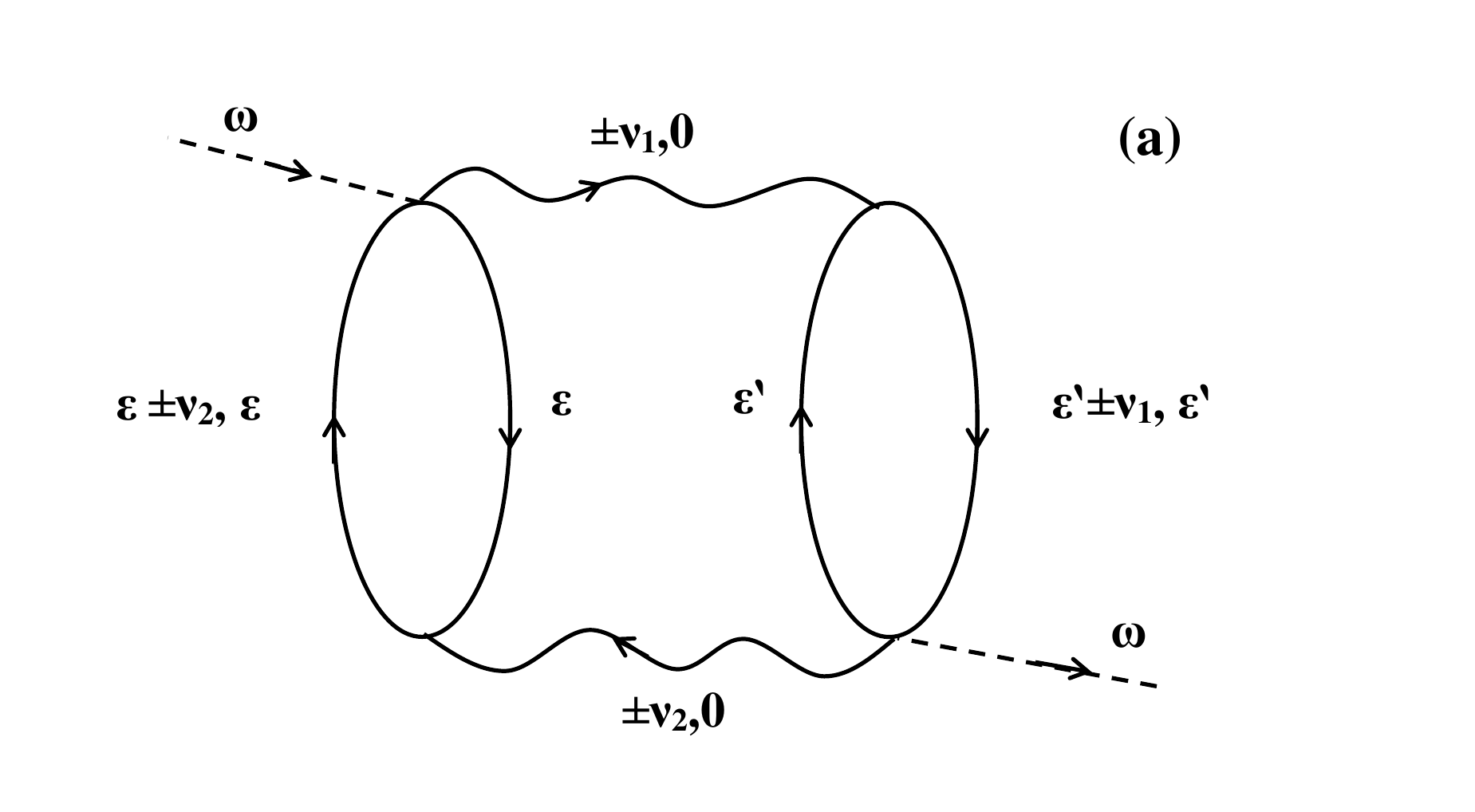}
\includegraphics [width=.52\textwidth]{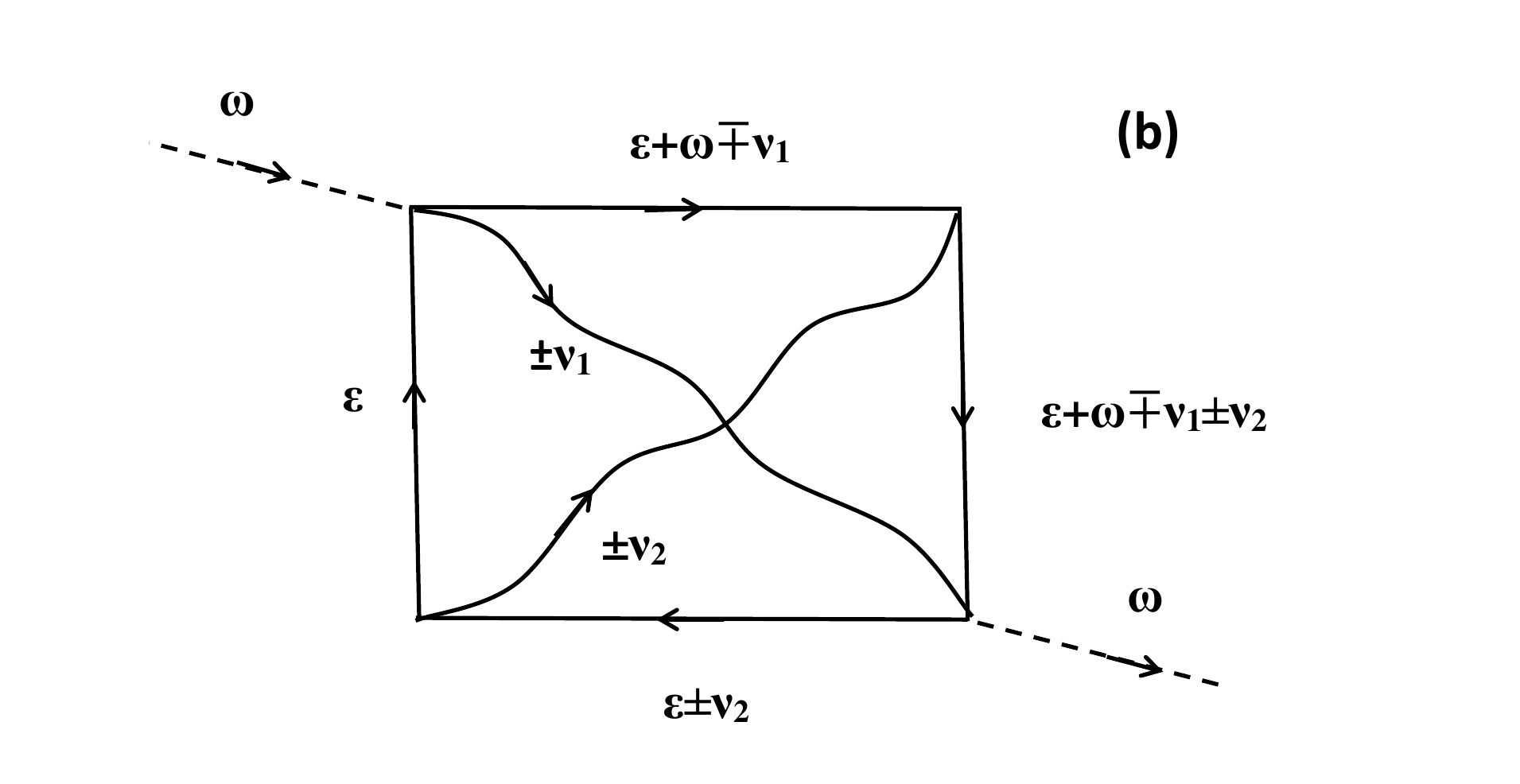}
\caption { Skeleton diagrams: solid straight lines are electron propagators on either right or left electrode (being interchanged at each vertex), wavy lines represent spin propagator with poles at the Larmor frequency $\nu_1$ or $\nu_2$ or at 0 ($S_z$ propagator), dashed lines are external current sources. (a) Conservation of frequency at either current vertex leads to resonances
$\delta(\omega\mp\nu_1),\delta(\omega\mp\nu_2),\delta(\omega\mp|\nu_1-\nu_2|)$ and $\delta(\omega)$. (b) A diagram that does not show a resonance.}
\label{diagrams}
\end{figure}

We proceed now to our diagrammatic expansion. First, consider skeleton diagrams, i.e. without Keldysh indices, that show readily which type of diagrams to 4th order can produce a resonance. Fig. \ref{diagrams}a shows a typical diagram that has a resonance, i.e. frequency conservation at each vertex yields readily $\delta$ function resonances at $\nu_1,\nu_2,|\nu_1-\nu_2|$ and $0$. In contrast, Fig. \ref{diagrams}b shows that all spin frequencies merely shift an electron energy which is being integrated. Hence a weak $\omega$ dependence, i.e. a non-resonant effect. [We note that similar skeleton diagrams can be constructed also for our single dot model Eq. \eqref{e03}, one spin line is eliminated while its vertices remain as the direct tunneling $W$ term.]

We present a detailed Keldysh diagrammatic expansion \cite{SM}.
The results are consistent with the reasoning above and are summarized in the 4th column of table I. We also solve \cite{SM} a Lindblad equation for this case to identify the various linewidths $\Gamma(\omega_{res})$ of the various resonances $\omega_{res}$, leading to
\beq{08}
&&\Gamma=16\pi eVN^2(0)(J_1^2+J_2^2)\qquad  \omega_{res}=\nu_1,\,\nu_2,\, 0 \\&&
\Gamma(|\nu_1-\nu_2|)=16\pi eVN^2(0)(J_1^2+J_2^2-J_1J_2\cos\half\theta\cos\phi)\nonumber
\eeq
This result is similar to that of the single spin case obtained from Eq. \eqref{e05}, except that each spin is affected also by the longitudinal relaxation of the other spin; furthermore, the $\nu_1-\nu_2$ resonance includes a non-secular $\sim J_1J_2$ term \cite{SM}.

The DC current is given in table I, as expected it is $\approx e\Gamma$, so that the background shot noise is $\approx e^2\Gamma$. The resonance signal at maximum is obtained from the discussion following Eq. \eqref{e07} and is confirmed by the diagrammatic expansion \cite{SM}, with the replacement $\delta(\omega-\omega_{res})\rightarrow 1/\pi\Gamma(\omega_{res})$. The ratio of this peak value and that of the background is given for $\nu_1,\,\nu_2$ in the table while for the other resonances it is $F(\omega_{res})$,
\beq{09}
F(0)&&=\frac{4\pi J_1^2J_2^2}{3(J_1^2+J_2^2)^2}\cos^2\half\theta\cos^2\phi\\
F(|\nu_1-\nu_2|)&&=\frac{2\pi J_1^2J_2^2\cdot \cos^2\half\theta(2+\tanh\frac{\nu_1}{2T} \tanh\frac{\nu_2}{2T})}{3(J_1^2+J_2^2)(J_1^2+J_2^2-J_1J_2\cos\theta\cos\phi)}\nonumber
\eeq
 We note that the resonance at $|\nu_1-\nu_2|$ is strongest without spin-orbit coupling, i.e. $\theta=\phi=0$, with a relatively narrow linewidth.

We consider next the relevance of our results to the experimental situation \cite{manassen1,balatsky1,manassen2}. First, we note that the data shows a sharp resonance even at room temperature. This is fully consistent with our results since the linewidth is dominated by the voltage with $eV\gg k_BT$. Second, we note that the linewidth of 25MHz at $I=0.1$nA \cite{manassen3} implies that the DC current via the spins (table I) is $\frac{3}{8}e\Gamma\approx 10^{-4}$nA, much smaller than the total current. We expect then that most of the current tunnels directly between the tip and substrate, indeed a dominant tunneling as it is not Coulomb blockaded. We expect that this current is incoherent with those via the spins, otherwise it would lead to a large shift $\delta\nu$ in the resonance frequency (see paragraph below Eq. \eqref{e05}). Finally, the ratio of peak noise power to that of the shot noise has been estimated \cite{manassen2} as $O(1)$, however, since the power spectrum is measured via modulation of the magnetic field its absolute value has not been so far directly measured.

These experiments \cite{manassen1,balatsky1,manassen2} aim to probe a known spin site on a surface. We propose that a second spin is present, allowing for the observed strong signal. The most likely location for the second spin is on the STM tip which is usually made of a heavy metal with significant spin-orbit coupling. Indeed, the presence of dangling bond surface states in various tip materials is known \cite{chen}, such states are candidates for spin sites.
By extending the measured frequency range, we predict the observation of a second Larmor frequency $\nu_2$ as well as a signal at a lower frequency $|\nu_1-\nu_2|$. The latter in fact may well be stronger than those at either $\nu_1$ or $\nu_2$ if the spin-orbit effect is weak, i.e. small $\theta$.  We note that preliminary data shows a strong signal at low frequency for either defects on a SiC surface or for Tempo molecules on Au substrate \cite{manassen4}.

We note finally that the second type of ESR-STM, i.e. enhanced DC current at resonance with an applied AC voltage \cite{baumann,willke} involves a magnetized Fe atom on the tip. While this is superficially similar to our 2-spin scenario, it is a fundamentally different mechanism, being based on a permanently strong magnetic atom. In our scenario both spin sites exhibit spin fluctuations, in fact even the average spin of each site is extremely weak \cite{SM} $\approx \nu_i/eV\ll 1 (i=1,2)$; yet, data on noise in the spin current might show similarities.

In conclusion we have solved a number of models showing an ESR-STM phenomenon, concluding that the model of two spins with strong on-site Coulomb interactions is the most likely to account for the data. Observation of our prediction for additional magnetic field dependent frequencies in the power spectrum would be the clearest support for our mechanism.\\

\acknowledgements
We thank Y. Manassen for illuminating discussions on his data and the experimental setup.  We also thank C. P. Moca, G. Zar\'and and Y. Meir for stimulating discussions.  BH also gratefully acknowledges funding by the German DFG through the DIP programme [FO703/2-1].

\begin{widetext}

\bigskip

\bigskip

\begin{large}
\begin{center}

{\bf Double Quantum Dot scenario for spin resonance in current noise\\ \vspace{.5cm}
Supplementary Material}

\end{center}
\end{large}

\bigskip

We present in this supplementary details of our derivation for tunneling in presence of spin-orbit coupling (section (i)) and our derivation for the current noise with sections on (ii) noninteracting case, (iii) Schrieffer-Wolff transformation, (iv) double QD case, (v) single QD case, and finally (vi) evaluation of spin relaxations. While these details are fairly lengthy, we emphasize that the main text supplies convincing arguments for essentially all the results, using simple insights. The present methodological presentation is then necessary to make sure that no significant factor is overlooked.

\section{Spin orbit matrix $\hat u$}

In this section we first extend Bardeen's formula \cite{bardeen} for tunneling to include spin-orbit and then we use this formula to estimate the parameters of the spin rotation matrix $\hat u$. In particular we consider tunneling from the STM tip to a spin site and show the dependence of the SU(2) matrix $\hat u$ on the position of the spin site.

The extension of Bardeen's formula to include spin-orbit \cite{caso} is derived here in more detail.
Consider a curved surface $S$, labeled here as a coordinate $x=0$, that separates Hamiltonians ${\cal H}_L,\,{\cal H}_R$ with potentials $U_L(\xx),\,U_R(\xx)$ for the left and right terminals of the tunneling, respectively, e.g. $R$ is the tip and $L$ is a spin site,
\beq{01}
{\cal H}_R&=&\frac{{\hat \pp}^2}{2m}+\tilde U_R(\xx,\hat \pp)\theta(x),\qquad\qquad  \,\,\,\,
\tilde U_R(\xx,\hat \pp)=U_R(\xx)+\alpha{\hat \pp}\cdot {\bf E}(\xx)\times{\bm\sigma}\nonumber\\
{\cal H}_L&=&\frac{{\hat \pp}^2}{2m}+\tilde U_L(\xx,\hat \pp)\theta(-x),\qquad  \qquad
\tilde U_L(\xx,\hat \pp)=U_L(\xx)+\alpha{\hat \pp}\cdot {\bf E}(\xx)\times{\bm \sigma}
\eeq
where $\alpha$ is the spin-orbit coupling, ${\bf E}(\xx)$ is the local electric field and ${\bm\sigma}$ is the spin operator, e.g. Pauli matrices for spin $\half$.

Consider spinor wavefunctions $\varphi_R(\xx), \varphi_L(\xx)$  which solve  ${\cal H}_R\varphi_R(\xx)=\epsilon\varphi_R(\xx),\,{\cal H}_L\varphi_L(\xx)=\epsilon\varphi_L(\xx)$ with degenerate eigenvalue $\epsilon$, describing elastic tunneling. The separating surface $S$ is chosen such that $\varphi_R(\xx), \varphi_L(\xx)$  are localized on the $R,L$ sides, respectively, i.e. decay exponentially into the opposite sides and their overlap is small. Perturbation theory for weak tunneling considers the eigenstate of the full Hamiltonian in the form $\varphi_L(\xx)+\eta\varphi_R(\xx)$ with a shifted eigenvalue $\epsilon+\delta\epsilon$, hence
\beq{02}
&&[\frac{p^2}{2m}+\tilde U_L(\xx,p)\theta(-x)+\tilde U_R(\xx,\pp)\theta(x)][\varphi_L(\xx)+\eta\varphi_R(\xx)]=
(\epsilon+\delta\epsilon)[\varphi_L(\xx)+\eta\varphi_R(\xx)]
\nonumber\\
&&\Rightarrow\qquad \tilde U_R(\xx,\pp)\theta(x)\varphi_L(\xx)+\eta\tilde U_L(\xx,\pp)\theta(-x)\varphi_R(\xx)=
\delta\epsilon[\varphi_L(\xx)+\eta\varphi_R(\xx)]\nonumber\\
&&\int_\xx\varphi^\dagger_R(\xx)\Rightarrow \,\, \eta\delta\epsilon\approx \int_\xx\varphi_R^\dagger(\xx)\tilde U_R(\xx,\pp)\theta(x)\varphi_L(\xx)\equiv t_{L\rightarrow R}\nonumber\\
&&\int_x\varphi^\dagger_L(\xx)\Rightarrow \,\, \delta\epsilon \approx \eta\int_\xx\varphi^\dagger_L(\xx)\tilde U_L(\xx,\pp)\theta(-x)\varphi_R(\xx)\equiv \eta t_{R\rightarrow L}
\eeq
where we neglect terms with two decaying eigenfunctions $\int_\xx\varphi_R^\dagger(\xx)\tilde U_L(\xx,\pp)\theta(-x)\varphi_R(\xx),\,\int_\xx\varphi^\dagger_L(\xx)\tilde U_R(\xx,\pp)\theta(x)\varphi_L(\xx)$ as well as a product with an overlap term $\delta\epsilon \int_\xx \varphi_R^\dagger(\xx)\varphi_L(\xx)$ which is of higher order. Defining the tunneling elements as
$t_{L\rightarrow R}=|t|\eexp{i\gamma}$, then as shown below $t_{R\rightarrow L}=|t|\eexp{-i\gamma}$  and the equation above determines $\delta\epsilon=\pm |t|$ and $\eta=\pm\eexp{i\gamma}$.

We proceed to evaluate $t_{L\rightarrow R}$. Taking the Hermitian conjugate of ${\cal H}_R\varphi_R(\xx)=\epsilon\varphi_R(\xx)$ we can rewrite, avoiding here partial integration,
\beq{03}
\int_{x>0}\varphi^\dagger_R(\xx)U_R(\xx)\varphi_L(\xx)=\int_{x>0}\{[\epsilon-\frac{{\hat \pp}^2}{2m}]\varphi^\dagger_R(\xx)\}\varphi_L(\xx)+\alpha\int_{x>0}[{\hat \pp}\varphi^\dagger_R(\xx)\cdot {\bf E}(\xx)\times{\bm\sigma}]\varphi_L(\xx)
\eeq
Using $\epsilon\varphi_L(\xx)=[\frac{\pp^2}{2m}]\varphi_L(\xx)$  for $x>0$ and ${\bm\nabla}\times{\bf E}(\xx)=0$ (static fields),
\beq{04}
t_{L\rightarrow R}=&&\int_{x>0} \{\varphi^\dagger_R(\xx)\frac{\pp^2}{2m}\varphi_L(\xx)-
[\frac{\pp^2}{2m}\varphi_R^\dagger(\xx)]\varphi_L(\xx)\}+
\alpha\int_{x>0}{\hat \pp}[\varphi^\dagger_R(\xx)\cdot {\bf E}(\xx)\times{\bm\sigma}\varphi_L(\xx)]
\equiv  t_0+t_1
\eeq
The first term gives Bardeen's formula
\beq{05}
t_0=-\frac{\hbar^2}{2m}\int_{x>0}{\bm \nabla}\{\varphi^\dagger_R(\xx){\bm \nabla}\varphi_L(\xx)-
[{\bm \nabla}\varphi^\dagger_R(\xx)]\varphi_L(\xx)\}
=-\frac{\hbar^2}{2m}\int_S\{\varphi^\dagger_R(\xx){\bm \nabla}\varphi_L(\xx)-
[{\bm \nabla}\varphi^\dagger_R(\xx)]\varphi_L(\xx)\}\cdot {\bf dS}
\eeq
The spin-orbit term reduces also to a surface term
\beq{06}
t_1=-i\alpha \cdot\int_S\varphi^\dagger_R(\xx){\bf E}(\xx)\times{\bm\sigma}\varphi_L(\xx)\cdot{\bf dS}
\eeq

To obtain the $R\rightarrow L$ tunneling we need repeat the previous calculation with $L\leftrightarrow R$ and integrating on $x<0$. Hence $t_{R\rightarrow L}=t_0'+t_1'$ with
\beq{07}
t_0'=-\frac{\hbar^2}{2m}\int_S\{\varphi^\dagger_L(\xx){\bm \nabla}\varphi_R(\xx)-
[{\bm \nabla}\varphi^\dagger_L(\xx)]\varphi_R(\xx)\}\cdot {\bf dS}'=t_0^*
\eeq
since the normal to the surface is now in the opposite direction ${\bf dS}'=-{\bf dS}$.
The spin-orbit term is
\beq{08}
t_1'=&&-i\alpha \cdot\int_S\varphi^\dagger_L(\xx){\bf E}(\xx)\times{\bm\sigma}\varphi_R(\xx)\cdot{\bf dS}'
=t_1^*
\eeq
Hence $t_{R\rightarrow L}=t_{L\rightarrow R}^*$.

We next estimate the spin rotation angle $\theta$ as defined by $\hat u=\eexp{i\sigma_z\phi}\eexp{\half i\sigma_y\theta}$. We note first that the tip wave function is localized at the tip apex, hence the surface $S$ varies on an atomic scale \cite{chen} so as to minimize the overlap between the states $\varphi_R(\xx),\,\varphi_L(\xx))$; the surface $S$ can be taken as a semi-circle around the tip apex \cite{chen}. Furthermore,
The surface integrals for both $t_0,\,t_1$ are dominated by a point where the overlap of the localized wavefunctions is maximized. This point $\xx_0$ is the closest one to the position of the spin site. Hence
\beq{09}
\frac{t_1}{t_0}\approx -\frac{2m\xi\alpha}{\hbar^2}{\bf E}(\xx_0)\times {\bm\sigma}\cdot \hat{\bf n}(\xx_0)
\eeq
where $\hat{\bf n}(\xx_0)$ is a unit vector perpendicular to the surface $S$ at $\xx_0$ and $\xi$ is a combined decay length of the localized states, an atomic scale.

A further input are the equipotential lines in an STM set up \cite{devel}. The STM tip consists of an atomically sharp apex that is supported by a tip body of a much larger scale, i.e. 20-100nm. The equipotential lines have therefore a sharp component as well as a smooth one on the larger scale. Therefore ${\bf E}(\xx_0)$ has a component $E_0$ in the $z$ direction, perpendicular to the substrate, while
$\hat{\bf n}(\xx_0)$, chosen to be in an $y,z$ plane, has an angle $\beta$ relative to the $z$ axis, pointing towards the spin site. This estimate yields
\beq{10}
\tan\half\theta\approx \frac{2m\xi\alpha}{\hbar^2}E_0\sin\beta
\eeq

Finally, we need to estimate the spin-orbit coupling $\alpha$. We consider Tungsten (W) as a representative material for an STM tip. Data \cite{shikin} on clean W(110) and on one monolayer H on W(110) show a spin-orbit splitting of $0.5$eV at wavevector $k\approx 0.3\AA^{-1}$. Further data \cite{rotenberg} on W(110) shows a spin-orbit splitting of $\sim 0.2$eV, increasing to $0.5$eV with 0.5 monolayers of Li, reflecting an enhancement of the W spin-orbit by the electric field induced by the Li coverage \cite{rotenberg}. We infer that $\alpha E_0\approx 10^{-8}$eV$\cdot$cm, a value that is likely to be enhanced by the strong electric field in an STM setup. Taking $\xi\approx$1nm we obtain with the latter $\alpha E_0$ that $\tan\half\theta\approx 3\sin\beta$. Hence the spin flip angle, as well as the matrix $\hat u$ have a strong dependence on the location of the spin. In the case of two spin sites, their different locations would lead to different matrices $\hat u_1\neq \hat u_2$. If the spin sites correspond to different types of atoms or molecules, then the different decay length $\xi$ would also cause a difference between their matrices $\hat u_1,\,\hat u_2$.

\vspace{2cm}

\section{Current noise -- noninteracting case}

We study here the 2-spin system as given in Eq. (2) of the main text. We use the rotated Keldysh basis \cite{kamenev} so that the action of the right (R) and left (L) leads is
\begin{equation}\label{s0}
    S_0=\int dt\sum_{l,k}c^{\dagger}_{l,k}g^{-1}_{l,k}c_{l,k}\qquad l=R,L
\end{equation}
where the Greens functions (GF) involve the retarded $g^r$, advanced $g^a$ and Keldysh components $g^K$, in units of the right or left density of states $N(0)$ (assumed equal for simplicity), are
\begin{eqnarray}\label{g}
{\bar g}^{r}&=&\frac{1}{2\pi N(0)}\sum_k\frac{1}{\epsilon-\epsilon_{l,k}+i\delta}=-\half i \qquad
{\bar g}^{a}=\frac{1}{2\pi N(0)}\sum_k\frac{1}{\epsilon-\epsilon_{l,k}-i\delta}=+\half i
\nonumber\\
{\bar g}_l^{K}(\epsilon)&=&-if_l(\epsilon)\equiv -i\tanh\frac{\epsilon\mp \half eV}{2T}
\,\,\Rightarrow\qquad \bar g(\epsilon)=\left(\begin{array}{cc} \bar g^r & \bar g^K(\epsilon)\\ 0 &  \bar g^a
\end{array}\right)=-i\left(\begin{array}{cc}\half & f_l(\epsilon)\\ 0 & -\half \end{array}\right)
\end{eqnarray}
where $l=R,L$ corresponds to $+,-$, respectively. [We note, however, that care is needed if the combination $g_+= g^r+ g^a$ appears, e.g. $\int_\epsilon g_+(\epsilon)g_+(\epsilon+\omega)\neq 0$].

The tunneling part of the action includes a quantum source that couples to the current operator and to a Pauli matrix $\tau_x$ in the rotated Keldysh basis
\begin{eqnarray}
  S_{tun}^{(2)}&=&-\int dt\sum_{k}[c^{\dagger}_{L,k}(1+\alpha \tau_x)(t_1d_1+t_2d_2) +\nonumber\\
 &&\{c^{\dagger}_{R,k}(1-\alpha \tau_x)(t'_1 d_1 +t'_2 \hat u d_2)+h.c ]
 \end{eqnarray}
 The  source $\alpha(t) $ represents  the quantum source \cite{kamenev} field $\alpha(t)=-\alpha^*(t)$  which is set to zero after variations that define either the current or the noise power.
 (The source here couples symmetrically to the left and right lead currents, hence a factor $\half$ is inserted in the following for each current).

 The $c$ electron operators of the leads can be integrated out, leading to an effective  action in terms of dot electrons, represented as a spinor $\hat d^\dagger=(d_1^\dagger,d_2^\dagger)$,
\begin{eqnarray}
  S_{eff} &=& \int dtdt'\hat{d}^{\dagger}(t) \left(
                                          \begin{array}{cc}
                                            G^{-1r}_0 -M^{11} &  -M^{12} \\
                                            -M^{21} &  G^{-1a}_0 -M^{22} \\
                                          \end{array}
                                        \right)(tt')\,\,
  \hat{d}(t') \label{invG}\\
 &&G^{-1\alpha\beta}|_{\alpha\rightarrow 0}=G^{-1\alpha\beta}_0-M^{\alpha\beta}|_{\alpha\rightarrow 0};\,\,\,\,\,\, G^{-1r,a}_0=\left(
                                                    \begin{array}{cc}
                                                      q_1^{r,a} & 0 \\
                                                      0 & q_2^{r,a} \\
                                                    \end{array}
                                                  \right)\label{G}\\
    q_1^{r,a}&=& [(i\frac{\partial}{\partial t}\pm i\delta) \sigma_0 +\half \nu_1  \sigma_z]\delta(t-t'),\,\,
 q_2^{r,a}= [(i\frac{\partial}{\partial t}\pm i\delta+\Delta)\sigma_0 +\half \nu_2  \sigma_z]\delta(t-t')\nonumber
\end{eqnarray}
where $r,a$ correspond to $+,-$ respectively.
Each Keldysh element of the self energy $M^{\alpha\beta}$ is a $2\times 2$ matrix in the spin site index, resulting in
\begin{eqnarray}
M^{\alpha \beta}(t,t')&=&\sum_{j=1,2}\left(
      \begin{array}{cc}
       \Gamma_1x^{\alpha \beta}_j(t,t')\sigma_0 & \Gamma x^{\alpha \beta}_j(t,t')(\delta_{j1}\sigma_0+\delta_{j2}\hat u)   \\
         \Gamma x^{\alpha \beta}_j(t,t')(\delta_{j1}\sigma_0+\delta_{j2}\hat u^{\dagger}) & \Gamma_2x^{\alpha \beta}_j(t,t')\sigma_0  \label{M}\\
      \end{array}
    \right)
\end{eqnarray}
 Here $x_j(t,t')=\gamma^j_{+}(t)g_j(t,t')\gamma^j_{-}(t')$; $\gamma^j_{\pm}(t)=1\pm (-1)^j \alpha(t)\tau_x$ and j=1(2) correspond to L(R) lead. Note that $x_j$ is a matrix in Keldysh space and is obtained by products of $\tau_x$ matrices and the GF of the leads, Eq. \eqref{g}.
 We assume here $t_1=t_1'$ and $t_2=t_2'$ for simplicity, and define $\Gamma_i=2\pi N(0)|t_i|^2,\,i=1,2$ as well as
  $\Gamma=\sqrt{\Gamma_1 \Gamma_2}$.

  Taking variation with respect to the quantum source $\alpha(t)$ we find the transport current I  and the noise power spectrum $S=S_1+S_2$,
 \begin{eqnarray}
  I(t) &=& \frac{ i}{2}e Tr[G(t_1t_3)\frac{\delta M(t_3t_1)}{\delta\alpha_{t}}]_{\alpha\rightarrow 0}\label{Jvar}\\
 S_1(t,t') &=& \frac{ i}{4}e^2 Tr[G(t_1t_3)\frac{\delta^2 M(t_3t_1)}{\delta\alpha_{t}\delta\alpha_{t^{\prime}}}]_{\alpha\rightarrow 0} \label{S1var}  \\
  S_2(t,t') &=&  \frac{i}{4}e^2 Tr[G(t_1t_3)\frac{\delta M(t_3t_4)}{\delta\alpha_{t}}  G(t_4t_5)\frac{\delta M(t_5t_1)}{\delta\alpha_{t^{\prime}}}]_{\alpha\rightarrow 0} \label{s2}
\end{eqnarray}
In these formulae integration over each time variable is implied except of $t_0$ and $t^{\prime}$. From equations (\ref{G}, \ref{M}) the inverse retarded Green function  $G^{-1r}$ is identified as a $4\times4$ matrix
 $G^{-1r}=G^{-1r}_0 -M^{11}|_{\alpha \rightarrow 0} $ or explicitly
\begin{equation}\label{r}
G^{-1r} =\left(
          \begin{array}{cc}
            (\epsilon+i\Gamma_1)\sigma_0-\half \nu_1\sigma_z & i\Gamma(\sigma_0+u)/2 \\
            i\Gamma(\sigma_0+u^{\dagger})/2 & (\epsilon+\Delta+i\Gamma_2)\sigma_0-\half \nu_2\sigma_z \\
          \end{array}
        \right)
\end{equation}
The Keldysh components of the GF are identified by taking the inverse of \eqref{G} at $\alpha \rightarrow 0$, which implies that $G^{r,a}$ are inverses of $G^{-1r,a}$, respectively, while
\begin{eqnarray}
G^{K}(\epsilon) &=& -iG^r(\epsilon)g_+^K(\epsilon)G^a(\epsilon) \\
  g^{K}_+(\epsilon) &=&\left(
                                          \begin{array}{cc}
                                            \Gamma_1\sigma_0 [f_L(\epsilon)+f_R(\epsilon)] &
                                            \Gamma [f_L(\epsilon)\sigma_0+f_R(\epsilon)\hat u ]\\
                                            \Gamma [f_L(\epsilon)\sigma_0+f_R(\epsilon)\hat u^\dagger] &  \Gamma_2\sigma_0 [f_L(\epsilon)+f_R(\epsilon)] \\
                                          \end{array}
                                        \right)
\end{eqnarray}

The first variation of the self energy M takes the form
\begin{eqnarray}
  \frac{\delta M(t_3t_4)}{\delta\alpha_{t}} &=& \left(
                                          \begin{array}{cc}
                                            -(a-\tilde{a})\Gamma_1 & \Gamma(-a_u+\tilde{a}_ u) \\
                                            \Gamma(-a_{u^+}+\tilde{a}_{u^+}) &  -(a-\tilde{a})\Gamma_2 \\
                                          \end{array}
                                        \right) \label{Mdot} \\
  a &=& \tau_x\sigma_0 (g_L-g_R)_{tt_4}\delta(t-t_3);\,\,\,\tilde{a}=\sigma_0 (g_L-g_R)_{t_3t}\tau_x \delta(t-t_4)\nonumber\\
  a_u&=&\tau_x (\sigma_0g_L-g_R \hat u)_{tt_4}\delta(t-t_3);\,\,\,\tilde{a}_ u=\sigma_0 (g_L-g_R\hat u)_{t _3t}\tau_x \delta(t-t_4)\nonumber\\
a_{u^+}&=&\tau_x (\sigma_0g_L-g_R \hat u^+)_{tt_4}\delta(t-t_3);\,\,\,\, \tilde{a}_{u^+}=\sigma_0 (g_L-g_R\hat u^+)_{t_3 t}\tau_x \delta(t-t_4)\nonumber
\end{eqnarray}
By  taking the trace in Eq.(\ref{Jvar}) over Keldysh space  we obtain  the current
\begin{eqnarray}
I &=& \frac{e}{2} \int\frac{d\epsilon}{2\pi}Tr[(G^r(\epsilon)
\hat{g }^K(\epsilon+\omega)-G^a(\epsilon)\hat{g }^K(\epsilon-\omega))-G^K \hat{m}]\\
 \hat{g }^K(\epsilon) &=& -i\left(
                     \begin{array}{cc}
                       \Gamma_1\sigma_0(f_L-f_R )\,\,&  \Gamma(f_L\sigma_0-f_R \hat u) \\
                       \Gamma( f_L\sigma_0-f_R \hat u^{\dagger})\,\, & \Gamma_2 \sigma_0(f_L-f_R) \\
                     \end{array}
                   \right)
   \\
 \hat{m} &=& i\left(
                \begin{array}{cc}
                  0 & -\sigma_0+u \\
                  -\sigma_0+u^{\dagger} & 0 \\
                \end{array}
              \right)
\end{eqnarray}
The  $S_1$ part of noise power incudes second variation of self energy
\begin{eqnarray}
 \frac{\delta^2 M(t_1t_2)}{\delta\alpha_{t}\delta\alpha_{t'}} &=&- \tau_x\left(
                                          \begin{array}{cc}
                                           \Gamma_1 (b+\tilde{b})& \Gamma(b_u+\tilde{b}_ u) \\
                                           \Gamma( b_{u^+}+\tilde{b}_{u^+}) &  \Gamma_2(b+\tilde{b}) \\
                                          \end{array}
                                        \right)\tau_x \label{Mdot} \\
  b &=& \sigma_0 (g_L+g_R)_{tt'}\delta(t'-t_2)\delta(t-t_1);\,\,\,\tilde{b}=\sigma_0 (g_L+g_R)_{t't} \delta(t-t_2)\delta(t'-t_1)\nonumber\\
  b_u&=&(\sigma_0g_L+g_R u)_{tt'}\delta(t-t_1)\delta(t'-t_2);\,\,\,\tilde{b}_ u=\sigma_0 (g_L+g_Ru)_{t' t} \delta(t-t_2)\delta(t'-t_1)\nonumber\\
b_{u^+}&=&(\sigma_0 g_L+g_R u^+)_{tt'}\delta(t-t_1)\delta(t'-t_2);\,\,\,\, \tilde{b}_{u^+}=\sigma_0 (g_L+g_R u^+)_{t' t} \delta(t-t_2)\delta(t'-t_1)\nonumber
\end{eqnarray}
The trace of $S_1$ in Keldysh space may be written, with a shorthand notation $\ddot M^{\alpha\beta}=\delta^2 M^{\alpha\beta}/\delta\alpha\delta\alpha$, as $Tr=G^r \ddot M^{11}+G^a \ddot M^{22}+G^K \ddot M^{21}$. Using  explicit expression for Keldysh components of M we calculate $S_1$:
\begin{eqnarray}
S_1(\omega) &=& \frac{-i e^2\Gamma}{4} \int\frac{d\epsilon}{2\pi}Tr[G^K(\epsilon) (g_+^K(\epsilon+\omega)+g_+^K(\epsilon-\omega)) -2(G^r(\epsilon)-G^a(\epsilon))g_+  ] \label{S1}\\
  g_+&=& \left(
                    \begin{array}{cc}
                      \Gamma_1/\Gamma\sigma_0 &  (\sigma_0+u)/2 \\
                      (\sigma_0+u^{\dagger})/2 & \Gamma_2/\Gamma\sigma_0 \\
                    \end{array}
                  \right)
\end{eqnarray}
Our main interest is the experimental situation with large voltages $eV\gg \nu_1,\nu_2,T$, then $S_1$  does not depend on frequency.

We consider next the noise $S_2$ using \eqref{s2}. With a shorthand notation $\dot{M}=\frac{\delta M}{\delta\alpha}$ the trace becomes
\begin{eqnarray}
  Tr &=& G^a\dot{M}^{21}G^r\dot{M}^{12}+G^r\dot{M}^{12}G^a\dot{M}^{21}+G^a\dot{M}^{22}G^a\dot{M}^{22}+
  G^r\dot{M}^{11}G^r\dot{M}^{11}+ \\
  &&G^K\dot{M}^{21}G^r\dot{M}^{11}+G^r\dot{M}^{11}G^K\dot{M}^{21}+G^a\dot{M}^{21}G^K\dot{M}^{22}
  +G^K\dot{M}^{22}G^a\dot{M}^{21}+G^K\dot{M}^{21}G^K\dot{M}^{21}\nonumber\label{Tr}
\end{eqnarray}
Decomposing $S_2=S_a+S_b+S_c$ we define terms with the same type of two GFs (advanced or retarded) as $S_a(\omega)$, terms that include two Keldysh GFs and also terms which have one advanced and one retarded dot GF as $S_b(\omega)$, while the remaining terms stand for $S_c(\omega)$. Thus we obtain
\begin{eqnarray}
S_a(\omega)  &=& \frac{e^2\Gamma^2}{4}\int\frac{d\epsilon}{2\pi}Tr[\hat{g}^K(\epsilon+\omega)
  [G^a(\epsilon+\omega)\hat{g}^K(\epsilon) G^a(\epsilon)
  +G^r(\epsilon)\hat{g}^K(\epsilon)G^r(\epsilon+\omega)]\label{Sa}\\
 S_b (\omega) &=& \frac{e^2\Gamma^2}{4}\int\frac{d\epsilon}{2\pi}Tr[\hat{m}G^K(\epsilon+\omega) m\hat{}G^K(\epsilon)
 +\hat m G^-(\epsilon+\omega)\hat{m} G^-(\epsilon)]\label{Sb}\\
  S_c(\omega) &=& \frac{e^2\Gamma^2}{4}\int\frac{d\epsilon}{2\pi}Tr[G^a(\epsilon)\hat{m}
  G^K(\epsilon+\omega)\hat{g}^K(\epsilon) -
   G^r(\epsilon)\hat{g}^K(\epsilon) G^K(\epsilon+\omega)\hat{m}]+(\omega \rightarrow -\omega)
   \nonumber\\\label{Sc}
  \end{eqnarray}
  where $G^-(\epsilon)=G^r(\epsilon)-G^a(\epsilon)$.

For large voltage we can simplify expressions for current and $S_1$. To do this we notice that in the limit $V>>T,\omega$ and $\Gamma_1=\Gamma_2$
 \begin{eqnarray}
   g_+^K(\epsilon) &=& -i \hat{m};\qquad \,\,\hat{g}^K(\epsilon)=2ig_+\\
  G^K(\epsilon) &=& \Gamma G^r(\epsilon )\hat{m}G^a(\epsilon)
 \end{eqnarray}
 Note that $g_+^K(\epsilon)$ becomes constant so that $S_a(\omega)=0$. We also find
 \begin{eqnarray}
  I &=& i e\Gamma \int\frac{d\epsilon}{2\pi}Tr[(G^-(\epsilon)g_+
+\frac{1}{2}G^- \hat{m}G^- \hat{m}]\label{jj}\\
S_1(\omega) &=& \frac{e^2\Gamma}{4} \int\frac{d\epsilon}{2\pi}Tr[G^-(\epsilon)\hat{m}G^-(\epsilon) \hat{m} -2G^-(\epsilon)g_+  \label{ss1}]
 \end{eqnarray}
 We note that in the absence of spin orbit scattering ($\hat u=1$) the  matrix $\hat{m}=0$ and noise is just the $\omega$ independent  $S_1$, i.e. no resonance as expected.

  We wrote a Mathematica program for evaluating the Fano factor
  \begin{equation}\label{P}
  F(\omega)  =\frac{S_b(\omega)+S_c(\omega)}{S_1}\,.
  \end{equation}
  First we show in Fig. 1 the saturation of the Fano factor at small $\Gamma$ occuring at $\frac{\Gamma}{\nu_1}\lesssim 0.01$ for both high energy resonances $\omega=\nu_1\pm\Delta$ and the low energy resonance at $\omega=\Delta$. Fig. 2 shows the resonance splitting due to $\Delta$, Fig. 3 shows the splitting due to $\nu_1\neq\nu_2$ while Fig. 4 shows splittings due to both effects, using a smaller $\theta=\pi/2$ where the high frequency resonances are weaker.\\

    \begin{figure}  \centering
\includegraphics [width=.5\textwidth]{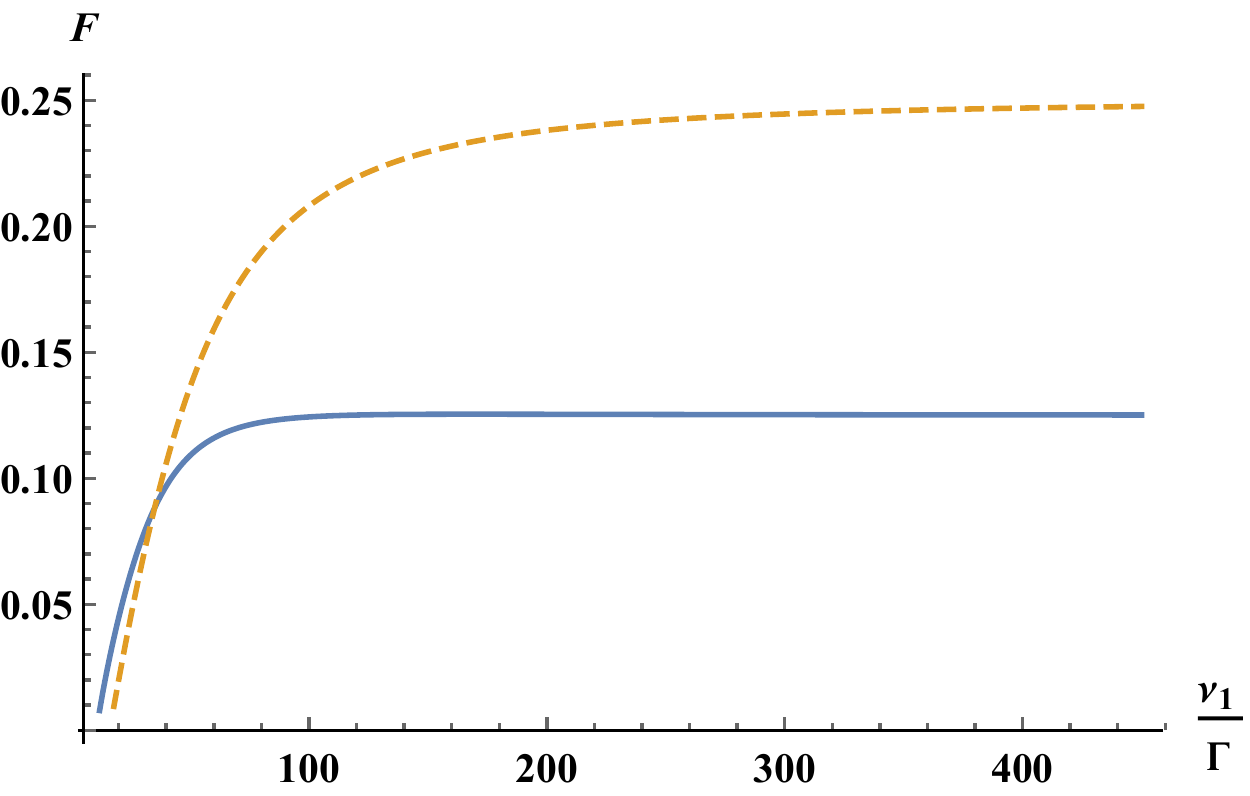}
\caption{Fano factor $F$ for $\nu_1=\nu_2,\, \Gamma=\Gamma_1=\Gamma_2,\,\theta=\pi,\,\phi=\pi/2,\, \Delta/\nu_1=0.05$ as function of $\nu_1/\Gamma$. Full line (blue) is $F(\omega)$ at the resonance  $\omega=\nu_1+\Delta$ while the dashed line (yellow) is $F(\omega)$ at the resonance $\omega=\Delta$}.
\label{Fig1.pdf}
\end{figure}

    \begin{figure}  \centering
\includegraphics [width=.5\textwidth]{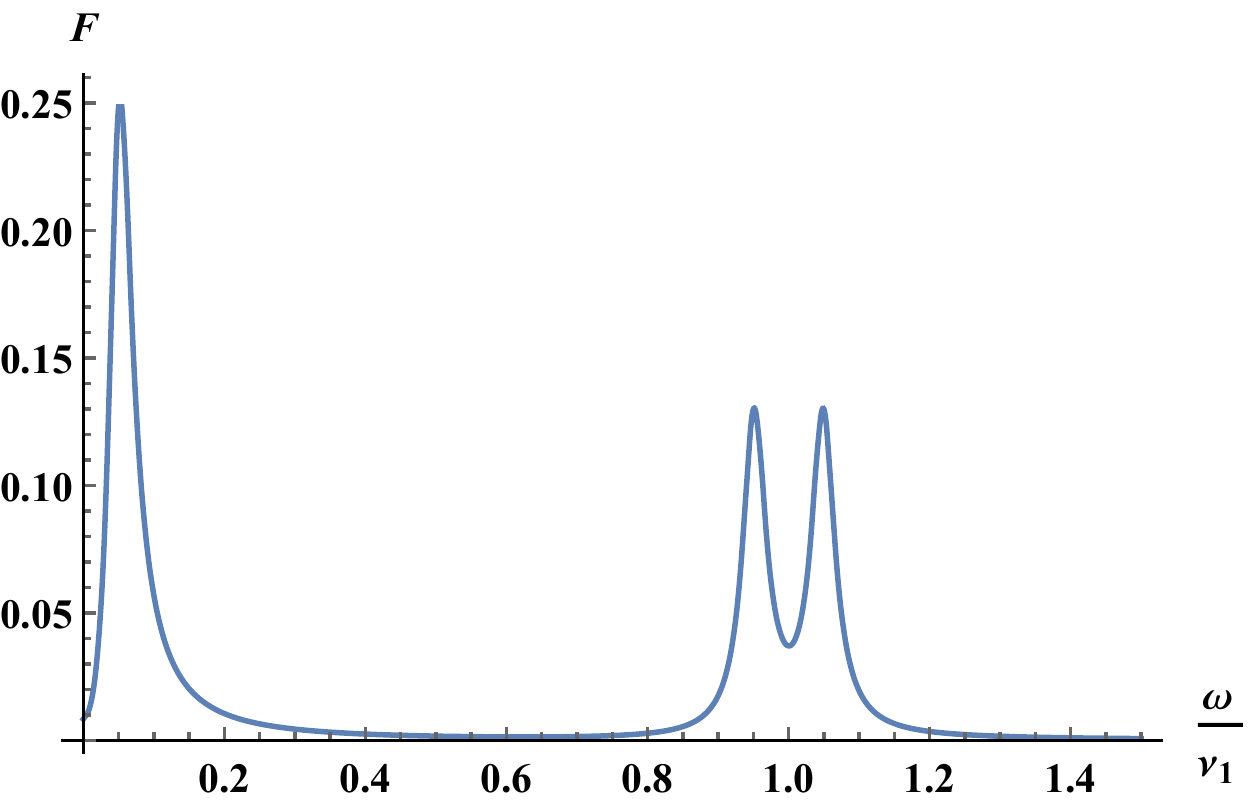}
\caption{Fano factor $F(\omega)$ for $\nu_1=\nu_2,\, \Gamma=\Gamma_1=\Gamma_2,\,\frac{\Gamma}{\nu_1}=0.01,\,\theta=\pi,\,\phi=\pi/2,\, \Delta=0.05\nu_1, $.}
\label{Fig2.pdf}
\end{figure}

\begin{figure}  \centering
\includegraphics [width=.5\textwidth]{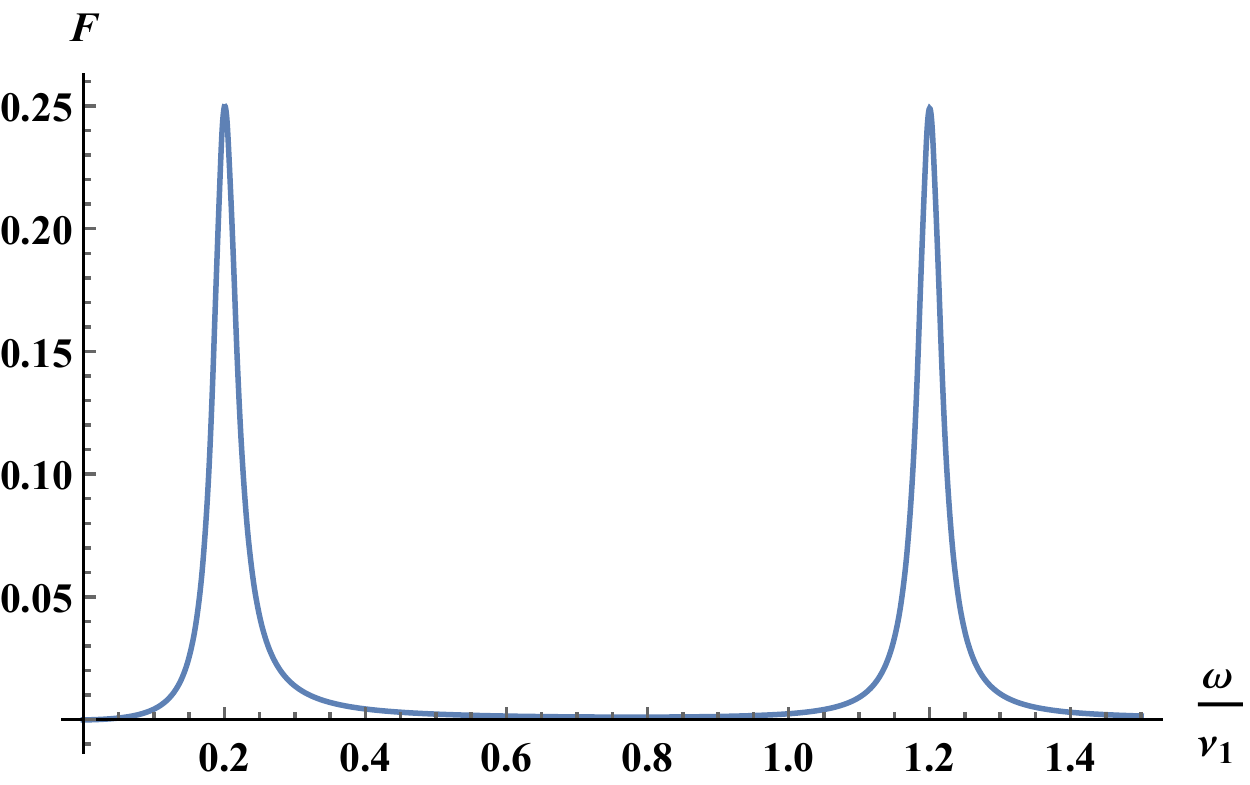}
\caption{Fano factor $F(\omega)$ for $\nu_2=1.4\nu_1,\, \Gamma=\Gamma_1=\Gamma_2,\,\frac{\Gamma}{\nu_1}=0.01,\,\theta=\pi,\,\phi=\pi/2,\, \Delta=0, $.}
\label{Fig3.pdf}
\end{figure}

\begin{figure}  \centering
\includegraphics [width=.5\textwidth]{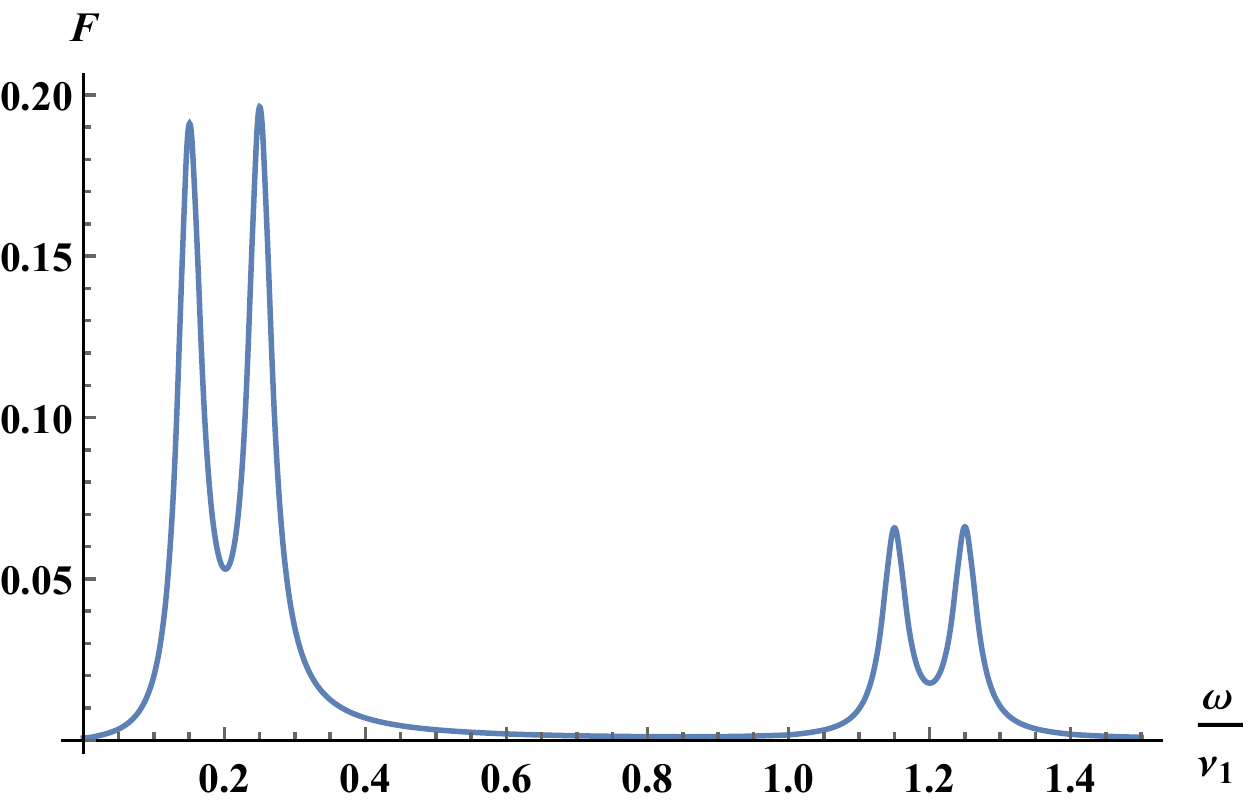}
\caption{Fano factor $F(\omega)$ for $\nu_2=1.4\nu_1,\, \Gamma=\Gamma_1=\Gamma_2,\,\frac{\Gamma}{\nu_1}=0.01,\,\theta=\pi/2,\,\phi=\pi/2,\, \Delta=0.05\nu_1,  $.}
\label{Fig3.pdf}
\end{figure}

\newpage
 Finally, we consider the limiting case of small $\Gamma$. Since $G^K(\epsilon)\sim \Gamma$ the dominant term is the second one in  Eq. \eqref{Sb} for $S_b(\omega)$, i.e.
 \begin{equation}\label{S}
  S_b(\omega) = \frac{e^2\Gamma^2}{4}\int\frac{d\epsilon}{2\pi}Tr[\bar{m}G^-(\epsilon+\omega)\bar{m}G^-(\epsilon)]
  \end{equation}
 In this limit we consider only diagonal terms of $G^{-1}$ in Eq. (\ref{r}) (in the diagonal terms $\Gamma$ is kept infinitesimal), hence
 \begin{eqnarray}
  G_{11}^-(\epsilon) &=& (-2i\Gamma )/((\epsilon - \half g_1\mu_B H)^2 + \Gamma^2),\,\,\,
  G_{22}^-(\epsilon)= (-2i\Gamma )/((\epsilon + \half g_1\mu_B H)^2 + \Gamma^2) \nonumber\\
  G_{33}^-(\epsilon)&=& (-2i\Gamma )/((\epsilon +\Delta- \half g_2\mu_B H)^2 + \Gamma^2),\,\,\,
  G_{44}^-(\epsilon)=(-2i\Gamma )/((\epsilon +\Delta+ \half g_2\mu_B H)^2 + \Gamma^2) \nonumber\\
  \label{Gmin}
\end{eqnarray}
 Using these GFs and taking the trace in Eqs.(\ref{jj}, \ref{ss1})
we obtain $J=4e\Gamma$ and $S_1=2e^2\Gamma$.

The trace  has a form:
\begin{eqnarray}
  Tr &=& \sin^2[\theta/2][G_{44}^-(\epsilon+\omega)G_{11}^-(\epsilon)+ G_{33}^- G_{22}^- + G_{22}^- G_{33}^- + G_{11}^- G_{44}^-\nonumber\\
   &+&(1 + \cos^2[\theta/2]  - 2\cos[\varphi] \cos[\theta/2])[G_{33}^- G_{11}^- + G_{44}^- G_{22}^- +
   G_{11}^- G_{33}^-+ G_{22}^- G_{44}^-]\nonumber\\
\end{eqnarray}
 with all first GFs depend on $\epsilon+\omega$ and second ones depend on $\epsilon$.
Hence
\begin{eqnarray}
 && S_{b}(\omega) =\frac{e^2\Gamma^2}{4}\int\frac{d\epsilon}{2\pi}\{\sin^2\half\theta\,[ G_{44}^-(\epsilon+\omega)G_{11}^-(\epsilon)+
  G_{33}^-(\epsilon+\omega)G_{22}^-(\epsilon)]+\nonumber \\
   && (1 + \cos^2\half\theta  - 2\cos\varphi \cos\half\theta)[G_{33}^-(\epsilon+\omega)G_{11}^-(\epsilon)+
  G_{44}^-(\epsilon+\omega)G_{22}^-(\epsilon)]\}+(\omega\rightarrow-\omega)
\end{eqnarray}
Performing the $\epsilon$ integration with $\Gamma\rightarrow 0$ we finally obtain
\begin{eqnarray}\label{result}
  S_{b}(\omega) &&= \half\pi e^2\Gamma^2\{\sin^2\half\theta[\delta (\half(g_1+g_2)\mu_B H + \Delta - \omega)+ \delta (\half(g_1+g_2)\mu_B H - \Delta - \omega)]+\nonumber\\
   &&  (1 + \cos^2\half\theta  - 2\cos\varphi \cos\half\theta) [\delta (\half(g_1-g_2)\mu_B H+ \Delta -\omega)+\delta(\half(g_1-g_2)\mu_BH-\Delta-\omega)]\}\nonumber\\&&
   +(\omega\rightarrow -\omega)
\end{eqnarray}
We note that the linewidth $\Gamma$ defines the width of the GFs in Eq. \eqref{Gmin}. Since the power spectrum is a convolution of two such Lorenzians its linewidth is $2\Gamma$, i.e. Eq. \eqref{result} involves at each resonance $\omega_{res}$
\begin{equation}
\delta(\omega-\omega_{res})\rightarrow\frac{1}{\pi}\frac{2\Gamma}{(\omega-\omega_{res})^2+(2\Gamma)^2}
\end{equation}
In table I of the main text the $\Gamma$ of this section is replaced by $\half\Gamma$ to agree with the definitions of the other cases.

The power spectrum at resonance is then $e^2\Gamma/4$ (apart of the $\theta,\phi$ dependent factors) so that the Fano factor, i.e. dividing by $S_1=2e^2\Gamma$, has $1/8$, as summarized in the 3rd column of table I in the main text.

\section{Schrieffer-Wolff transformation -- double QD}

We derive the Schrieffer-Wolff transformation (SW) by Hewson's method \cite{hewson} of performing a perturbation expansion directly on the Hamiltonian. We also assume that double occupancy of the dots is forbidden (infinite $U_1,U_2$) while the ionization potentials $\Delta_1,\Delta_2$ are finite and large. Transport is then allowed by first ionization and then recharging from the leads (co-tunneling). This simplifies the algebra, while capturing the essential form of the transformed Hamiltonian.

The 2-dot problem with 2 leads $\alpha=L,R$, including spin-orbit represented by an SU(2) matrix $\hat u$ is given by the Hamiltonian, with $d, \, c$ operators as spinors,
\beq{15}
\h_0&=&\sum_{k\alpha}(\epsilon_{k\alpha}+eV_\alpha)c_{k\alpha}^\dagger c_{k\alpha}\qquad  \alpha=R,L\nonumber\\
\h_d&=&d_1^\dagger(\Delta_1+g_1S_{1z})d_1+d_2^\dagger(\Delta_2+g_2S_{2z})d_2
+U_1n_{1\uparrow}n_{1\downarrow}+U_2n_{2\uparrow}n_{2\downarrow}\nonumber\\
\h_T&=&c_L^\dagger[v_{1L}d_1+v_{2L}d_2]+c_R^\dagger[v_{1R}d_1+v_{2R}\hat u d_2]+h.c.
\eeq
In the limit $U_1,U_2=\infty$
consider the subspace $|0\rangle$, $|10\rangle$ (2-spinor), $|01\rangle$ (2 spinor), $|11\rangle$ (4 spinor). The Hamiltonian has the form
\beq{16}
&&\left(\begin{array}{cccc} \h_0 & (v_{1L}c_L^\dagger+v_{1R}c_R^\dagger)d_1 & (v_{2L}c_L^\dagger+v_{2R}c_R^\dagger\hat u)d_2 & 0\\
d_1^\dagger(v_{1L}^*c_L+v_{1R}^*c_R)  &  \h_0+\Delta_1+g_1S_{1z} & 0 & (v_{2L}c_L^\dagger+v_{2R}c_R^\dagger\hat u)d_2\\
d_2^\dagger(v_{2L}^*c_L+v_{2R}^*\hat u^\dagger c_R) & 0 & \h_0+\Delta_2+g_2S_{2z} & (v_{1L}c_L^\dagger+v_{1R}c_R^\dagger)d_1\\
0 & \,\, d_2^\dagger( v_{2L}^*c_L+v_{2R}^*\hat u ^\dagger c_R) & \,\,d_1^\dagger (v_{1L}^*c_L+v_{1R}^*c_R) & \,\,\,\,\h_0'\end{array}\right)\nonumber\\
&&\times \left(\begin{array}{c} |0\rangle \\|10\rangle\\|01\rangle\\|11\rangle \end{array}\right)
=E \left(\begin{array}{c} |0\rangle \\|10\rangle\\|01\rangle\\|11\rangle \end{array}\right),\qquad \h_0'=\h_0+d_1^\dagger(\Delta_1+g_1S_{1z})d_1+ d_2^\dagger(\Delta_2 +g_2 S_{2z})d_2
\eeq

The 1st line of (\ref{e16}) yields $|0\rangle$ in terms of the other states,
\beq{17}
|0\rangle=\frac{1}{E-\h_0}\left[(v_{1L}c_L^\dagger+v_{1R}c_R^\dagger)d_1|10\rangle+(v_{2L}c_L^\dagger
+v_{2R}c_R^\dagger \hat u)d_2|01\rangle\right]
\eeq
Substituting in the 2nd line yields
\beq{18}
&&d_1^\dagger(v_{1L}^*c_L+v_{1R}^*c_R)\frac{1}{E-\h_0}\left[(v_{1L}c_L^\dagger+v_{1R}c_R^\dagger)
d_1|10\rangle
+(v_{2L}c_L^\dagger+v_{2R}c_R^\dagger \hat u)d_2|01\rangle\right]\nonumber\\&& +(\h_0+\Delta_1+g_1S_{1z})|10\rangle
+ (v_{2L}c_L^\dagger+v_{2R}c_R^\dagger\hat u)d_2|11\rangle=E|10\rangle
\eeq
The $d_1,d_2$ terms are of order $v^2/E\ll E$ and could be neglected in leading order in $v_{i\alpha}$, yet it is of some interest to keep the $d_2$ term as it describes induced tunneling between the spins
\beq{19}
&&\hat t\equiv (v_{1L}^*c_L+v_{1R}^*c_R)
\frac{1}{E-\h_0}(v_{2L}c_L^\dagger+v_{2R}c_R^\dagger \hat u)\nonumber\\
&&|10\rangle=\frac{1}{E-\Delta_1-\h_0} \left[ (v_{2L}c_L^\dagger+v_{2R}c_R^\dagger \hat u)d_2|11\rangle
+ d_1^\dagger \hat t d_2 |01\rangle\right]
\eeq
The 3rd line with solution (\ref{e17}) for $|0\rangle$ is
\beq{20}
&&d_2^\dagger (v_{2L}^*c_L+v_{2R}^*\hat u^\dagger c_R)\frac{1}{E-\h_0}\left[(v_{1L}c_L^\dagger+v_{1R}c_R^\dagger)d_1|10\rangle +(v_{2L}c_L^\dagger+v_{2R}c_R^\dagger \hat u)d_2|01\rangle\right]\nonumber\\&&+(\h_0+\Delta_2+g_2S_{2z})|01\rangle
+ (v_{1L}c_L^\dagger+v_{1R}c_R^\dagger)d_1|11\rangle=E|01\rangle
\eeq
Now ignore the $d_2$ term and keep $d_1$ representing $\hat t^\dagger$,
\beq{21}
|01\rangle=\frac{1}{E-\Delta_2-\h_0}\left[(v_{1L}c_L^\dagger+v_{1R}c_R^\dagger )d_1|11\rangle
+d_2^\dagger \hat t^\dagger d_1 |10\rangle\right]
\eeq
Finally $|10\rangle,|01\rangle$ can be written in terms of $|11\rangle$ using their leading terms and then
the 4th line of (\ref{e16}) can be written in terms of $|11\rangle$ to identify the effective Hamiltonian,
\beq{22}
&&\{d_2^\dagger (v_{2L}^*c_L+v_{2R}^*\hat u^\dagger c_R)\frac{1}{E-\Delta_1-\h_0}\left[  (v_{2L}c_L^\dagger+v_{2R}c_R^\dagger \hat u)d_2+d_1^\dagger \hat t d_2 \frac{1}{E-\Delta_2-\h_0}(v_{1L}c_L^\dagger+v_{1R}c_R^\dagger )d_1\right]
\nonumber\\
&&+d_1^\dagger(v_{1L}^*c_L+v_{1R}^*c_R) \frac{1}{E-\Delta_2-\h_0}\left[(v_{1L}c_L^\dagger+v_{1R}c_R^\dagger )d_1+d_2^\dagger \hat t^\dagger d_1\frac{1}{E-\Delta_1-\h_0}  (v_{2L}c_L^\dagger+v_{2R}c_R^\dagger \hat u)d_2\right] \nonumber\\
&&+[\h_0+d_1^\dagger(\Delta_1+g_1S_{1z})d_1+ d_2^\dagger (\Delta_2+g_2S_{2z})d_2]\}|11\rangle=E|11\rangle
\eeq
The $\hat t$ terms yield a product of ${\bf S}_1,{\bf S}_2$ and 4 fermion operators with a
coefficient $\sim \frac{v^4}{\Delta^3}$ which is smaller than even the 2nd order of the other terms, hence is neglected. If $\hat t\rightarrow t$, a  c-number tunneling term in the original Hamiltonian with $t\approx v$, then it yields $\frac{v^2t}{\Delta^2}c_L^\dagger{\bm \sigma}c_R\cdot {\bf S}_1\times
{\bf S}_2$ and similar terms with $\hat u$. To 4th order it is much smaller than the terms that we keep ($J_1^2J_2^2$), while to 2nd order it does not give a resonance.

We note that when all $U,|\Delta_1|,|\Delta_2|$ are finite and large, a process of tunneling from e.g. the first neutral dot to the second one yields a term $J_{12}{\bf S}_1\cdot{\bf S}_2$ with $J_{12}\approx \frac{t^2}{U+|\Delta_1|}$. This yields higher order corrections to the transport, however it may affect the eigenfrequencies of the 2-dot system. We assume here that this effect is negligible; in particular in the scenario that one spin is on the tip and the other on the surface we have $v_{1L}\approx v_{2R}\approx t\ll v_{2L}\approx v_{1R}$, hence $J_{12}\sim t^2\ll J_i\sim v^*_{iL}v_{iR}\,\, (i=1,2)$, $J_i$ being the exchange terms responsible for the resonances (see \eqref{e24} below).

To identify the result in terms of spin operators we note an identity for spinors $c_\alpha$ on either lead $\alpha=R,L$ and spinors $d_i$ on either dot $i=1,2$ with spin operators ${\bf S}_i=\half d_i^\dagger{\bm \sigma}d_i$ defined on singly occupied dots,
\beq{23}
(d_i^\dagger\hat u_1 c_\alpha)(c_\beta^\dagger \hat u_2 d_i)=-c_\beta^\dagger(\hat u_2{\bm \sigma}\hat u_1^\dagger\cdot {\bf S}_i+\half)c_\alpha +const
\eeq
which can be shown by using rotated fermions, e.g. $\tilde c=\hat u^\dagger c$, also $\hat u_1,\,\hat u_2$ are either $\hat u$ or $1 \,(i=1,2)$.
Using then $E=\h_0+\Delta_1+\Delta_2$ and assuming $|\Delta_1|,|\Delta_2|\gg eV_\alpha, \epsilon_{k\alpha}$ we obtain
\beq{24}
\h_{eff}&=&\h_0+g_1S_{1z}+g_2S_{z2}+J_{1L}c_L^\dagger({\bm \sigma}\cdot {\bf S}_1+\half)c_L+J_{1R}c_R^\dagger({\bm \sigma}\cdot {\bf S}_1+\half)c_R\nonumber\\
&&+J_{2L}c_L^\dagger({\bm \sigma}\cdot {\bf S}_2+\half)c_L+J_{2R}c_R^\dagger(\hat u{\bm \sigma}\hat u^\dagger\cdot {\bf S}_2+\half)c_R\nonumber\\
&&+[2J_1c_R^\dagger( {\bm \sigma}\cdot {\bf S}_1+\half)c_L +h.c.]+[2J_2c_R^\dagger(\hat u {\bm \sigma}\cdot {\bf S}_2+\half\hat u)c_L+h.c.]
\nonumber\\
 &&J_{iL}= \frac{|v_{iL}|^2}{\Delta_i},\qquad J_{iR}= \frac{|v_{iR}|^2}{\Delta_i},\qquad 2J_i=\frac{v_{iL}^*v_{iR}}{|\Delta_i|},\qquad\qquad i=1,2
\eeq
We note that $\Delta_i\,(i=1,2)$ multiply
$\sum_\sigma d_{i\sigma}^\dagger d_{i\sigma}=1$ in the $|11\rangle$ subspace, hence these are constant terms and do not participate in the excitation spectrum (\ref{e24}).

It is important to note that without the spin-orbit effect, i.e. if $\hat u=1$, then the total z component spin is conserved. This is seen by the scalar product ${\bm \sigma}\cdot {\bf S}_i$ that
for any spin flip of the dot involves an opposite spin flip of the tunneling electrons.
 In fact, the original form \eqref{e15} commutes with $S_z$, i.e. independent of the Scrieffer-Wolff procedure.

The conservation of $S_z$ implies no resonance at $g_1$ or $g_2$ in the noise spectrum since a resonance with a single spin flip implies that $S_z$ is not conserved. Hence a spin-orbit term is essential for the observation of these resonances. Two opposite spin flips are allowed, i.e. the resonance at $|g_1-g_2|$ can be seen even without spin-orbit.
\\
\section{Current Noise via Keldysh -- double QD}

\subsection{Tools}

Assume $V>0$ so that current flows to the right, i.e. electrons flow to the left and $\dot N_L>0$. With the choice $e=|e|>0$, the current operator is obtained by
\beq{25}
j&=&e\dot N_L=-ie[N_L,\h_{eff}]=-ie2J_2\sum [c_{Lk\sigma}^\dagger c_{Lk\sigma},c_{Rk'\sigma'}^\dagger c_{Lk''\sigma''}](\hat u {\bm \sigma}+\half\hat u)_{\sigma'\sigma''}\cdot {\bf S}_2+h.c.
\nonumber \\&&+(\hat u\rightarrow 1; 2\rightarrow 1)
=ie2J_2\sum c_{Rk'\sigma'}^\dagger c_{Lk\sigma}(\hat u {\bm \sigma}+\half\hat u)_{\sigma'\sigma''}\cdot {\bf S}_2\delta_{\sigma\sigma''}\delta{kk''}+h.c. +(\hat u\rightarrow 1; 2\rightarrow 1)\nonumber\\
&=&ie2J_2c_R^\dagger(\hat u {\bm \sigma}\cdot {\bf S}_2+\half\hat u)c_L-ie2J_1c_R^\dagger({\bm \sigma}\cdot {\bf S}_1+\half)c_L+h.c.
\eeq
Since spin operators do not satisfy Wick's theorem it is convenient to represent them
by Abrikosov pseudofermions \cite{abrikosov,moca} $f^\dagger_{i\gamma},\,f_{i\gamma}$ for spin site $i=1,2$ and spin component $\gamma=\pm$,
 \beq{26}
 {\bf S}_i\rightarrow \half\sum_{\gamma,\delta} f^\dagger_{i\gamma}\bm{\tau}_{\gamma,\delta}f_{i\delta},\qquad \sum_\gamma f^\dagger_{i\gamma}f_{i\gamma}=1
 \eeq
where $\bm\tau$ are the Pauli matrices. To the 4th order that we need one can in fact use spin propagators, since all diagrams involve pseudofermion closed loops (e.g. Fig. \ref{diagram1}), yet we present the calculation with pseudofermions, anticipating higher order extensions in the future.
The operators $c^\dagger,\,c$ are replaced on the Keldysh contour (e.g. Fig. \ref{diagram1})  by Grasmann variables $\bar\psi_{\alpha s}=(\bar\psi_{\alpha s+},\bar\psi_{\alpha s-})$ where $\alpha=R,L$, while $\bar f_{i \gamma}=(\bar f_{i \gamma+},\bar f_{i \gamma-})$. For evaluating the leading order in the noise we keep only the transfer terms with spin,
\beq{27}
{\cal H}_0&=& \sum_{ks\alpha}(\epsilon_{k\alpha}+eV_\alpha)c^\dagger_{ks\alpha}c_{ks\alpha}
+\sum_{i\gamma}\lambda_{i\gamma} f^\dagger_{i\gamma}f_{i\gamma}\qquad \lambda_{i\gamma}=\lambda_0+\half \gamma \nu_i\nonumber\\
S_0&=&\sum_{\alpha ks} \bar \psi_{ks\alpha } \hat G_{k\alpha}^{-1} \psi_{ks\alpha }+\sum_{i\gamma}\bar f_{i\gamma} \hat F^{-1}_{i\gamma}f_{i\gamma}\nonumber\\
S_{eff}&=& S_0+\sum_\pm \pm[J_1\bar \psi_{R\pm}{\bm\sigma}\psi_{L\pm}\cdot \bar f_{1\pm}\bm\tau f_{1\pm}
+J_2\bar \psi_{R\pm}\hat u{\bm\sigma}\psi_{L\pm}\cdot \bar f_{2\pm}\bm\tau f_{2\pm}+h.c.]
\eeq
where $\nu_1=g_1\mu_BB,\,\nu_2=g_2\mu_BB$ are the Larmor frequencies and $\psi_{\alpha \pm}$ in the interaction term are at $x=0$, i.e. $\psi_{\alpha \pm}= \sum_k \psi_{k\alpha \pm}$. Eventually $\lambda_0\rightarrow \infty$ to enforce single occupancy of each of the $f$ psudofermions. Note that $G_{k\alpha}$ are spin independent. The current is
\beq{28}
j=ie\sum_\pm [J_1\bar\psi_{R\pm}{\bm \sigma}\psi_{L\pm}\cdot \bar f_{1\pm}\bm\tau f_{1\pm}+ J_2\bar\psi_{R\pm}(\hat u {\bm \sigma})\psi_{L\pm}\cdot \bar f_{2\pm}\bm\tau f_{2\pm}]+h.c.
\eeq

The Green's functions on the Keldysh contour is given by \cite{kamenev}
\beq{30}
i\hat G(t)=\left(\begin{array}{cc} \langle T\psi_+\bar\psi_+\rangle &
\langle \psi_+\bar\psi_-\rangle \\ \langle \psi_-\bar\psi_+\rangle & \langle \bar T\psi_-\bar\psi_-\rangle
\end{array}\right)=i\left(\begin{array}{cc}G_t(t) & G^<(t)\\
G^>(t) &  G_{\bar t}(t) \end{array}\right)
\eeq
where, defining $t=t_1-t_2$,
\beq{31}
&& iG^>(t_1,t_2)=(1-f(\epsilon))\eexp{-i\epsilon t}\nonumber\\
&&iG^<(t_1,t_2)=-f(\epsilon)\eexp{-i\epsilon t}
\nonumber\\
&&iG^t(t_1,t_2)=\theta(t)(1-f(\epsilon))\eexp{-i\epsilon t}
-\theta(-t)f(\epsilon)\eexp{-i\epsilon t} \nonumber\\
&&iG^{\bar t}(t_1,t_2)=\theta(-t)(1-f(\epsilon))\eexp{-i\epsilon t}
-\theta(t)f(\epsilon)\eexp{-i\epsilon t}
\eeq
Similarly, for the pseudofermions,
\beq{32}
iF_{i\gamma}^>(t_1,t_2)&=&[1-f(\lambda_{i\gamma})]\eexp{-i\lambda_{i\gamma}t}=
\eexp{-i\lambda_{i\gamma}t}+O(\eexp{-\beta\lambda_0})\nonumber\\
iF^<_{i\gamma}(t_1,t_2)&=&-f(\lambda_{i\gamma})\eexp{-i\lambda_{i\gamma}t}
=-\eexp{-\beta\lambda_{i\gamma}}\eexp{-i\lambda_{i\gamma}t}+O(\eexp{-2\beta\lambda_0})\nonumber\\
iF_{i\gamma}^t(t_1,t_2)&=&\theta(t)\eexp{-i\lambda_{i\gamma}t}-\theta(-t)
\eexp{-\beta\lambda_{i\gamma}-i\lambda_{i\gamma}t}+O(\eexp{-2\beta\lambda_0})\nonumber\\
iF_{i\gamma}^{\bar t}(t_1,t_2)&=&\theta(-t)\eexp{-i\lambda_{i\gamma}t}-\theta(t)
\eexp{-\beta\lambda_{i\gamma}-i\lambda_{i\gamma}t}+O(\eexp{-2\beta\lambda_0})
\eeq
where in each term only the leading order in $\eexp{-\beta\lambda_0}$ is kept. We expect that eventually the kept terms will cancel with the normalization $\sum_{i\gamma}\eexp{-\beta\lambda_{i\gamma}}$.\\

\subsection{Diagram Rules}

Vertices: each vertex carries a factor of either $\pm ieJ_1\bm\sigma\cdot\bm\tau$ or
$\pm ieJ_2(\hat u \bm\sigma)\cdot\bm\tau$; The sign is $+$ for the current source of $c^\dagger_Rc_L$ and $-$ for $c^\dagger_Lc_R$ and is the same on both contours (to generate the quantum part), is $+$ for interaction vertex on the $+$ contour and $-$ on the $-$ contour (recall $Z=\eexp{iS}$).

 The n-th order has n+2 vertices, two of which correspond to current terms (sources) with $t_i$ external variables, while n vertices have internal $t_i$ that are integrated upon. The first external time $t$ is on the $+$ contour and the second $t'$ is on the $-$ contour; this generates $S(t,t')=\langle j_+(t)j_-(t')\rangle$. [This is sometimes \cite{moca} denoted as $S^<(t,t')$; the symmetrized form is obtained, after Fourier, by $S(\omega)+S(-\omega)$].
Since the interactions and current terms have the same form one can choose in each diagram various combinations for the external vertices. Hence it is more compact to define diagrams in real time and later identify the various options for the external frequency.

Lines: Greens' functions are full (dashed) lines for a $\psi\bar\psi$ ($f\bar f$) product, with the direction chosen \cite{kamenev} to be in the direction from $\bar\psi(t_1)$ to $\psi(t_2)$ ($\bar f(t_1)$ to $f(t_2)$), the Greens' function argument is $t_2-t_1$. The direction is conserved across each vertex (particle conservation). Each vertex connects to 4 lines -- 2 outgoing full and dashed, 2 incoming full and dashed. There are 4 types of Greens' functions:

A line connecting vertices within the $+$ contour is $-iG^t$ ($-iF^t$).

A line connecting vertices within the $-$ contour is $-iG^{\bar t}$ ($-iF^{\bar t}$).

A line connecting $+$ to $-$ contours, i.e. goes down, is $-iG^>$ ($-iF^>$).

A line connecting $-$ to $+$ contours, i.e. goes up, is $-iG^<$ ($-iF^<$).

A closed fermion or pseudofermion loop has an additional $-$ sign (one $\bar\psi\psi$ in the loop needs to be reordered).

Finally, to generate a single $\eexp{-\beta\lambda_{i\gamma}}$ factor (projection on a singly occupied pseudofermion space) need one, and only one, of 3 factors: a single $F^>$, a single $F^t$ with $t<0$ or a single $F^{\bar t}$ with $t>0$. (Note that $\eexp{-\beta\lambda_{i\gamma}}$ in $F^>$ can be neglected since to close a loop an $F^<$ must be present). Hence the pseudofermion lines must be time ordered along the Keldysh contour.

\subsection{Traces}

Consider here traces needed in the following for the quantity $\tilde I_4$ (which implicitly depends on $\gamma_1,\gamma_2,\gamma_3,\gamma_4$),
\beq{206}
&&\tilde I_4=\tr[\sigma^i\sigma^k\hat u^\dagger]\tr[ \sigma^j\sigma^l\hat u^\dagger]
\tau^i_{\gamma_1\gamma_2}\tau^j_{\gamma_2\gamma_1}\tau^k_{\gamma_3\gamma_4}\tau^l_{\gamma_4\gamma_3}
\eeq
The traces involve the electron spin operators $\sigma^i$ while $\tau^j$ are matrices for the psudospin operators. Using the definition $\hat u=\eexp{i\sigma_z\phi}\eexp{\half i\sigma_y \theta}$ we obtain
\beq{207}
&&\tr[(\sigma^i)^2\hat u^\dagger]=2\cos\half\theta\cos\phi\nonumber\\
&&\tr[\sigma^x\sigma^y\hat u^\dagger]=2\cos\half\theta\sin\phi\nonumber\\
&&\tr[\sigma^x\sigma^z\hat u^\dagger]=-2\sin\half\theta\cos\phi\nonumber\\
&&\tr[\sigma^y\sigma^z\hat u^\dagger]=2\sin\half\theta\sin\phi\nonumber\\
&&\tau^z_{\gamma_1\gamma_2}\tau^z_{\gamma_2\gamma_1}=\delta_{\gamma_1,\gamma_2}\nonumber\\
&&\tau^x_{\gamma_1\gamma_2}\tau^x_{\gamma_2\gamma_1}=\tau^y_{\gamma_1\gamma_2}\tau^y_{\gamma_2\gamma_1}
=\delta_{\gamma_1,-\gamma_2}\nonumber\\
&&\tau^z_{\gamma_1\gamma_2}\tau^x_{\gamma_2\gamma_1}=\tau^z_{\gamma_1\gamma_2}\tau^y_{\gamma_2\gamma_1}
=0\nonumber\\
&&\tau^x_{\gamma_1\gamma_2}\tau^y_{\gamma_2\gamma_1}=
-\tau^y_{\gamma_1\gamma_2}\tau^x_{\gamma_2\gamma_1}=i\delta_{\gamma_1,-\gamma_2}\mbox{sign}\gamma_1
\eeq
Define next $\tilde I_4^{ik}$ where indices $j,l$ are summed as indicated on the right,
\beq{208}
\tilde I^{xx}=&2\cos\half\theta\cos\phi[2\cos\half\theta\cos\phi\delta_{\gamma_1,-\gamma_2}
\delta_{\gamma_3,-\gamma_4}\qquad\qquad\qquad\qquad\qquad \qquad &j=x,l=x\nonumber\\
&+2\cos\half\theta\sin\phi
\delta_{\gamma_1,-\gamma_2}i\delta_{\gamma_3,-\gamma_4}\mbox{sign}\gamma_3
-2\cos\half\theta\sin\phi
i\delta_{\gamma_1,-\gamma_2}\mbox{sign}\gamma_1\delta_{\gamma_3,-\gamma_4}\qquad &j=x,l=y;
\,\,j=y,l=x\nonumber\\
&\qquad +2\cos\half\theta\cos\phi\, i\delta_{\gamma_1,-\gamma_2}\mbox{sign}\gamma_1
i\delta_{\gamma_3,-\gamma_4}\mbox{sign}\gamma_3]\qquad\qquad\qquad\qquad\qquad\qquad  &j=y,l=y\nonumber\\
\tilde I^{yy}=&2\cos\half\theta\cos\phi[2\cos\half\theta\cos\phi(-i)\delta_{\gamma_1,-\gamma_2}
\mbox{sign}\gamma_1(-i)\delta_{\gamma_3,-\gamma_4}\mbox{sign}\gamma_3\qquad\qquad\qquad  &j=x,l=x\nonumber\\
&+2\cos\half\theta\sin\phi(-i)
\delta_{\gamma_1,-\gamma_2}\mbox{sign}\gamma_1\delta_{\gamma_3,-\gamma_4}
-2\cos\half\theta\sin\phi
\delta_{\gamma_1,-\gamma_2}(-i)\delta_{\gamma_3,-\gamma_4}\mbox{sign}\gamma_3\qquad &j,l=x,y;
\,\,=y,x\nonumber\\
&\qquad +2\cos\half\theta\cos\phi\, \delta_{\gamma_1,-\gamma_2}
\delta_{\gamma_3,-\gamma_4}]=\tilde I_4^{xx}\qquad\qquad\qquad\qquad\qquad  &j=y,l=y\nonumber\\
\tilde I^{zz}=&2\cos\half\theta\cos\phi[2\cos\half\theta\cos\phi\,
\delta_{\gamma_1\gamma_2}\delta_{\gamma_3\gamma_4}]\qquad\qquad\qquad\qquad\qquad &j=z,l=z\nonumber\\
&\tilde I^{xy}=2\cos\half\theta\sin\phi[2\cos\half\theta\cos\phi\delta_{\gamma_1,-\gamma_2}(-i)
\delta_{\gamma_3,-\gamma_4}\mbox{sign}\gamma_3\qquad &j=x,l=x\nonumber\\
&+2\cos\half\theta\sin\phi\delta_{\gamma_1,-\gamma_2}\delta_{\gamma_3,-\gamma_4}-2\cos\half\theta\sin\phi\,
i\delta_{\gamma_1,-\gamma_2}\mbox{sign}\gamma_1(-i)\delta_{\gamma_3,-\gamma_4}\mbox{sign}\gamma_3\qquad &j,l=x,y;\,\,=y,x\nonumber\\
&+2\cos\half\theta\cos\phi\,i\delta_{\gamma_1,-\gamma_2}\mbox{sign}\gamma_1\delta_{\gamma_3,-\gamma_4}]
\qquad\qquad\qquad\qquad\qquad &j=y,l=y\nonumber\\
\tilde I^{yx}=&-2\cos\half\theta\sin\phi[2\cos\half\theta\cos\phi(-i)\delta_{\gamma_1,-\gamma_2}
\mbox{sign}\gamma_1 \delta_{\gamma_3,-\gamma_4}\qquad &j=x,l=x\nonumber\\
&+2\cos\half\theta\sin\phi(-i)\delta_{\gamma_1,-\gamma_2}\mbox{sign}\gamma_1\,i
\delta_{\gamma_3,-\gamma_4}\mbox{sign}\gamma_3-2\cos\half\theta\sin\phi\,
\delta_{\gamma_1,-\gamma_2}\delta_{\gamma_3,-\gamma_4}\qquad &j,l=x,y;\,\,=y,x\nonumber\\
&+2\cos\half\theta\cos\phi\,\delta_{\gamma_1,-\gamma_2}\,i\delta_{\gamma_3,-\gamma_4}
\mbox{sign}\gamma_3]=\tilde I^{xy}
\qquad\qquad\qquad\qquad\qquad &j=y,l=y\nonumber\\
\tilde I^{xz}=&-2\sin\half\theta\cos\phi[-2\sin\half\theta\cos\phi\delta_{\gamma_1,-\gamma_2}
\delta_{\gamma_3\gamma_4}
+2\sin\half\theta\sin\phi\,i\delta_{\gamma_1,-\gamma_2}\mbox{sign}\gamma_1
\delta_{\gamma_3\gamma_4}]\,\,\, &j,l=x,z;\,y,z\nonumber\\
\tilde I^{zx}=&2\sin\half\theta\cos\phi[2\sin\half\theta\cos\phi\delta_{\gamma_1\gamma_2}
\delta_{\gamma_3,-\gamma_4}
-2\sin\half\theta\sin\phi\,\delta_{\gamma_1\gamma_2}\,i
\delta_{\gamma_3,-\gamma_4}\mbox{sign}\gamma_3]\,\,\, &j,l=z,x;\,z,y\nonumber\\
\tilde I^{yz}=&2\sin\half\theta\sin\phi[-2\sin\half\theta\cos\phi(-i)\delta_{\gamma_1,-\gamma_2}
\mbox{sign}\gamma_1 \delta_{\gamma_3\gamma_4}
2\sin\half\theta\sin\phi\,\delta_{\gamma_1,-\gamma_2}\,
\delta_{\gamma_3\gamma_4}]\,\,\, &j,l=x,z;\,y,z\nonumber\\
\tilde I^{zy}=&-2\sin\half\theta\sin\phi[2\sin\half\theta\cos\phi\delta_{\gamma_1\gamma_2}
(-i) \delta_{\gamma_3,-\gamma_4}\mbox{sign}\gamma_3
-2\sin\half\theta\sin\phi\,\delta_{\gamma_1\gamma_2}\,
\delta_{\gamma_3,-\gamma_4}]\,\,\, &j,l=z,x;\,z,y\nonumber\\
\eeq
Collecting all terms, i.e. summation on $i,k$ yields
\beq{209}
&&\tilde I_4=4\delta_{\gamma_1\gamma_2}\delta_{\gamma_3,\gamma_4}\cos^2\half\theta\cos^2\phi
+4(\delta_{\gamma_1\gamma_2}\delta_{\gamma_3,-\gamma_4}+\delta_{\gamma_1,-\gamma_2}
\delta_{\gamma_3\gamma_4})\sin^2\half\theta\nonumber\\
&&+8\delta_{\gamma_1,-\gamma_2}\delta_{\gamma_3,-\gamma_4}(1-\mbox{sign}\gamma_1\mbox{sign}\gamma_3)
\cos^2\half\theta
\eeq

\subsection{Order n=0}

   \begin{figure}[b]  \centering
\includegraphics [width=.5\textwidth]{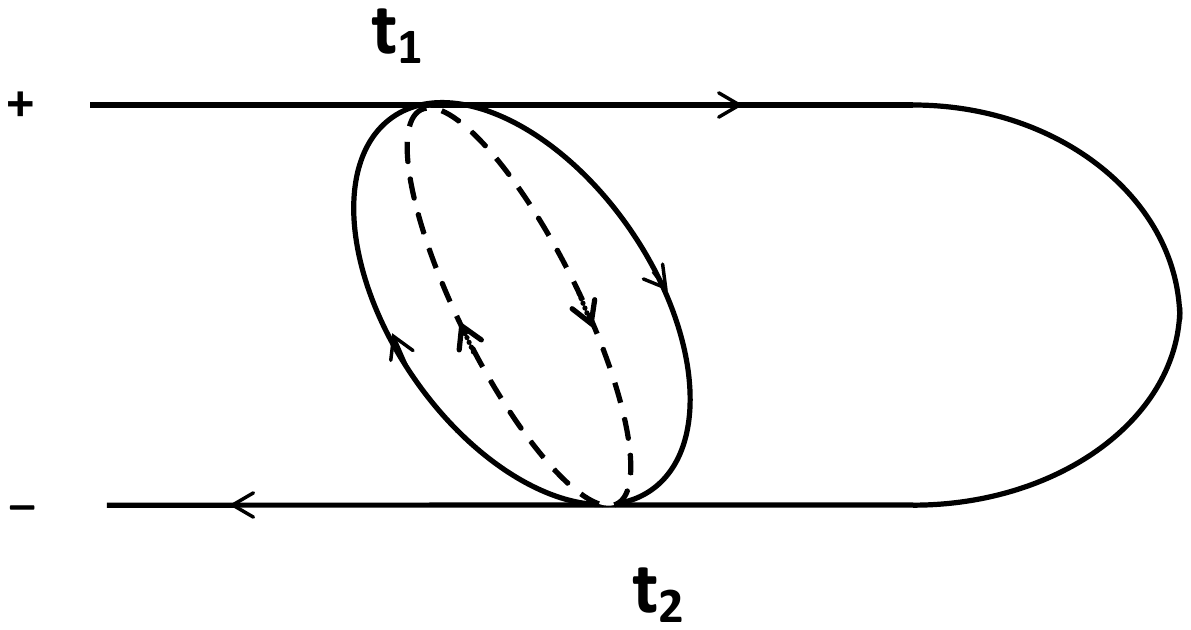}
\caption{Diagram 1}
\label{diagram1}
\end{figure}

To order the current operators we choose the first on contour $+$ at $t_1$, the second on contour $-$ at $t_2$, see Fig. \ref{diagram1}. The signs involve $-i$ on the $R\rightarrow L$ and $i$ for the $L\rightarrow R$ vertices.
Hence the noise is (the numbers on top of each field are paired to show contractions, terms with $R\leftrightarrow L$ and with $1\rightarrow 2$ to be added below))
\beq{33}
&& S_1(t_1,t_2)=\langle j_+(t_1)j_-(t_2)\rangle\nonumber\\
&& =e^2J_1^2\left\langle \stackrel{1}{\bar\psi}_{R+}(t_1)\bm\sigma_{\alpha\beta}
\stackrel{2}{\psi}_{L+}(t_1)\cdot \stackrel{3}{\bar f}_{1+}(t_1)\bm\tau_{\gamma\delta}\stackrel{4}{f}_{1+}(t_1)\stackrel{2}{\bar\psi}_{L-}(t_2)
\bm\sigma_{\alpha'\beta'}\stackrel{1}{\psi}_{R-}(t_2)
\stackrel{4}{\bar f}_{1-}(t_2)\bm\tau_{\gamma'\delta'}\stackrel{3}{f}_{1-}(t_2)\right\rangle\nonumber\\
&&=e^2J_1^2\sum_{\alpha,...,\delta'}(-i)G_R^>(t_2-t_1)iG_L^<(t_1-t_2)\delta_{\alpha\beta'}
\delta_{\beta\alpha'}(-i)F_{1\gamma}^>(t_2-t_1)iF_{1\delta}^<(t_1-t_2)\delta_{\gamma\delta'}
\delta_{\delta\gamma'}\times\nonumber\\
&& \bm\sigma_{\alpha\beta}\cdot\bm\tau_{\gamma\delta}
\bm\sigma_{\beta\alpha}\cdot\bm\tau_{\delta\gamma}
\eeq
since $G$ are diagonal and spin independent, while $F$ are diagonal and spin dependent.
Using Pauli matrix identities
\beq{34}
&&\sum_{\alpha\beta}\sigma^i_{\alpha\beta}\sigma^j_{\beta\alpha}=
\sum_\alpha(\sigma^i\sigma^j)_{\alpha\alpha}=2\delta_{ij}\,,\qquad
\tau^z_{\gamma\delta}\tau^z_{\delta\gamma}=\delta_{\gamma\delta},\qquad \tau^x_{\gamma\delta}\tau^x_{\delta\gamma}=\tau^y_{\gamma\delta}\tau^y_{\delta\gamma}
=\delta_{\gamma,-\delta}\qquad \Rightarrow\nonumber\\
&&\sum_{i\gamma\delta}\tau^i_{\gamma\delta}\tau^i_{\delta\gamma}F^>_{1\gamma}F^<_{1\delta}
=\sum_\gamma[2F^>_{1\gamma}F^<_{1,-\gamma}+F^>_{1\gamma}F^<_{1\gamma}]
\eeq
Therefore
\beq{35}
&& S_1(t=t_1-t_2)=e^2J_1^2G_R^>(-t)G^<_L(t)\sum_\gamma[2F^>_{1\gamma}(-t)F^<_{1,-\gamma}(t)
+F^>_{1\gamma}(-t)F^<_{1\gamma}(t)]\\
&&S_1(\omega)=2\pi e^2J_1^2\int_{\epsilon_R,\epsilon_L}[1-f_R(\epsilon_R)]f_L(\epsilon_L) \sum_\gamma[2\eexp{-\beta\lambda_{1,-\gamma}}\delta(\omega+\epsilon_R-\epsilon_L+\lambda_{1\gamma}-\lambda_{1,-\gamma})
+\eexp{-\beta\lambda_{1,\gamma}}\delta(\omega+\epsilon_R-\epsilon_L)]\nonumber
\eeq
where $S_1(\omega)=\int_tS_1(t)\eexp{i\omega t}$. The voltage is assumed large $eV\gg
\lambda_{i\gamma},\omega$ so that $\epsilon_L\approx\epsilon_R$ and the integration range for which $1-f_R(\epsilon_R)=1,\, f_L(\epsilon_L)=1$  gives $N(0)[eV+O(\lambda_{i\gamma},\omega)]$, where $N(0)$ is the density of states (per spin) in the leads. The term
with $R\leftrightarrow L$ has $[1-f_L(\epsilon)]f_R(\epsilon)$ is much smaller by order $k_BT/eV$ and is neglected. Finally, dividing by the normalization $\sum_\gamma \eexp{-\beta\lambda_{1,\gamma}}$ and then adding the term with $J_2^2$ we obtain
\beq{36}
S_1(\omega)=6\pi e^2(J_1^2+J_2^2)eVN^2(0)
\eeq
Note that the DC current is
\beq{361}
&&j_{DC}=ie\langle[J_1\bar\psi_{R+}{\bm \sigma}\psi_{L+}\cdot \bar f_{1+}\bm\tau f_{1+}+ J_2\bar\psi_{R+}(\hat u {\bm \sigma})\psi_{L+}\cdot \bar f_{2+}\bm\tau f_{2+}-h.c.]_{t_1}\times\nonumber\\
&&(-i)[J_1\bar\psi_{L-}{\bm \sigma}\psi_{R-}\cdot \bar f_{1-}\bm\tau f_{1-}+ J_2\bar\psi_{L-} {\bm \sigma}\hat u^\dagger\psi_{R-}\cdot \bar f_{2-}\bm\tau f_{2-}+h.c.]_{t_2}\rangle=S_1(0)/e
\eeq

\subsection{Order n=2}

We look for interference between two spin, the lowest order is $n=2$, i.e. 4 vertices with at least one on each contour. We keep only diagrams that have separate electron loops, other types do not lead to resonances (see Fig. 2b in the main text), thus the diagrams in Fig. \ref{diagram2} are neglected.
      \begin{figure*}  \centering
\includegraphics [width=.45\textwidth]{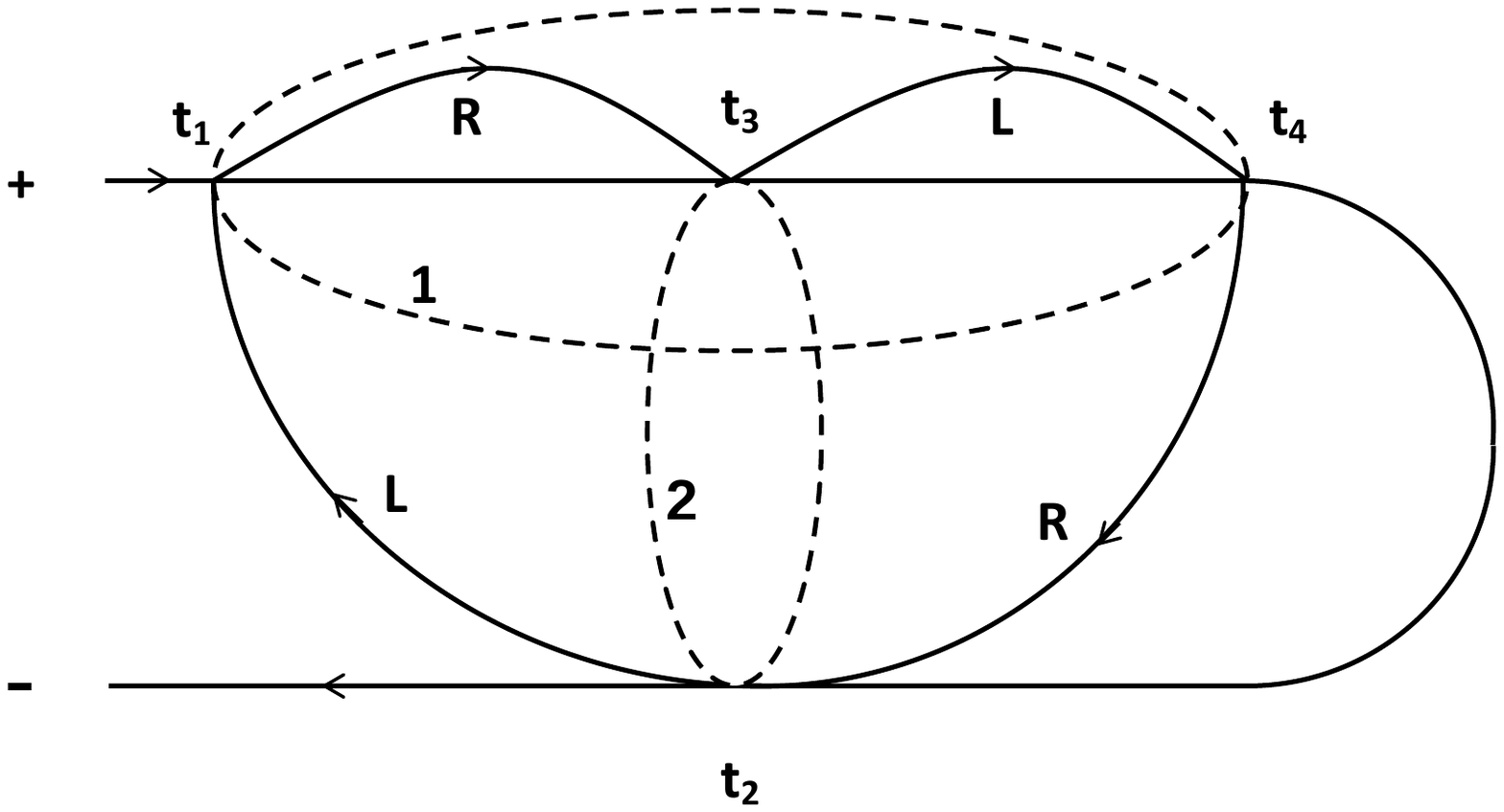}
\includegraphics [width=.45\textwidth]{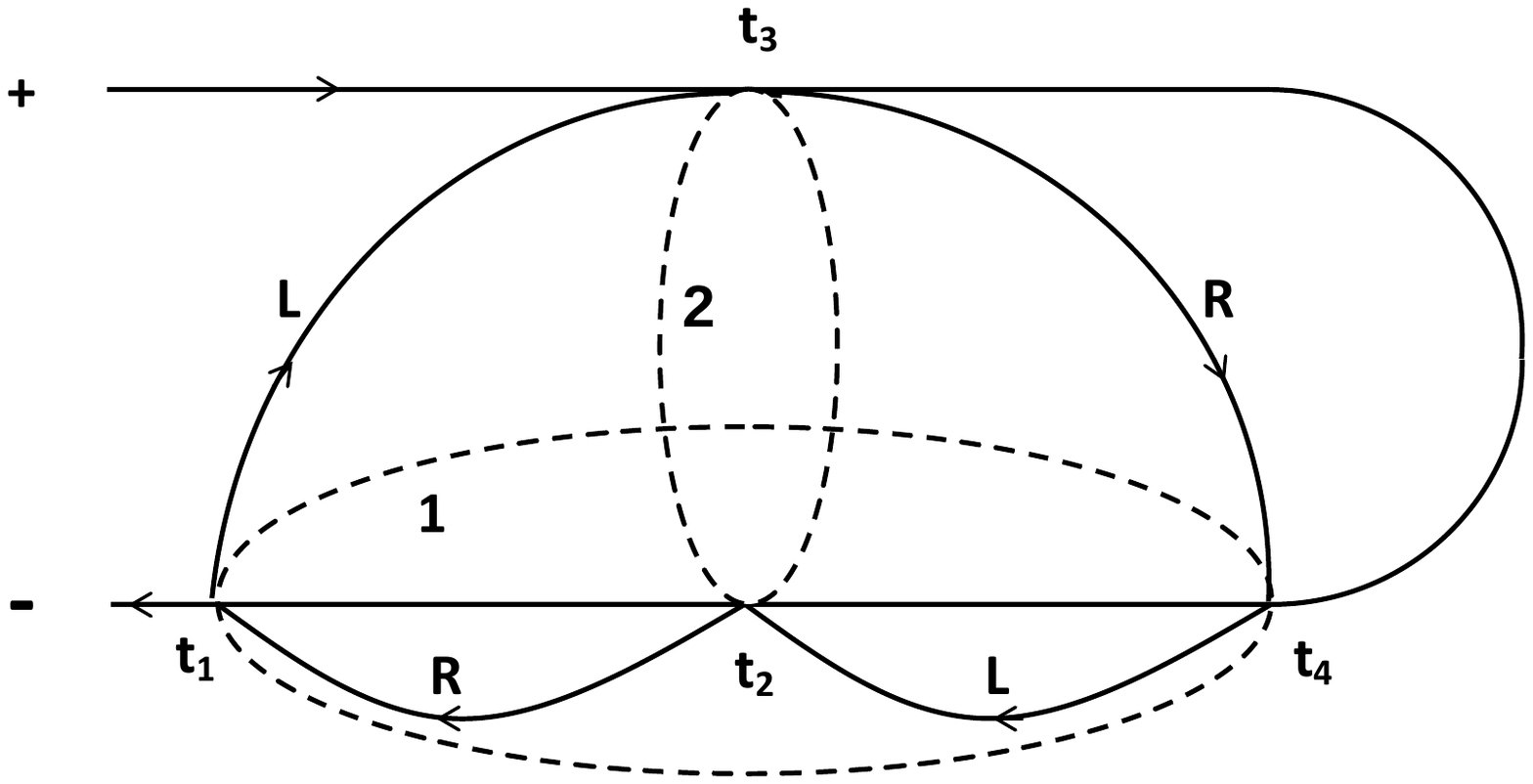}
\caption{Diagrams 2 and 3}
\label{diagram2}
\end{figure*}

     \begin{figure}  \centering
\includegraphics [width=.5\textwidth]{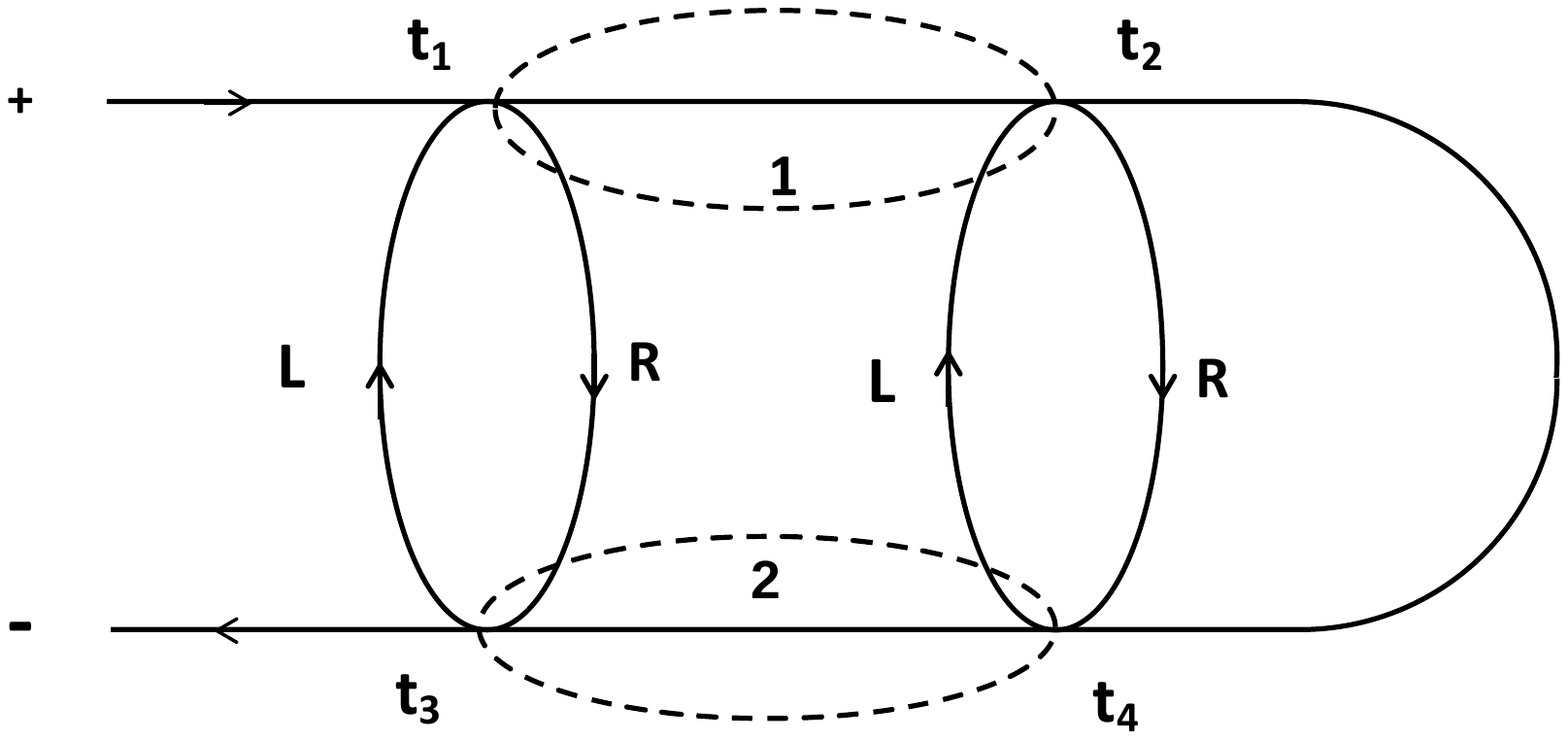}
\caption{Diagram 4}
\label{diagram4}
\end{figure}

Consider first the diagram of Fig. \ref{diagram4}; overall sign is $-i^2$ from currents and $-i^2$ from interactions,
\beq{54}
&&S_4=i^4e^2J_1^2J_2^2\left\langle [\stackrel{1}{\bar\psi}_{R+}\bm\sigma_{\alpha_1\beta_1}\stackrel{2}{\psi}_{L+}\cdot \stackrel{3}{\bar f}_{1+}\bm\tau_{\gamma_1\delta_1}\stackrel{4}{f}_{1+}]_{t_1}[\stackrel{5}{\bar\psi}_{R+}\bm
\sigma_{\alpha_2\beta_2}\stackrel{6}{\psi}_{L+}\cdot\stackrel{4}{\bar f}_{1+}\bm\tau_{\gamma_2\delta_2}\stackrel{3}{f}_{1+}]_{t_2}\right.\nonumber\\
&&\left.\qquad [\stackrel{2}{\bar\psi}_{L-}\bm(\sigma \hat u^\dagger)_{\alpha_3\beta_3}\stackrel{1}{\psi}_{R-}\cdot\stackrel{7}{\bar f}_{2-}\bm\tau_{\gamma_3\delta_3}\stackrel{8}{f}_{2-}]_{t_3} [\stackrel{6}{\bar\psi}_{L-}(\bm\sigma \hat u^\dagger)_{\alpha_4\beta_4}\cdot\stackrel{5}{\psi}_{R-}\stackrel{8}{\bar f}_{2-}\bm\tau_{\gamma_4\delta_4} \stackrel{7}{f}_{2-}]_{t_4}\right\rangle\nonumber\\
&&=e^2J_1^2J_2^2(-i)G^>_R(t_3-t_1)iG_L^<(t_1-t_3)(-i)G_R^>(t_4-t_2)iG_L^<(t_2-t_4)\times\nonumber\\
&&\qquad (-i)F^t_{1\gamma_1}(t_2-t_1)iF^t_{1\gamma_2}(t_1-t_2)(-i)F^{\bar t}_{2\gamma_3}(t_4-t_3)
iF_{2\gamma_4}^{\bar t}(t_3-t_4)\times\nonumber\\
&& \delta_{\alpha_1\beta_3}\delta_{\alpha_3\beta_1}\delta_{\alpha_2\beta_4}\delta_{\alpha_4\beta_2}
\delta_{\gamma_1\delta_2}\delta_{\gamma_2\delta_1}\delta_{\gamma_3\delta_4}\delta_{\gamma_4\delta_3}
\sigma^i_{\alpha_1\beta_1}\tau^i_{\gamma_1\delta_1}
\sigma^j_{\alpha_2\beta_2}\tau^j_{\gamma_2\delta_2}(\sigma^k \hat u^\dagger)_{\alpha_3\beta_3}\tau^k_{\gamma_3\delta_3}
(\sigma^l\hat u^\dagger)_{\alpha_4\beta_4}\tau^l_{\gamma_4\delta_4}\nonumber\\
&&=e^2J_1^2J_2^2\tr[\sigma^i\sigma^k\hat u^\dagger]\tr[ \sigma^j\sigma^l\hat u^\dagger]
\tau^i_{\gamma_1\gamma_2}\tau^j_{\gamma_2\gamma_1}\tau^k_{\gamma_3\gamma_4}\tau^l_{\gamma_4\gamma_3}
\times\nonumber\\
&&(1-f_R(\epsilon_R))\eexp{-i\epsilon_R(t_3-t_1)}(-)f_L(\epsilon_L)\eexp{-i\epsilon_L(t_1-t_3)}
(1-f_R(\epsilon'_R))\eexp{-i\epsilon'_R(t_4-t_2)}(-)f_L(\epsilon'_L)\eexp{-i\epsilon'_L(t_2-t_4)}
\nonumber\\
&& [\theta(t_2-t_1)-\theta(t_1-t_2)\eexp{-\beta\lambda_1}]\eexp{-i\lambda_1(t_2-t_1)}
[\theta(t_1-t_2)-\theta(t_2-t_1)\eexp{-\beta\lambda_2}]\eexp{-i\lambda_2(t_1-t_2)}\nonumber\\
&&[\theta(t_3-t_4)-\theta(t_4-t_3)\eexp{-\beta\lambda_3}]\eexp{-i\lambda_3(t_4-t_3)}
[\theta(t_4-t_3)-\theta(t_3-t_4)\eexp{-\beta\lambda_4}]\eexp{-i\lambda_4(t_3-t_4)}
\eeq
Here $\lambda_1=\lambda^{(1)}_{\gamma_1},\lambda_2=\lambda^{(1)}_{\gamma_2}$ belong to spin 1 while
$\lambda_3=\lambda^{(2)}_{\gamma_3},\lambda_4=\lambda^{(2)}_{\gamma_4}$ belong to spin 2.
\beq{55}
&&I_4=e^2J_1^2J_2^2\tr[\sigma^i\sigma^k\hat u^\dagger]\tr[ \sigma^j\sigma^l\hat u^\dagger]
\tau^i_{\gamma_1\gamma_2}\tau^j_{\gamma_2\gamma_1}\tau^k_{\gamma_3\gamma_4}\tau^l_{\gamma_4\gamma_3}
(1-f_R(\epsilon_R))(1-f_R(\epsilon'_R))f_L(\epsilon_L)f_L(\epsilon'_L)\nonumber\\
&&S_4=I_4\eexp{-i(\epsilon_R-\epsilon_L)(t_3-t_1)-i(\epsilon'_R-\epsilon'_L)(t_4-t_2)-i\lambda_{12}(t_2-t_1)
-i\lambda_{34}(t_4-t_3)}\nonumber\\
&&i\int_{\omega_1}[\frac{\eexp{-\beta\lambda_2}}{\omega_1+i\eta}+
\frac{\eexp{-\beta\lambda_1}}{-\omega_1+i\eta}]\eexp{-i\omega_1(t_2-t_1)}
i\int_{\omega_2}[\frac{\eexp{-\beta\lambda_4}}{\omega_2+i\eta}+
\frac{\eexp{-\beta\lambda_3}}{-\omega_2+i\eta}]\eexp{-i\omega_2(t_3-t_4)}
\eeq
Consider $S_{4a}$ with external $t_1,t_4$ so $\int_{t_1}\eexp{i\omega(t_1-t_4)},\,\int_{t_2,t_3}$ give 3 independent $\delta$ functions with $\epsilon'_{RL}-\lambda_{12}-\omega_1=
-\epsilon_{RL}+\lambda_{34}-\omega_2=\omega+\epsilon_{RL}+\lambda_{12}+\omega_1=0$
where $\epsilon_{RL}=\epsilon_R-\epsilon_L,\,\epsilon'_{RL}=\epsilon'_R-\epsilon'_L$, hence
\beq{57}
&&S_{4a}(\omega)=-I_4 2\pi\delta(\omega+\epsilon_{RL}+\epsilon'_{RL})[\frac{\eexp{-\beta\lambda_2}}
{\epsilon'_{RL}-\lambda_{12}+i\eta}+\frac{\eexp{-\beta\lambda_1}}
{-\epsilon'_{RL}+\lambda_{12}+i\eta}]
[\frac{\eexp{-\beta\lambda_4}}
{-\epsilon_{RL}+\lambda_{34}+i\eta}+\frac{\eexp{-\beta\lambda_3}}
{\epsilon_{RL}-\lambda_{34}+i\eta}]\nonumber\\
&&=-I_4 2\pi\delta(\omega+\epsilon_{RL}+\epsilon'_{RL})[P(\frac{1}{\epsilon'_{RL}-\lambda_{12}})
(\eexp{-\beta\lambda_2}-\eexp{-\beta\lambda_1})-i\pi\delta(\epsilon'_{RL}-\lambda_{12})
(\eexp{-\beta\lambda_2}+\eexp{-\beta\lambda_1})]\nonumber\\
&&\times [P(\frac{1}{-\epsilon_{RL}+\lambda_{34}})
(\eexp{-\beta\lambda_4}-\eexp{-\beta\lambda_3})-i\pi\delta(-\epsilon_{RL}+\lambda_{34})
(\eexp{-\beta\lambda_4}+\eexp{-\beta\lambda_3})]
\eeq
The product of the principal parts is weakly $\omega$ dependent, the imaginary part cancels with the $1\leftrightarrow 2$ term $S_{4a}'$ (see below), hence its real part is
\beq{58}
&&S_{4a}(\omega)=I_42\pi^3\delta(\omega+\epsilon_{RL}+\epsilon'_{RL})\delta(\epsilon'_{RL}-\lambda_{12})
\delta(-\epsilon_{RL}+\lambda_{34})
(\eexp{-\beta\lambda_2}+\eexp{-\beta\lambda_1})(\eexp{-\beta\lambda_4}+\eexp{-\beta\lambda_3})
\nonumber\\
&&=I_42\pi^3\delta(\omega+\lambda_{12}+\lambda_{34})\delta(\epsilon'_{RL}-\lambda_{12})
\delta(-\epsilon_{RL}+\lambda_{34})
(\eexp{-\beta\lambda_2}+\eexp{-\beta\lambda_1})(\eexp{-\beta\lambda_4}+\eexp{-\beta\lambda_3})
\eeq
Integrating the four electron energies for $eV\gg\lambda_{12},\lambda_{34}$
\beq{59}
&&\tilde I_4=\tr[\sigma^i\sigma^k\hat u^\dagger]\tr[ \sigma^j\sigma^l\hat u^\dagger]
\tau^i_{\gamma_1\gamma_2}\tau^j_{\gamma_2\gamma_1}\tau^k_{\gamma_3\gamma_4}\tau^l_{\gamma_4\gamma_3}
\nonumber\\
&&\int_\epsilon S_{4a}(\omega)=2\pi^3e^2J_1^2J_2^2(eV)^2N^4(0)\tilde I_4\delta(\omega+\lambda_{12}+\lambda_{34})
(\eexp{-\beta\lambda_2}+\eexp{-\beta\lambda_1})(\eexp{-\beta\lambda_4}+\eexp{-\beta\lambda_3})
\nonumber\\  \eeq
where $\tilde I_4$ is evaluated above in Eq. \ref{e209}.
As argued for Fig. 2a of the main text, this diagram indeed shows resonances.

Next is $S_4$ with external currents in $t_2,t_4$, $\int_{t_2}\eexp{i\omega(t_2-t_4)},\,\int_{t_1,t_3}$ give 3 independent $\delta$ functions
with $\omega+\epsilon'_{RL}-\lambda_{12}-\omega_1=-\epsilon_{RL}+\lambda_{34}-\omega_2=
\epsilon_{RL}+\lambda_{12}+\omega_1=0$.
This is very similar to $S_{4a}$,
\beq{61}
&&S_{4b}(\omega)=-I_4 2\pi\delta(\omega+\epsilon_{RL}+\epsilon'_{RL})[\frac{\eexp{-\beta\lambda_2}}
{\omega+\epsilon'_{RL}-\lambda_{12}+i\eta}+\frac{\eexp{-\beta\lambda_1}}
{-\omega-\epsilon'_{RL}+\lambda_{12}+i\eta}]
[\frac{\eexp{-\beta\lambda_4}}
{-\epsilon_{RL}+\lambda_{34}+i\eta}+\frac{\eexp{-\beta\lambda_3}}
{\epsilon_{RL}-\lambda_{34}+i\eta}]\nonumber\\
&&=-I_42\pi\delta(\omega+\epsilon_{RL}+\epsilon'_{RL})[P(\frac{1}{\omega+\epsilon'_{RL}-\lambda_{12}})
(\eexp{-\beta\lambda_2}-\eexp{-\beta\lambda_1})-i\pi\delta(\omega+\epsilon'_{RL}-\lambda_{12})
(\eexp{-\beta\lambda_2}+\eexp{-\beta\lambda_1})]\nonumber\\
&&\times [P(\frac{1}{-\epsilon_{RL}+\lambda_{34}})
(\eexp{-\beta\lambda_4}-\eexp{-\beta\lambda_3})-i\pi\delta(-\epsilon_{RL}+\lambda_{34})
(\eexp{-\beta\lambda_4}+\eexp{-\beta\lambda_3})]
\eeq
The product of the principal parts is weakly $\omega$ dependent, the imaginary part cancels with the $1\leftrightarrow 2$ term $S_4'$ (see below), hence its real part is
\beq{62}
&&S_{4b}(\omega)=I_42\pi^3\delta(\omega+\epsilon_{RL}+\epsilon'_{RL})
\delta(\omega+\epsilon'_{RL}-\lambda_{12})
\delta(-\epsilon_{RL}+\lambda_{34})
(\eexp{-\beta\lambda_2}+\eexp{-\beta\lambda_1})(\eexp{-\beta\lambda_4}+\eexp{-\beta\lambda_3})
\nonumber\\
&&=I_4 2\pi^3\delta(\lambda_{12}+\lambda_{34})\delta(\omega+\epsilon'_{RL}-\lambda_{12})
\delta(-\epsilon_{RL}+\lambda_{34})
(\eexp{-\beta\lambda_2}+\eexp{-\beta\lambda_1})(\eexp{-\beta\lambda_4}+\eexp{-\beta\lambda_3})
\eeq
The divergence due to $\delta(\lambda_{12}+\lambda_{34})$ is studied in the summary below.
 Other options for external currents are included in the 2 cases above by smoothly interchanging the two bubbles, which is included in the time integrations.

Consider next $S_4'$ where spins $1\leftrightarrow 2$. The interchange implies that the spin-orbit matrix $\hat u$ is now in vertices $t_1,t_2$ with $\sigma^i,\sigma^j$ (instead of $\hat u^\dagger$ at $t_3,t_4$ with $\sigma^k,\sigma^l$), hence the traces become $\tr[\hat u\sigma^i\sigma^k]\tr[\hat u\sigma^j\sigma^l]$. Also now $\gamma_1,\gamma_2$ belong to spin 2 while $\gamma_3,\gamma_4$ belong to spin 1. To revert to the previous notation relabel
$\gamma_1,\gamma_2\leftrightarrow \gamma_3,\gamma_4$, hence need $i\leftrightarrow k$ and $j\leftrightarrow l$ as well as $\lambda_1,\lambda_2\leftrightarrow \lambda_3,\lambda_4$. This yields the same spin indices as in Eq. (\ref{e55}) with $I_4\rightarrow I_4'$ and the traces replaced by
$ \tr[\hat u\sigma^k\sigma^i]\tr[\hat u\sigma^l\sigma^j]$. Note $\tr[\hat u\sigma^k\sigma^i]=\tr[\sigma^i\sigma^k\hat u^\dagger]^*=\tr[\sigma^i\sigma^k\hat u^\dagger]$ which happens to be real (subsection C), therefore $I_4'=I_4$. Now change in $S_4$ the last form in Eq. \ref{e54} $t_2\leftrightarrow t_3$ and $t_1\leftrightarrow t_4$ so that all $\theta$ functions
regain the form as in \ref{e54} and
\beq{63}
&&S_4'=I_4'[\theta(t_2-t_1)-\theta(t_1-t_2)\eexp{-\beta\lambda_1}]
[\theta(t_1-t_2)-\theta(t_2-t_1)\eexp{-\beta\lambda_2}]
[\theta(t_3-t_4)-\theta(t_4-t_3)\eexp{-\beta\lambda_3}]\times\nonumber\\
&&[\theta(t_4-t_3)-\theta(t_3-t_4)\eexp{-\beta\lambda_4}]\eexp{-i\epsilon_{RL}(t_2-t_4)
-i\epsilon'_{RL}(t_1-t_3)
-i\lambda_{34}(t_3-t_4)-i\lambda_{12}(t_1-t_2)}=S_4^*(\epsilon_{RL}\leftrightarrow\epsilon'_{RL})
\eeq
Furthemore, $\omega$ is defined as an ingoing frequency on the upper times $t_1,t_2$, with the change of time variables it needs $\rightarrow -\omega$, hence for both $S_{4a}',S_{4b}'$ we have $S_4'(-\omega)=
\int_tS_4^*(t)\eexp{-i\omega t}=S_4^*(\omega)$.
 Interchanging the integration variables $\epsilon_{RL}\leftrightarrow \epsilon'_{RL}$, and noting that $I_4$ is invariant under this, shows that $\int_\epsilon S_{4a}'=\int_\epsilon S_{4a}^*$,
  $\int_\epsilon S_{4b}'=\int_\epsilon S_{4b}^*$

     \begin{figure}  \centering
\includegraphics [width=.5\textwidth]{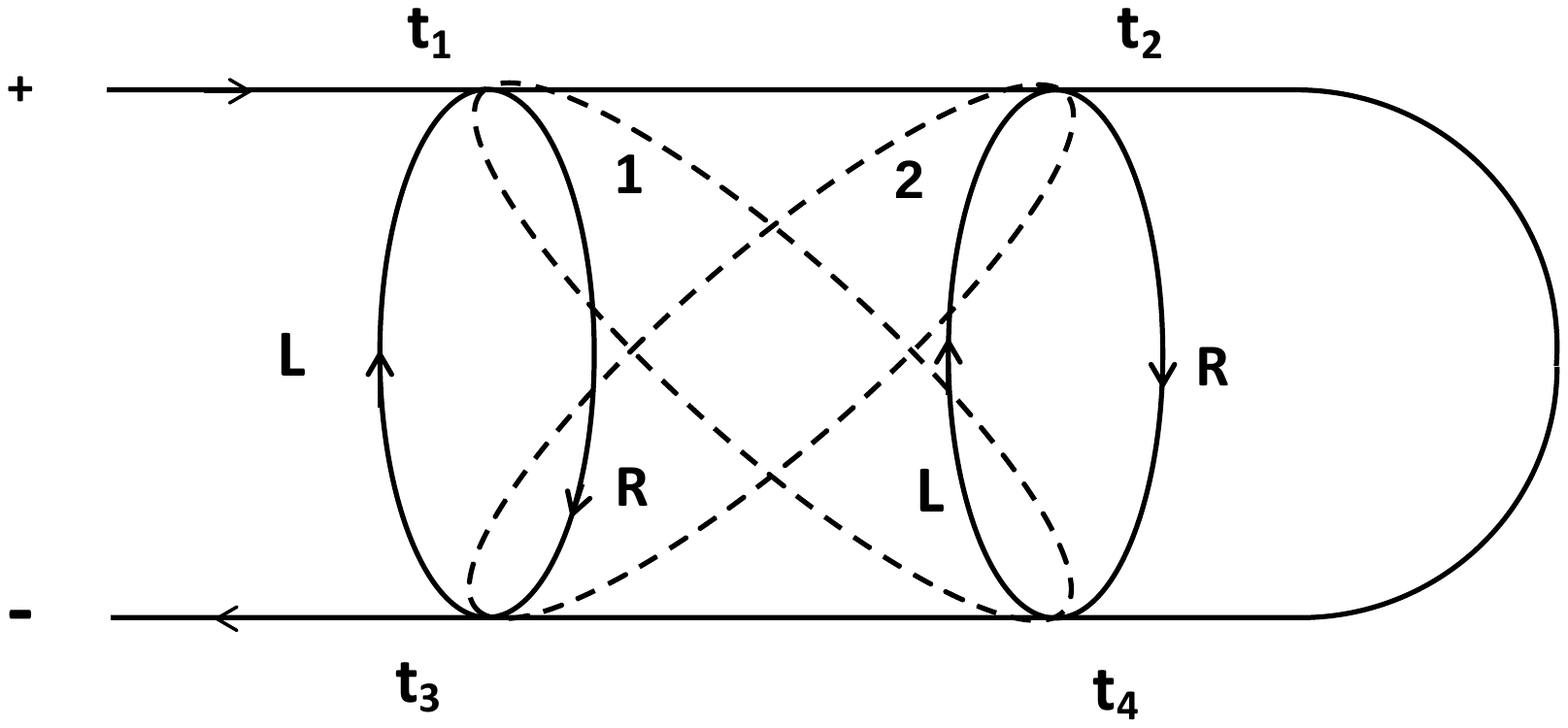}
\caption{Diagram 5}
\label{diagram5}
\end{figure}
Consider now the crossed diagram Fig. \ref{diagram5},
\beq{64}
&&S_5=i^4e^2J_1^2J_2^2\left\langle [\stackrel{1}{\bar\psi}_{R+}\bm\sigma_{\alpha_1\beta_1}\stackrel{2}{\psi}_{L+}\cdot \stackrel{3}{\bar f}_{1+}\bm\tau_{\gamma_1\delta_1}\stackrel{4}{f}_{1+}]_{t_1}[\stackrel{5}{\bar\psi}_{R+}(\hat u \bm
\sigma)_{\alpha_2\beta_2}\stackrel{6}{\psi}_{L+}\cdot\stackrel{7}{\bar f}_{2+}\bm\tau_{\gamma_4\delta_4}\stackrel{8}{f}_{2+}]_{t_2}\right.\nonumber\\
&&\left.\qquad [\stackrel{2}{\bar\psi}_{L-}(\bm\sigma \hat u^\dagger)_{\alpha_3\beta_3}\stackrel{1}{\psi}_{R-}\cdot\stackrel{8}{\bar f}_{2-}\bm\tau_{\gamma_3\delta_3}\stackrel{7}{f}_{2-}]_{t_3} [\stackrel{6}{\bar\psi}_{L-}\bm\sigma_{\alpha_4\beta_4}\cdot\stackrel{5}{\psi}_{R-}\stackrel{4}{\bar f}_{1-}\bm\tau_{\gamma_2\delta_2} \stackrel{3}{f}_{1-}]_{t_4}\right\rangle\nonumber\\
&&=e^2J_1^2J_2^2(-i)G^>_R(t_3-t_1)iG_L^<(t_1-t_3)(-i)G_R^>(t_4-t_2)iG_L^<(t_2-t_4)\times\nonumber\\
&&\qquad (-i)F^>_{1\gamma_1}(t_4-t_1)iF^<_{1\gamma_2}(t_1-t_4)(-i)F^>_{2\gamma_4}(t_3-t_2)
iF_{2\gamma_3}^<(t_2-t_3)\times\nonumber\\
&& \delta_{\alpha_1\beta_3}\delta_{\alpha_3\beta_1}\delta_{\alpha_2\beta_4}\delta_{\alpha_4\beta_2}
\delta_{\gamma_1\delta_2}\delta_{\gamma_2\delta_1}\delta_{\gamma_4\delta_3}\delta_{\gamma_3\delta_4}
\sigma^i_{\alpha_1\beta_1}\tau^i_{\gamma_1\delta_1}
(\hat u\sigma^j)_{\alpha_2\beta_2}\tau^j_{\gamma_4\delta_4}(\sigma^k \hat u^\dagger)_{\alpha_3\beta_3}\tau^k_{\gamma_3\delta_3}
\sigma^l_{\alpha_4\beta_4}\tau^l_{\gamma_2\delta_2}\nonumber\\
&&=e^2J_1^2J_2^2\tr[\sigma^i\sigma^k\hat u^\dagger]\tr[ \hat u\sigma^j\sigma^l]
\tau^i_{\gamma_1\gamma_2}\tau^j_{\gamma_4\gamma_3}\tau^k_{\gamma_3\gamma_4}\tau^l_{\gamma_2\gamma_1}
\times\nonumber\\
&&(1-f_R(\epsilon_R))\eexp{-i\epsilon_R(t_3-t_1)}(-)f_L(\epsilon_L)\eexp{-i\epsilon_L(t_1-t_3)}
(1-f_R(\epsilon'_R))\eexp{-i\epsilon'_R(t_4-t_2)}(-)f_L(\epsilon'_L)\eexp{-i\epsilon'_L(t_2-t_4)}
\nonumber\\
&&\eexp{-i\lambda_{12}(t_4-t_1)}\eexp{-\beta\lambda_2}\eexp{+i\lambda_{34}(t_3-t_2)}\eexp{-\beta\lambda_3}
\eeq
Interchanging $j\leftrightarrow l$ and noting $\tr[\hat u\sigma^l\sigma^j]=\tr[\sigma^j\sigma^l\hat u^\dagger]^*=\tr[\sigma^j\sigma^l\hat u^\dagger]$ as all traces are real (subsection C) the coefficient is the same $I_4$ as above. Consider $S_{5a}$ with external $t_1,t_4$ so $\int_{t_1}\eexp{i\omega(t_1-t_4)},\,\int_{t_2,t_3}$ give 3 independent $\delta$ functions,
\beq{65}
&&S_{5a}(\omega)=I_4 (2\pi)^3\delta(\epsilon'_{RL}-\lambda_{34})\delta(-\epsilon_{RL}+\lambda_{34})
\delta(\omega+\epsilon_{RL}+\lambda_{12})\nonumber\\
&&\int_\epsilon S_{5a}(\omega)=(2\pi)^3e^2J_1^2J_2^2(eV)^2N^4(0)\tilde I_4\delta(\omega+\lambda_{12}+\lambda_{34})\eexp{-\beta\lambda_2-\beta\lambda_3}
\eeq
Consider $S_{5b}$ with external $t_2,t_4$ so $\int_{t_2}\eexp{i\omega(t_2-t_4)},\,\int_{t_1,t_3}$ give 3 independent $\delta$ functions,
\beq{66}
&&S_{5b}(\omega)=I_4(2\pi)^3 \delta(\lambda_{12}+\lambda_{34})\delta(\epsilon_{RL}+\lambda_{12})
\delta(\omega+\epsilon'_{RL}-\lambda_{34})
\nonumber\\
&&\int_\epsilon S_{5b}(\omega)=(2\pi)^3e^2J_1^2J_2^2(eV)^2N^4(0)\tilde I_4\delta(\lambda_{12}+\lambda_{34})\eexp{-\beta\lambda_2-\beta\lambda_3}
\eeq
 Next is $S_{5c}$ with external $t_1,t_3$ so $\int_{t_1}\eexp{i\omega(t_1-t_3)},\,\int_{t_2,t_4}$ give 3 independent $\delta$ functions,
\beq{67}
&&S_{5c}(\omega)=I_4 (2\pi)^3\delta(\lambda_{12}+\lambda_{34})\delta(\omega+\epsilon_{RL}+\lambda_{12})
\delta(\epsilon'_{RL}-\lambda_{34})
\nonumber\\
&&\int_\epsilon S_{5c}(\omega)=(2\pi)^3e^2J_1^2J_2^2(eV)^2N^4(0)\tilde I_4\delta(\lambda_{12}+\lambda_{34})\eexp{-\beta\lambda_2-\beta\lambda_3}
\eeq
Finally $S_{5d}$ with external $t_2,t_3$ so $\int_{t_2}\eexp{i\omega(t_2-t_3)},\,\int_{t_1,t_4}$ give 3 independent $\delta$ functions,
\beq{68}
&&S_{5d}(\omega)=I_4 (2\pi)^3\delta(\omega-\lambda_{12}-\lambda_{34})\delta(\epsilon_{RL}+\lambda_{12})
\delta(\omega+\epsilon'_{RL}-\lambda_{34})
\nonumber\\
&&\int_\epsilon S_{5a}(\omega)=(2\pi)^3e^2J_1^2J_2^2(eV)^2N^4(0)\tilde I_4\delta(\omega-\lambda_{12}-\lambda_{34})\eexp{-\beta\lambda_2-\beta\lambda_3}
\eeq
Interchanging 1,2 is not needed since it is equivalent to shifting the two electron bubbles and reversing their positions which is included in the time integration.

    \begin{figure}[b]  \centering
\includegraphics [width=.5\textwidth]{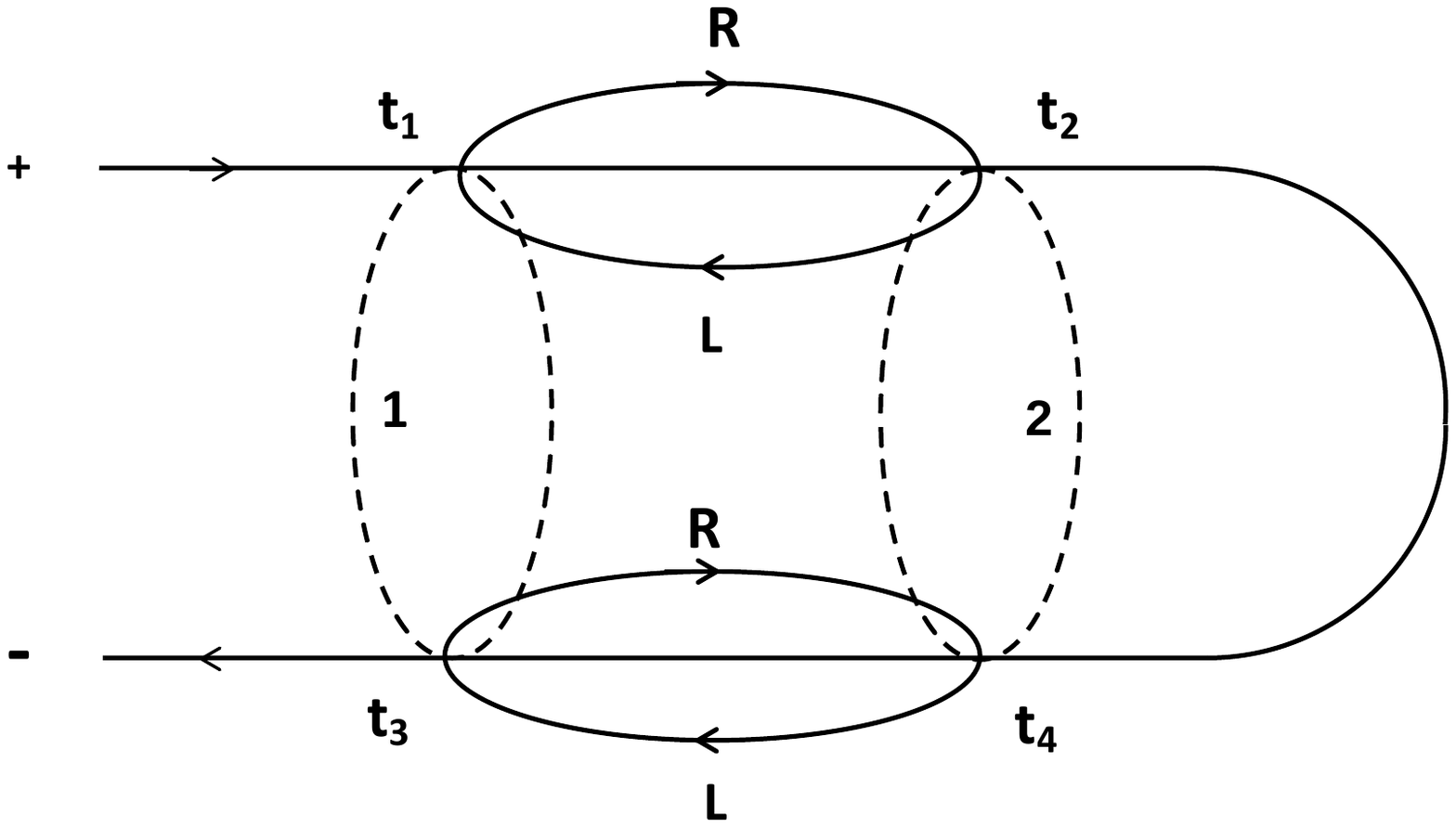}
\caption{Diagram 6}
\label{diagram6}
\end{figure}

Consider $S_6$ in Fig. \ref{diagram6}, sign assumes that external currents are at $t_1$ ($i$) and at
$t_4$ ($-i$)
\beq{69}
&&S_6=e^2J_1^2J_2^2\left\langle [\stackrel{1}{\bar\psi}_{R+}\bm\sigma_{\alpha_1\beta_1}\stackrel{2}{\psi}_{L+}\cdot \stackrel{3}{\bar f}_{1+}\bm\tau_{\gamma_1\delta_1}\stackrel{4}{f}_{1+}]_{t_1}[\stackrel{2}{\bar\psi}_{L+}(\bm
\sigma \hat u^\dagger)_{\alpha_2\beta_2}\stackrel{1}{\psi}_{R+}\cdot\stackrel{5}{\bar f}_{2+}\bm\tau_{\gamma_3\delta_3}\stackrel{6}{f}_{2+}]_{t_2}\right.\nonumber\\
&&\left.\qquad [\stackrel{7}{\bar\psi}_{R-}\bm\sigma _{\alpha_3\beta_3}\stackrel{8}{\psi}_{L-}\cdot\stackrel{4}{\bar f}_{1-}\bm\tau_{\gamma_2\delta_2}\stackrel{3}{f}_{1-}]_{t_3} [\stackrel{8}{\bar\psi}_{L-}(\bm\sigma \hat u^\dagger)_{\alpha_4\beta_4}\cdot\stackrel{7}{\psi}_{R-}\stackrel{6}{\bar f}_{2-}\bm\tau_{\gamma_4\delta_4} \stackrel{5}{f}_{2-}]_{t_4}\right\rangle\nonumber\\
&&=e^2J_1^2J_2^2(-i)G^t_R(t_2-t_1)iG_L^t(t_1-t_2)(-i)G_R^{\bar t}(t_4-t_3)iG_L^{\bar t}(t_3-t_4)\times\nonumber\\
&&\qquad (-i)F^>_{1\gamma_1}(t_3-t_1)iF^<_{1\gamma_2}(t_1-t_3)(-i)F^>_{2\gamma_3}(t_4-t_2)
iF_{2\gamma_4}^<(t_2-t_4)\times\nonumber\\
&& \delta_{\alpha_1\beta_2}\delta_{\alpha_2\beta_1}\delta_{\alpha_3\beta_4}\delta_{\alpha_4\beta_3}
\delta_{\gamma_1\delta_2}\delta_{\gamma_2\delta_1}\delta_{\gamma_3\delta_4}\delta_{\gamma_4\delta_3}
\sigma^i_{\alpha_1\beta_1}\tau^i_{\gamma_1\delta_1}
(\sigma^j\hat u^\dagger)_{\alpha_2\beta_2}\tau^j_{\gamma_3\delta_3}\sigma^k _{\alpha_3\beta_3}\tau^k_{\gamma_2\delta_2}
(\sigma^l\hat u^\dagger)_{\alpha_4\beta_4}\tau^l_{\gamma_4\delta_4}\nonumber\\
&&=e^2J_1^2J_2^2\tr[\sigma^i\sigma^j\hat u^\dagger]\tr[ \sigma^k\sigma^l\hat u^\dagger]
\tau^i_{\gamma_1\gamma_2}\tau^j_{\gamma_3\gamma_4}\tau^k_{\gamma_2\gamma_1}\tau^l_{\gamma_4\gamma_3}
\times\nonumber\\
&& [\theta(t_2-t_1)(1-f_R(\epsilon_R))-\theta(t_1-t_2)f_R(\epsilon_R)]\eexp{-i\epsilon_{RL}(t_2-t_1))}
[\theta(t_1-t_2)(1-f_L(\epsilon_L)))-\theta(t_2-t_1)f_L(\epsilon_L)]\nonumber\\
&&[\theta(t_4-t_3)(1-f_R(\epsilon'_R))-\theta(t_3-t_4)f_R(\epsilon'_R)]\eexp{-i\epsilon'_{RL}(t_4-t_3)}
[\theta(t_3-t_4)(1-f_L(\epsilon'_L))-\theta(t_4-t_3)f_L(\epsilon'_L)]\nonumber\\
&&\eexp{-i\lambda_{12}(t_3-t_1)}\eexp{-\beta\lambda_2}\eexp{-i\lambda_{34}(t_4-t_2)}\eexp{-\beta\lambda_4}
\eeq
Defining $I_6$,
\beq{70}
&&I_6=e^2J_1^2J_2^2\tr[\sigma^i\sigma^j\hat u^\dagger]\tr[ \sigma^k\sigma^l\hat u^\dagger]
\tau^i_{\gamma_1\gamma_2}\tau^j_{\gamma_3\gamma_4}\tau^k_{\gamma_2\gamma_1}\tau^l_{\gamma_4\gamma_3}
\nonumber\\
&&S_6=I_6(-i)^2\int_{\omega_1}[\frac{(1-f_R(\epsilon_R))f_L(\epsilon_L)}{\omega_1+i\eta}+
\frac{(1-f_L(\epsilon_L))f_R(\epsilon_R)}{-\omega_1+i\eta}]\eexp{-i\omega_1(t_2-t_1)}
\int_{\omega_2}[\frac{(1-f_R(\epsilon'_R))f_L(\epsilon'_L)}{\omega_2+i\eta}+
\frac{(1-f_L(\epsilon'_L))f_R(\epsilon'_R)}{-\omega_2+i\eta}]\times\nonumber\\&&
\eexp{-i\omega_2(t_4-t_3)}
\eexp{-i\epsilon_{RL}(t_2-t_1)-i\epsilon'_{RL}(t_4-t_3)-i\lambda_{12}(t_3-t_1)-i\lambda_{34}(t_4-t_2)}
\eexp{-\beta\lambda_2-\beta\lambda_4}
\eeq
Consider $S_{6a}$ with $\int_{t_1}\eexp{i\omega(t_1-t_4)}\int_{t_2,t_3}$ yielding 3 independent
$\delta$ functions with $\omega+\omega_1+\epsilon_{RL}+\lambda_{12}=
-\omega_1-\epsilon_{RL}+\lambda_{34}=\omega_2+\epsilon'_{RL}-\lambda_{12}=0$, hence
\beq{71}
&&S_{6a}=-I_62\pi\delta(\omega+\lambda_{12}+\lambda_{34})[\frac{(1-f_R(\epsilon_R))f_L(\epsilon_L)}
{-\epsilon_{RL}+\lambda_{34}+i\eta}+\frac{(1-f_L(\epsilon_L))f_R(\epsilon_R)}
{\epsilon_{RL}-\lambda_{34}+i\eta}]\times\nonumber\\&&[\frac{(1-f_R(\epsilon'_R))f_L(\epsilon'_L)}
{-\epsilon'_{RL}+\lambda_{12}+i\eta}+\frac{(1-f_L(\epsilon'_L))f_R(\epsilon'_R)}
{\epsilon'_{RL}-\lambda_{12}+i\eta}]\eexp{-\beta\lambda_2-\beta\lambda_4}
\eeq
Consider next $S_{6b}$ with $\int_{t_2}\eexp{i\omega(t_2-t_4)}\int_{t_1,t_3}$ yielding 3 independent
$\delta$ functions with $\omega_1+\epsilon_{RL}+\lambda_{12}=\omega-\omega_1-\epsilon_{RL}+\lambda_{34}=
\omega_2+\epsilon'_{RL}-\lambda_{12}=0$
(note opposite sign as current has now $c_L^\dagger c_R$), hence
\beq{72}
&&S_{6b}=+I_62\pi\delta(\omega+\lambda_{12}+\lambda_{34})[\frac{(1-f_R(\epsilon_R))f_L(\epsilon_L)}
{-\epsilon_{RL}-\lambda_{12}+i\eta}+\frac{(1-f_L(\epsilon_L))f_R(\epsilon_R)}
{\epsilon_{RL}+\lambda_{12}+i\eta}]\times\nonumber\\&&[\frac{(1-f_R(\epsilon'_R))f_L(\epsilon'_L)}
{-\epsilon'_{RL}+\lambda_{12}+i\eta}+\frac{(1-f_L(\epsilon'_L))f_R(\epsilon'_R)}
{\epsilon'_{RL}-\lambda_{12}+i\eta}]\eexp{-\beta\lambda_2-\beta\lambda_4}
\eeq
Consider next $S_{6c}$ with $\int_{t_1}\eexp{i\omega(t_1-t_3)}\int_{t_2,t_4}$ yielding 3 independent
$\delta$ functions with $\omega+\omega_1+\epsilon_{RL}+\lambda_{12}=
-\omega_1-\epsilon_{RL}+\lambda_{34}=-\omega+\omega_2+\epsilon'_{RL}-\lambda_{12}=0$, hence
\beq{73}
&&S_{6c}=+I_62\pi\delta(\omega+\lambda_{12}+\lambda_{34})[\frac{(1-f_R(\epsilon_R))f_L(\epsilon_L)}
{-\epsilon_{RL}+\lambda_{34}+i\eta}+\frac{(1-f_L(\epsilon_L))f_R(\epsilon_R)}
{\epsilon_{RL}-\lambda_{34}+i\eta}]\times\nonumber\\&&[\frac{(1-f_R(\epsilon'_R))f_L(\epsilon'_L)}
{-\epsilon'_{RL}-\lambda_{34}+i\eta}+\frac{(1-f_L(\epsilon'_L))f_R(\epsilon'_R)}
{\epsilon'_{RL}+\lambda_{34}+i\eta}]\eexp{-\beta\lambda_2-\beta\lambda_4}
\eeq
Consider next $S_{6d}$ with $\int_{t_2}\eexp{i\omega(t_2-t_3)}\int_{t_1,t_4}$ yielding 3 independent
$\delta$ functions $\omega_1+\epsilon_{RL}+\lambda_{12}=\omega -\omega_1-\epsilon_{RL}+\lambda_{34}=
-\omega+\omega_2+\epsilon'_{RL}-\lambda_{12}=0$, hence
\beq{73}
&&S_{6d}=-I_62\pi\delta(\omega+\lambda_{12}+\lambda_{34})[\frac{(1-f_R(\epsilon_R))f_L(\epsilon_L)}
{-\epsilon_{RL}-\lambda_{12}+i\eta}+\frac{(1-f_L(\epsilon_L))f_R(\epsilon_R)}
{\epsilon_{RL}+\lambda_{12}+i\eta}]\times\nonumber\\&&[\frac{(1-f_R(\epsilon'_R))f_L(\epsilon'_L)}
{-\epsilon'_{RL}-\lambda_{34}+i\eta}+\frac{(1-f_L(\epsilon'_L))f_R(\epsilon'_R)}
{\epsilon'_{RL}+\lambda_{34}+i\eta}]\eexp{-\beta\lambda_2-\beta\lambda_4}
\eeq
Add 2 terms
\beq{751}
&&S_{6a}+S_{6b}=I_62\pi\delta(\omega+\lambda_{12}+\lambda_{34})[\frac{(1-f_R(\epsilon'_R))f_L(\epsilon'_L)}
{-\epsilon'_{RL}+\lambda_{12}+i\eta}+\frac{(1-f_L(\epsilon'_L))f_R(\epsilon'_R)}
{\epsilon'_{RL}-\lambda_{12}+i\eta}]\eexp{-\beta\lambda_2-\beta\lambda_4}\nonumber\\&&
[\frac{-f_L(\epsilon_L)}
{-\epsilon_{RL}+\lambda_{34}+i\eta}+\frac{-f_R(\epsilon_R)}
{\epsilon_{RL}-\lambda_{34}+i\eta}+f_R(\epsilon_R)f_L(\epsilon_L)(\frac{1}
{-\epsilon_{RL}+\lambda_{34}+i\eta}+\frac{1}{\epsilon_{RL}-\lambda_{34}+i\eta})\nonumber\\&&
\frac{f_L(\epsilon_L)}
{-\epsilon_{RL}-\lambda_{12}+i\eta}+\frac{f_R(\epsilon_R)}
{\epsilon_{RL}+\lambda_{12}+i\eta}-f_R(\epsilon_R)f_L(\epsilon_L)(\frac{1}
{-\epsilon_{RL}-\lambda_{12}+i\eta}+\frac{1}{\epsilon_{RL}+\lambda_{12}+i\eta})]\nonumber
\eeq
All terms with a single fermi function can be integrated on either $\epsilon_R$ or $\epsilon_L$ leading to $-i\pi$ which cancel, hence ignoring these terms
\beq{75}
&&S_{6a}+S_{6b}=I_62\pi\delta(\omega+\lambda_{12}+\lambda_{34})[\frac{(1-f_R(\epsilon'_R))f_L(\epsilon'_L)}
{-\epsilon'_{RL}+\lambda_{12}+i\eta}+\frac{(1-f_L(\epsilon'_L))f_R(\epsilon'_R)}
{\epsilon'_{RL}-\lambda_{12}+i\eta}]\eexp{-\beta\lambda_2-\beta\lambda_4}\times\nonumber\\&&
f_R(\epsilon_R)f_L(\epsilon_L)[-2\pi i\delta(\epsilon_{RL}-\lambda_{34})+2\pi i\delta(\epsilon_{RL}+\lambda_{12})]
\eeq
The next 2 terms add up to
\beq{771}
&&S_{6c}+S_{6d}=I_62\pi\delta(\omega+\lambda_{12}+\lambda_{34})
[\frac{(1-f_R(\epsilon'_R))f_L(\epsilon'_L)}
{-\epsilon'_{RL}-\lambda_{34}+i\eta}+\frac{(1-f_L(\epsilon'_L))f_R(\epsilon'_R)}
{\epsilon'_{RL}+\lambda_{34}+i\eta}]\eexp{-\beta\lambda_2-\beta\lambda_4}\nonumber\\&&
[\frac{f_L(\epsilon_L)}
{-\epsilon_{RL}+\lambda_{34}+i\eta}+\frac{f_R(\epsilon_R)}
{\epsilon_{RL}-\lambda_{34}+i\eta}-f_R(\epsilon_R)f_L(\epsilon)
(\frac{1}{-\epsilon_{RL}+\lambda_{34}+i\eta}+\frac{1}{\epsilon_{RL}-\lambda_{34}+i\eta})\nonumber\\&&
\frac{-f_L(\epsilon_L)}
{-\epsilon_{RL}-\lambda_{12}+i\eta}+\frac{-f_R(\epsilon_R)}
{\epsilon_{RL}+\lambda_{12}+i\eta}+f_R(\epsilon_R)f_L(\epsilon)
(\frac{1}{-\epsilon_{RL}-\lambda_{12}+i\eta}+\frac{1}{\epsilon_{RL}+\lambda_{12}+i\eta})]\nonumber
\eeq
Again, all terms with a single fermi function can be integrated on either $\epsilon_R$ or $\epsilon_L$ leading to $-i\pi$ which cancel, hence ignoring these terms
\beq{76}
&&S_{6c}+S_{6d}=I_62\pi\delta(\omega+\lambda_{12}+\lambda_{34})
[\frac{(1-f_R(\epsilon'_R))f_L(\epsilon'_L)}
{-\epsilon'_{RL}-\lambda_{34}+i\eta}+\frac{(1-f_L(\epsilon'_L))f_R(\epsilon'_R)}
{\epsilon'_{RL}+\lambda_{34}+i\eta}]\eexp{-\beta\lambda_2-\beta\lambda_4}\times\nonumber\\&&
f_R(\epsilon_R)f_L(\epsilon_L)[2\pi i\delta(\epsilon_{RL}-\lambda_{34})-2\pi i\delta(\epsilon_{RL}+\lambda_{12})]
\eeq
The same separation can be done on the $\epsilon'$ variables, ignoring terms that eventually cancel in the $\epsilon'$ integrations ($\pm i\pi$ terms)
\beq{77}
&&S_{6a}+S_{6b}+S_{6c}+S_{6d}=I_6 2\pi\delta(\omega+\lambda_{12}+\lambda_{34})
f_R(\epsilon_R)f_L(\epsilon_L)f_R(\epsilon'_R)f_L(\epsilon'_L)\nonumber\\&&
[2\pi i\delta(\epsilon_{RL}'-\lambda_{12})-2\pi i\delta(\epsilon'_{RL}+\lambda_{34})]
[2\pi i\delta(\epsilon_{RL}+\lambda_{12})-2\pi i\delta(\epsilon_{RL}-\lambda_{34})]
\eeq
The $\epsilon$ integration has $f_R(\epsilon_R)[f_L(\epsilon_R-\lambda_{34})-f_L(\epsilon_R+\lambda_{12})]\approx
-f_R(\epsilon_R)\frac{\partial f_L(\epsilon_R)}{\partial\epsilon_R}(\lambda_{34}+\lambda_{12})$ and similarly for the $\epsilon'$ integration. These terms are negligible for $\lambda_{12},\lambda_{34}\ll eV$.

Finally for $S_6$, in each fermion bubble one can change $R\leftrightarrow L$ independently (without changing spin label), which changes the sign of $\epsilon_{RL}$ and/or $\epsilon'_{RL}$ in the result \eqref{e77}. All these 3 versions are negligible in the same way.

   \begin{figure}  \centering
\includegraphics [width=.5\textwidth]{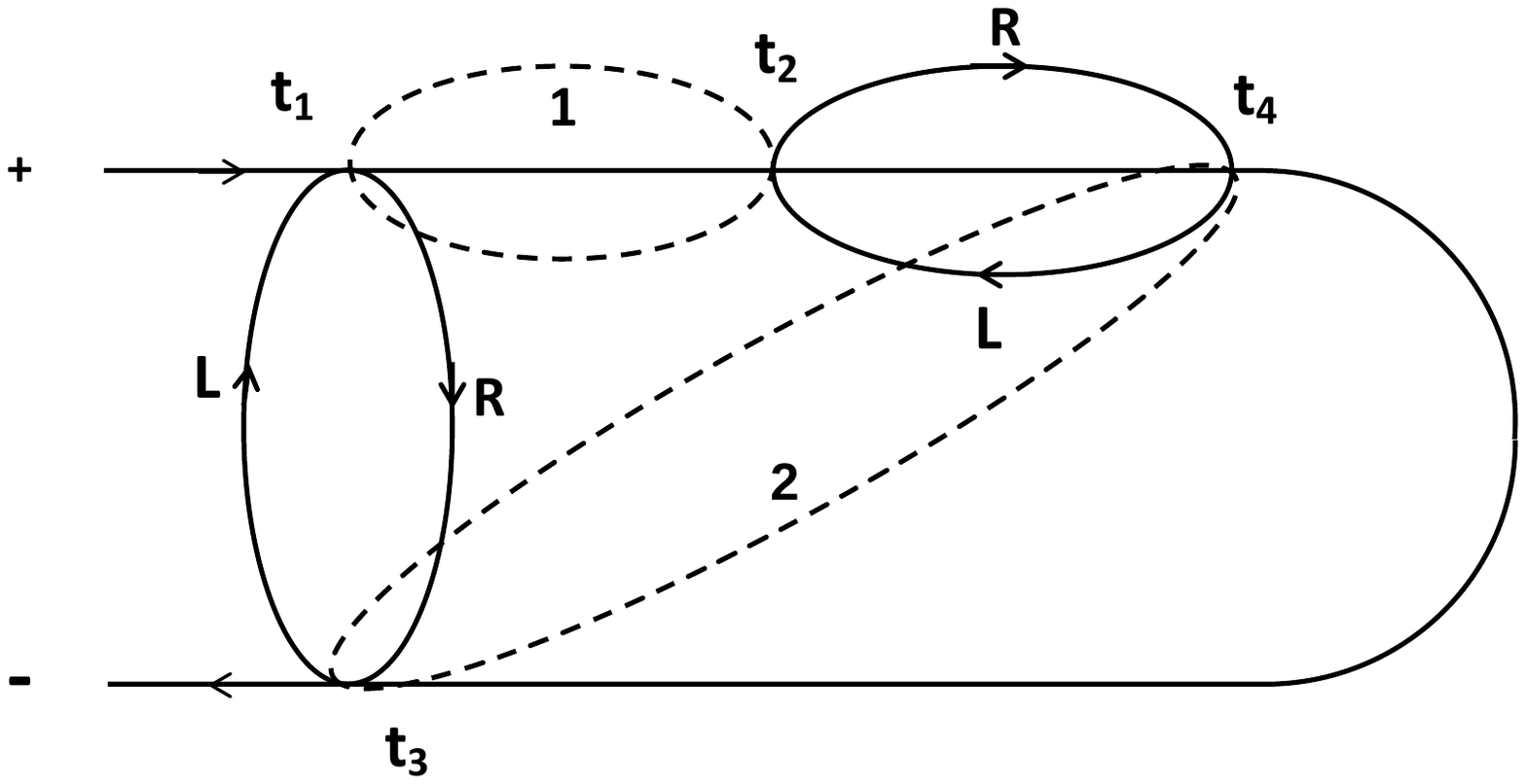}
\caption{Diagram 7}
\label{diagram7}
\end{figure}

Consider now $S_7$, Fig. \ref{diagram7}, sign assumes that external currents are at $t_1,t_3$,
\beq{78}
&&S_7=-i^4e^2J_1^2J_2^2\left\langle [\stackrel{1}{\bar\psi}_{R+}\bm\sigma_{\alpha_1\beta_1}\stackrel{2}{\psi}_{L+}\cdot \stackrel{3}{\bar f}_{1+}\bm\tau_{\gamma_1\delta_1}\stackrel{4}{f}_{1+}]_{t_1}[\stackrel{5}{\bar\psi}_{R+}\bm
\sigma_{\alpha_2\beta_2}\stackrel{6}{\psi}_{L+}\cdot\stackrel{4}{\bar f}_{1+}\bm\tau_{\gamma_2\delta_2}\stackrel{3}{f}_{1+}]_{t_2}\right.\nonumber\\
&&\left.\qquad [\stackrel{2}{\bar\psi}_{L-}\bm(\sigma \hat u^\dagger)_{\alpha_3\beta_3}\stackrel{1}{\psi}_{R-}\cdot\stackrel{7}{\bar f}_{2-}\bm\tau_{\gamma_3\delta_3}\stackrel{8}{f}_{2-}]_{t_3} [\stackrel{6}{\bar\psi}_{L+}(\bm\sigma \hat u^\dagger)_{\alpha_4\beta_4}\cdot\stackrel{5}{\psi}_{R+}\stackrel{8}{\bar f}_{2+}\bm\tau_{\gamma_4\delta_4} \stackrel{7}{f}_{2+}]_{t_4}\right\rangle\nonumber\\
&&=-e^2J_1^2J_2^2(-i)G^>_R(t_3-t_1)iG_L^<(t_1-t_3)(-i)G_R^t(t_4-t_2)iG_L^t(t_2-t_4)\times\nonumber\\
&&\qquad (-i)F^t_{1\gamma_1}(t_2-t_1)iF^t_{1\gamma_2}(t_1-t_2)(-i)F^{<}_{2\gamma_3}(t_4-t_3)
iF_{2\gamma_4}^{>}(t_3-t_4)\times\nonumber\\
&& \delta_{\alpha_1\beta_3}\delta_{\alpha_3\beta_1}\delta_{\alpha_2\beta_4}\delta_{\alpha_4\beta_2}
\delta_{\gamma_1\delta_2}\delta_{\gamma_2\delta_1}\delta_{\gamma_3\delta_4}\delta_{\gamma_4\delta_3}
\sigma^i_{\alpha_1\beta_1}\tau^i_{\gamma_1\delta_1}
\sigma^j_{\alpha_2\beta_2}\tau^j_{\gamma_2\delta_2}(\sigma^k \hat u^\dagger)_{\alpha_3\beta_3}\tau^k_{\gamma_3\delta_3}
(\sigma^l\hat u^\dagger)_{\alpha_4\beta_4}\tau^l_{\gamma_4\delta_4}\nonumber\\
&&=-e^2J_1^2J_2^2\tr[\sigma^i\sigma^k\hat u^\dagger]\tr[ \sigma^j\sigma^l\hat u^\dagger]
\tau^i_{\gamma_1\gamma_2}\tau^j_{\gamma_2\gamma_1}\tau^k_{\gamma_3\gamma_4}\tau^l_{\gamma_4\gamma_3}
\times\nonumber\\
&&(1-f_R(\epsilon_R))\eexp{-i\epsilon_R(t_3-t_1)}(-)f_L(\epsilon_L)\eexp{-i\epsilon_L(t_1-t_3)}
[\theta(t_4-t_2)(1-f_R(\epsilon_R'))-\theta(t_2-t_4)f_R(\epsilon_R')]\eexp{-i\epsilon_R'(t_4-t_2)}
\nonumber\\&&
[\theta(t_2-t_4)(1-f_L(\epsilon_L'))-\theta(t_4-t_2)f_L(\epsilon_L')]\eexp{-i\epsilon_L'(t_2-t_4)}
(-)\eexp{-\beta\lambda_3}\eexp{-i\lambda_3(t_4-t_3)}\eexp{-i\lambda_4(t_3-t_4)}
\nonumber\\&&
[\theta(t_2-t_1)-\theta(t_1-t_2)\eexp{-\beta\lambda_1}]\eexp{-i\lambda_1(t_2-t_1)}
[\theta(t_1-t_2)-\theta(t_2-t_1)\eexp{-\beta\lambda_2}]\eexp{-i\lambda_2(t_1-t_2)}\nonumber\\
\eeq
Multiply first the two $G^t$ and then Fourier,
\beq{79}
&& I_7=\tilde I_4(1-f_R(\epsilon_R))f_L(\epsilon_L)\nonumber\\
&& S_7=-I_7(-i)\int_{\omega_1}[\frac{(1-f_R(\epsilon_R'))f_L(\epsilon_L')}{\omega_1+i\eta}
+\frac{(1-f_L(\epsilon_L'))f_R(\epsilon_R')}{-\omega_1+i\eta}]\eexp{-i\omega_1(t_4-t_2)
-i\epsilon_{RL}'(t_4-t_2)-i\epsilon_{RL}(t_3-t_1)}\nonumber\\&&
(-i)\int_{\omega_2}[\frac{\eexp{-\beta\lambda_2}}{\omega_2+i\eta}+
\frac{\eexp{-\beta\lambda_1}}{-\omega_2+i\eta}]\eexp{-i\omega_2(t_2-t_1)-i\lambda_{12}(t_2-t_1)
-i\lambda_{34}(t_4-t_3)}\eexp{-\beta\lambda_3}
\eeq
For $S_{7a}$ need $\int_{t_1}\eexp{-\omega(t_1-t_3)}\int_{t_2,t_4}$ yielding 3 independent $\delta$ functions with $\omega+\omega_2+\epsilon_{RL}+\lambda_{12}=\omega_1-\omega_2+\epsilon_{RL}'-\lambda_{12}=
-\omega-\epsilon_{RL}+\lambda_{34}=0$
\beq{80}
&& S_{7a}=I_72\pi\delta(\omega+\epsilon_{RL}-\lambda_{34})
[\frac{(1-f_R(\epsilon_R'))f_L(\epsilon_L')}{-\epsilon_{RL}'-\lambda_{34}+i\eta}
+\frac{(1-f_L(\epsilon_L'))f_R(\epsilon_R')}{\epsilon_{RL}'+\lambda_{34}+i\eta}]\nonumber\\
&&\qquad [\frac{\eexp{-\beta\lambda_2}}{-\lambda_{12}-\lambda_{34}+i\eta}
+\frac{\eexp{-\beta\lambda_1}}{\lambda_{12}+\lambda_{34}+i\eta}]\eexp{-\beta\lambda_3}
\eeq
The 1st P.P. vanishes since $P.P.\int_{\epsilon'}\frac{-f_L(\epsilon_L')+f_R(\epsilon_R')}{\epsilon_R'-\epsilon_L'}=0$
by integrating either $\epsilon_R'$ or $\epsilon_L'$.
The 2nd  P.P. is then an imaginary term that cancels (see below), so remains the product of $\delta$ functions
\beq{81}
&&S_{7a}=-I_72\pi^3[(1-f_R(\epsilon_R'))f_L(\epsilon_L')+(1-f_L(\epsilon_L'))f_R(\epsilon_R')]
(\eexp{-\beta\lambda_2}+\eexp{-\beta\lambda_1})\eexp{-\beta\lambda_3}\times\nonumber\\
&&\qquad \delta(\omega+\epsilon_{RL}-\lambda_{34})\delta(\epsilon_{RL}'+\lambda_{34})
\delta(\lambda_{12}+\lambda_{34})
\eeq
which is divergent when $\lambda_{12}+\lambda_{34}=0$.

For $S_{7b}$ need $\int_{t_1}\eexp{-\omega(t_2-t_3)}\int_{t_1,t_4}$ yielding 3 independent $\delta$ functions with $\omega_2+\epsilon_{RL}+\lambda_{12}= \omega+\omega_1-\omega_2+\epsilon_{RL}'-\lambda_{12}=-\omega-\epsilon_{RL}+\lambda_{34}=0$, hence
\beq{82}
&& S_{7b}=I_72\pi\delta(\omega+\epsilon_{RL}-\lambda_{34})
[\frac{(1-f_R(\epsilon_R'))f_L(\epsilon_L')}{-\epsilon_{RL}'-\lambda_{34}+i\eta}
+\frac{(1-f_L(\epsilon_L'))f_R(\epsilon_R')}{\epsilon_{RL}'+\lambda_{34}+i\eta}]\nonumber\\
&&\qquad [\frac{\eexp{-\beta\lambda_2}}{-\epsilon_{RL}-\lambda_{12}+i\eta}
+\frac{\eexp{-\beta\lambda_1}}{\epsilon_{RL}+\lambda_{12}+i\eta}]\eexp{-\beta\lambda_3}
\eeq
The 1st P.P. vanishes as in \eqref{e80}, while the 2nd  P.P. is then an imaginary term that cancels (see below), so remains the product of $\delta$ functions
\beq{83}
&&S_{7b}=-I_72\pi^3[(1-f_R(\epsilon_R'))f_L(\epsilon_L')+(1-f_L(\epsilon_L'))f_R(\epsilon_R')]
(\eexp{-\beta\lambda_2}+\eexp{-\beta\lambda_1})\eexp{-\beta\lambda_3}\times\nonumber\\
&&\qquad \delta(\omega+\epsilon_{RL}-\lambda_{34})\delta(\epsilon_{RL}'+\lambda_{34})
\delta(\epsilon_{RL}+\lambda_{12})
\eeq

For $S_{7c}$ need $\int_{t_4}\eexp{i\omega(t_4-t_3)}\int_{t_1,t_2}$, it has a $-$ relative to $S_{7b}$ since current couples to $c_L^\dagger c_R$. It yields 3 independent $\delta$ functions with
$\omega_2+\epsilon_{RL}+\lambda_{12}=\omega_1-\omega_2+\epsilon_{RL}'-\lambda_{12}=
-\omega-\epsilon_{RL}+\lambda_{34}=0 $
\beq{84}
 S_{7c}=&&-I_72\pi\delta(\omega+\epsilon_{RL}-\lambda_{34})
[\frac{(1-f_R(\epsilon_R'))f_L(\epsilon_L')}{\omega-\epsilon_{RL}'-\lambda_{34}+i\eta}
+\frac{(1-f_L(\epsilon_L'))f_R(\epsilon_R')}{-\omega+\epsilon_{RL}'+\lambda_{34}+i\eta}]\times\nonumber\\&&
 [\frac{\eexp{-\beta\lambda_2}}{-\epsilon_{RL}-\lambda_{12}+i\eta}
+\frac{\eexp{-\beta\lambda_1}}{\epsilon_{RL}+\lambda_{12}+i\eta}]\eexp{-\beta\lambda_3}
\eeq
The 1st P.P. vanishes as in \eqref{e80}, while the 2nd  P.P. is then an imaginary term that cancels (see below), so remains the product of $\delta$ functions
\beq{85}
&&S_{7c}=+I_72\pi^3[(1-f_R(\epsilon_R'))f_L(\epsilon_L')+(1-f_L(\epsilon_L'))f_R(\epsilon_R')]
(\eexp{-\beta\lambda_2}+\eexp{-\beta\lambda_1})\eexp{-\beta\lambda_3}\times\nonumber\\
&&\qquad \delta(\omega+\epsilon_{RL}-\lambda_{34})\delta(-\omega+\epsilon_{RL}'+\lambda_{34})
\delta(\epsilon_{RL}+\lambda_{12})
\eeq
Both $S_{7b},S_{7c}$ have resonance $\delta(\omega-\lambda_{12}-\lambda_{34})$, however for $eV\gg\omega$ they cancels.

We note that $R\leftrightarrow L$ in the $t$ ordered fermion bubble is identical to the result above except for $\epsilon_{RL}'\rightarrow-\epsilon_{RL}'$, and a sign change in $S_{7b},S_{7c}$ that have a current vertex on this bubble. Hence for large $eV$ this yields a factor 2 in $S_{7a}$ while both $S_{7b},\, S_{7c}$ cancel. Before considering spin exchange $1\leftrightarrow 2$, we study Fig. \ref{diagram8}.

   \begin{figure}  \centering
\includegraphics [width=.5\textwidth]{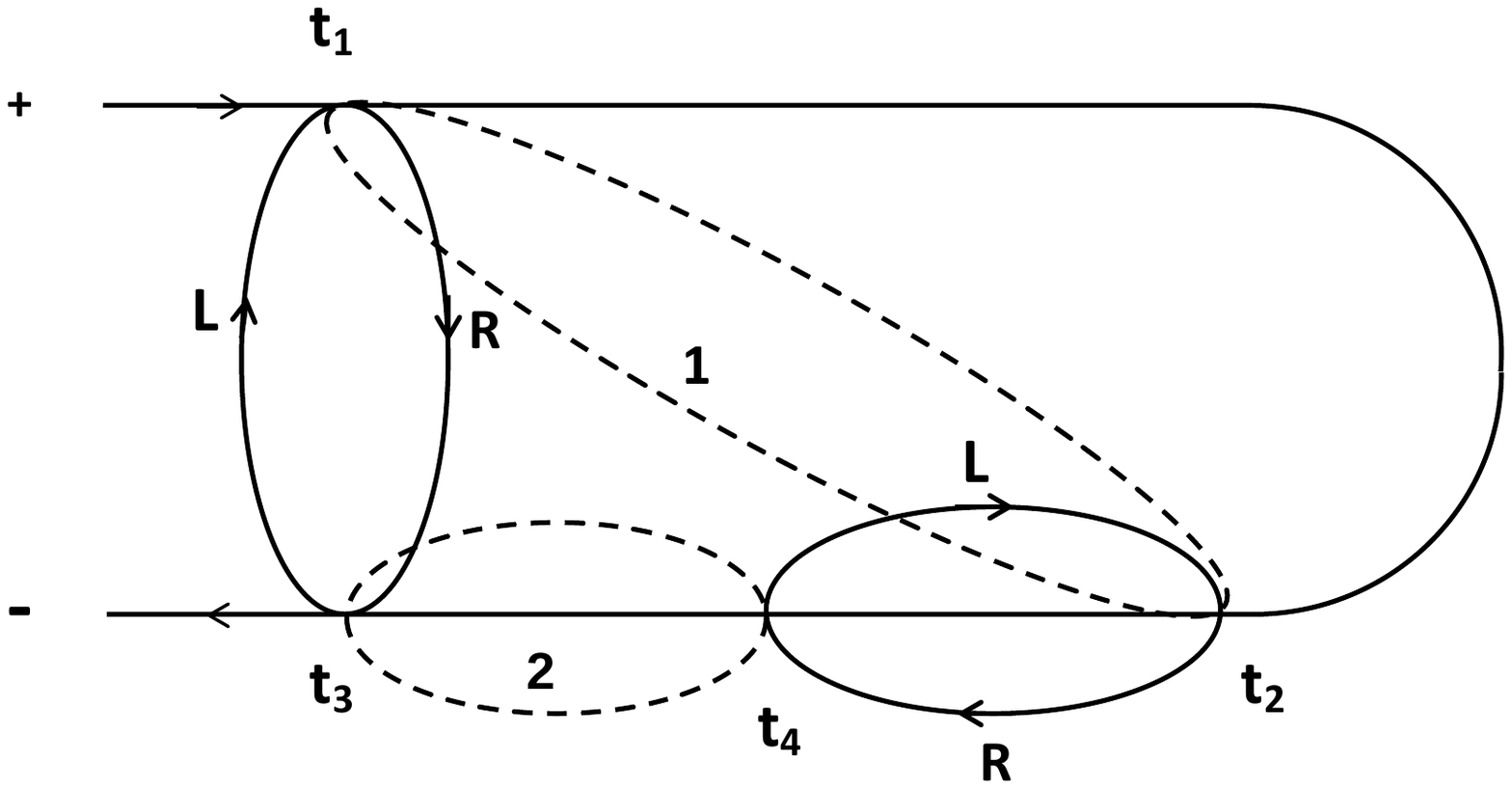}
\caption{Diagram 8}
\label{diagram8}
\end{figure}

The sign for $S_8$ assumes external currents at $t_1,t_3$.
\beq{86}
&&S_8=-i^4e^2J_1^2J_2^2\left\langle [\stackrel{1}{\bar\psi}_{R+}\bm\sigma_{\alpha_1\beta_1}\stackrel{2}{\psi}_{L+}\cdot \stackrel{3}{\bar f}_{1+}\bm\tau_{\gamma_1\delta_1}\stackrel{4}{f}_{1+}]_{t_1}[\stackrel{5}{\bar\psi}_{R-}\bm
\sigma_{\alpha_2\beta_2}\stackrel{6}{\psi}_{L-}\cdot\stackrel{4}{\bar f}_{1-}\bm\tau_{\gamma_2\delta_2}\stackrel{3}{f}_{1-}]_{t_2}\right.\nonumber\\
&&\left.\qquad [\stackrel{2}{\bar\psi}_{L-}\bm(\sigma \hat u^\dagger)_{\alpha_3\beta_3}\stackrel{1}{\psi}_{R-}\cdot\stackrel{7}{\bar f}_{2-}\bm\tau_{\gamma_3\delta_3}\stackrel{8}{f}_{2-}]_{t_3} [\stackrel{6}{\bar\psi}_{L-}(\bm\sigma \hat u^\dagger)_{\alpha_4\beta_4}\cdot\stackrel{5}{\psi}_{R-}\stackrel{8}{\bar f}_{2-}\bm\tau_{\gamma_4\delta_4} \stackrel{7}{f}_{2-}]_{t_4}\right\rangle\nonumber\\
&&=-e^2J_1^2J_2^2(-i)G^>_R(t_3-t_1)iG_L^<(t_1-t_3)(-i)G_R^{\bar t}(t_4-t_2)iG_L^{\bar t}(t_2-t_4)\times\nonumber\\
&&\qquad (-i)F^>_{1\gamma_1}(t_2-t_1)iF^<_{1\gamma_2}(t_1-t_2)(-i)F^{\bar t}_{2\gamma_3}(t_4-t_3)
iF_{2\gamma_4}^{\bar t}(t_3-t_4)\times\nonumber\\
&& \delta_{\alpha_1\beta_3}\delta_{\alpha_3\beta_1}\delta_{\alpha_2\beta_4}\delta_{\alpha_4\beta_2}
\delta_{\gamma_1\delta_2}\delta_{\gamma_2\delta_1}\delta_{\gamma_3\delta_4}\delta_{\gamma_4\delta_3}
\sigma^i_{\alpha_1\beta_1}\tau^i_{\gamma_1\delta_1}
\sigma^j_{\alpha_2\beta_2}\tau^j_{\gamma_2\delta_2}(\sigma^k \hat u^\dagger)_{\alpha_3\beta_3}\tau^k_{\gamma_3\delta_3}
(\sigma^l\hat u^\dagger)_{\alpha_4\beta_4}\tau^l_{\gamma_4\delta_4}\nonumber\\&&
=-e^2J_1^2J_2^2\tr[\sigma^i\sigma^k\hat u^\dagger]\tr[ \sigma^j\sigma^l\hat u^\dagger]
\tau^i_{\gamma_1\gamma_2}\tau^j_{\gamma_2\gamma_1}\tau^k_{\gamma_3\gamma_4}\tau^l_{\gamma_4\gamma_3}
\times\nonumber\\
&&(1-f_R(\epsilon_R))\eexp{-i\epsilon_R(t_3-t_1)}(-)f_L(\epsilon_L)\eexp{-i\epsilon_L(t_1-t_3)}
[\theta(t_2-t_4)(1-f_R(\epsilon_R'))-\theta(t_4-t_2)f_R(\epsilon_R')]\eexp{-i\epsilon_R'(t_4-t_2)}
\nonumber\\&&
[\theta(t_4-t_2)(1-f_L(\epsilon_L'))-\theta(t_2-t_4)f_L(\epsilon_L')]\eexp{-i\epsilon_L'(t_2-t_4)}
\eexp{-i\lambda_1(t_2-t_1)}(-)\eexp{-\beta\lambda_2}\eexp{-i\lambda_2(t_1-t_2)}
\nonumber\\&&
[\theta(t_3-t_4)-\theta(t_4-t_3)\eexp{-\beta\lambda_3}]\eexp{-i\lambda_{34}(t_4-t_3)}
[\theta(t_4-t_3)-\theta(t_3-t_4)\eexp{-\beta\lambda_4}]\nonumber\\&&
=-I_7(-i)\int_{\omega_1}[\frac{(1-f_R(\epsilon_R'))f_L(\epsilon_L')}{\omega_1+i\eta}
+\frac{(1-f_L(\epsilon_L'))f_R(\epsilon_R')}{-\omega_1+i\eta}]\eexp{-i\omega_1(t_2-t_4)
-i\epsilon_{RL}'(t_4-t_2)}\nonumber\\&&
(-i)\int_{\omega_2}[\frac{\eexp{-\beta\lambda_4}}{\omega_2+i\eta}
+\frac{\eexp{-\beta\lambda_3}}{-\omega_2+i\eta}]\eexp{-i\omega_2(t_3-t_4)
-i\epsilon_{RL}(t_3-t_1)}
\eexp{-\beta\lambda_2}\eexp{-i\lambda_{12}(t_2-t_1)-i\lambda_{34}(t_4-t_3)}
\eeq
For $S_{8a}$ need $\int_{t_1}\eexp{i\omega(t_1-t_2)}\int_{t_2,t_4}$, it yields 3 independent $\delta$ functions with $\omega+\epsilon_{RL}+\lambda_{12}=-\omega_1+\epsilon_{RL}'-\lambda_{12}=
-\omega-\epsilon_{RL}-\omega_2+\lambda_{34}=0$, hence
\beq{87}
&& S_{8a}=I_72\pi\delta(\omega+\epsilon_{RL}+\lambda_{12})
[\frac{(1-f_R(\epsilon_R'))f_L(\epsilon_L')}{\epsilon_{RL}'-\lambda_{12}+i\eta}
+\frac{(1-f_L(\epsilon_L'))f_R(\epsilon_R')}{-\epsilon_{RL}'+\lambda_{12}+i\eta}]\nonumber\\
&&\qquad [\frac{\eexp{-\beta\lambda_4}}{-\omega-\epsilon_{RL}+\lambda_{34}+i\eta}
+\frac{\eexp{-\beta\lambda_3}}{\omega+\epsilon_{RL}-\lambda_{34}+i\eta}]\eexp{-\beta\lambda_2}
\eeq
The 1st P.P. vanishes as in \eqref{e80}, while the 2nd  P.P. is then an imaginary term that cancels (see below), so remains the product of $\delta$ functions
\beq{88}
&&S_{8a}=-I_72\pi^3[(1-f_R(\epsilon_R'))f_L(\epsilon_L')+(1-f_L(\epsilon_L'))f_R(\epsilon_R')]
(\eexp{-\beta\lambda_3}+\eexp{-\beta\lambda_4})\eexp{-\beta\lambda_2}\times\nonumber\\
&&\qquad \delta(\omega+\epsilon_{RL}+\lambda_{12})\delta(\epsilon_{RL}'-\lambda_{12})
\delta(\omega+\epsilon_{RL}-\lambda_{34})
\eeq
This has a divergent $\delta(\lambda_{12}+\lambda_{34})$.

For $S_{8b}$ need $\int_{t_1}\eexp{i\omega(t_1-t_4)}\int_{t_2,t_3}$, it yields 3 independent $\delta$ functions with $\omega+\epsilon_{RL}+\lambda_{12}=-\omega_1+\epsilon_{RL}'-\lambda_{12}=
-\epsilon_{RL}-\omega_2+\lambda_{34}=0$, hence
\beq{89}
&& S_{8b}=I_72\pi\delta(\omega+\epsilon_{RL}+\lambda_{12})
[\frac{(1-f_R(\epsilon_R'))f_L(\epsilon_L')}{\epsilon_{RL}'-\lambda_{12}+i\eta}
+\frac{(1-f_L(\epsilon_L'))f_R(\epsilon_R')}{-\epsilon_{RL}'+\lambda_{12}+i\eta}]\nonumber\\
&&\qquad [\frac{\eexp{-\beta\lambda_4}}{-\epsilon_{RL}+\lambda_{34}+i\eta}
+\frac{\eexp{-\beta\lambda_3}}{\epsilon_{RL}-\lambda_{34}+i\eta}]\eexp{-\beta\lambda_2}\nonumber\\&&
=-I_72\pi^3[(1-f_R(\epsilon_R'))f_L(\epsilon_L')+(1-f_L(\epsilon_L'))f_R(\epsilon_R')]
(\eexp{-\beta\lambda_3}+\eexp{-\beta\lambda_4})\eexp{-\beta\lambda_2}\times\nonumber\\
&&\qquad \delta(\omega+\epsilon_{RL}+\lambda_{12})\delta(\epsilon_{RL}'-\lambda_{12})
\delta(\epsilon_{RL}-\lambda_{34})
\eeq
The 1st P.P. vanishes as in \eqref{e80}, while the 2nd  P.P. is then an imaginary term that cancels (see below), so remains the product of $\delta$ functions.

For $S_{8c}$ need $\int_{t_1}\eexp{i\omega(t_1-t_2)}\int_{t_3,t_4}$, with a $-$ relative to $S_{8b}$, it yields 3 independent $\delta$ functions with $\omega+\epsilon_{RL}+\lambda_{12}=
-\omega-\omega_1+\epsilon_{RL}'-\lambda_{12}=-\epsilon_{RL}-\omega_2+\lambda_{34}=0$, hence
\beq{90}
&& S_{8c}=-I_72\pi\delta(\omega+\epsilon_{RL}+\lambda_{12})
[\frac{(1-f_R(\epsilon_R'))f_L(\epsilon_L')}{-\omega+\epsilon_{RL}'-\lambda_{12}+i\eta}
+\frac{(1-f_L(\epsilon_L'))f_R(\epsilon_R')}{\omega-\epsilon_{RL}'+\lambda_{12}+i\eta}]\nonumber\\
&&\qquad [\frac{\eexp{-\beta\lambda_4}}{-\epsilon_{RL}+\lambda_{34}+i\eta}
+\frac{\eexp{-\beta\lambda_3}}{\epsilon_{RL}-\lambda_{34}+i\eta}]\eexp{-\beta\lambda_2}\nonumber\\&&
=+I_72\pi^3[(1-f_R(\epsilon_R'))f_L(\epsilon_L')+(1-f_L(\epsilon_L'))f_R(\epsilon_R')]
(\eexp{-\beta\lambda_3}+\eexp{-\beta\lambda_4})\eexp{-\beta\lambda_2}\times\nonumber\\
&&\qquad \delta(\omega+\epsilon_{RL}+\lambda_{12})\delta(-\omega+\epsilon_{RL}'-\lambda_{12})
\delta(\epsilon_{RL}-\lambda_{34})
\eeq
The 1st P.P. vanishes as in \eqref{e80}, while the 2nd  P.P. is then an imaginary term that cancels (see below), so remains the product of $\delta$ functions.
Both $S_{8b},S_{8c}$ have resonance $\delta(\omega-\lambda_{12}-\lambda_{34})$, however for $eV\gg\omega$ they cancel. As for diagram 7, $R\leftrightarrow L$ in the $t$ ordered fermion bubble is identical to the result above except for $\epsilon_{RL}'\rightarrow-\epsilon_{RL}'$ and a sign change in $S_{8b},\,S_{8c}$, hence a factor 2 for $S_{8a}$ while both $S_{8b},\,S_{8c}$ cancel.

Consider next $S_7'$ where spins $1\leftrightarrow 2$. Exactly as for $S_4'$ below Eq. \eqref{e62}, the interchange implies that the spin-orbit matrix $\hat u$ is now in vertices $t_1,t_2$ with $\sigma^i,\sigma^j$ (instead of $\hat u^\dagger$ at $t_3,t_4$ with $\sigma^k,\sigma^l$), hence the traces become $\tr[\hat u\sigma^i\sigma^k]\tr[\hat u\sigma^j\sigma^l]$. Also now $\gamma_1,\gamma_2$ belong to spin 2 while $\gamma_3,\gamma_4$ belong to spin 1. To revert to the previous notation relabel $\gamma_1,\gamma_2\leftrightarrow \gamma_3,\gamma_4$, hence need $i\leftrightarrow k$ and $j\leftrightarrow l$ as well as $\lambda_1,\lambda_2\leftrightarrow \lambda_3,\lambda_4$. This yields the same spin indices as in Eq. (\ref{e78}) with $I_7\rightarrow I_7'$ and the traces replaced by
$ \tr[\hat u\sigma^k\sigma^i]\tr\hat u\sigma^l\sigma^j]$. Note $\tr[\hat u\sigma^k\sigma^i]=\tr[\sigma^i\sigma^k\hat u^\dagger]^*=\tr[\sigma^i\sigma^k\hat u^\dagger]$ which happens to be real (subsection C), therefore $I_7'=I_7$.

Consider $S_{7a}'$ from \eqref{e80} with $\lambda_1,\lambda_2\leftrightarrow \lambda_3,\lambda_4$,
\beq{91}
S_{7a}'=&&I_72\pi\delta(\omega+\epsilon_{RL}-\lambda_{12})
[\frac{(1-f_R(\epsilon_R'))f_L(\epsilon_L')}{-\epsilon_{RL}'-\lambda_{12}+i\eta}
+\frac{(1-f_L(\epsilon_L'))f_R(\epsilon_R')}{\epsilon_{RL}'+\lambda_{12}+i\eta}]\nonumber\\
&&\qquad [\frac{\eexp{-\beta\lambda_4}}{-\lambda_{12}-\lambda_{34}+i\eta}
+\frac{\eexp{-\beta\lambda_3}}{\lambda_{12}+\lambda_{34}+i\eta}]\eexp{-\beta\lambda_1}
\eeq
From \eqref{e88}
$S_{7a}'=S_{8a}^*(\lambda_1\leftrightarrow\lambda_2,\lambda_3\leftrightarrow\lambda_4)$, which is just relabeling indices of the same spin, hence summation on spin indices allows keeping only the real parts.
Similarly $S_{7b}'=S_{8b}^*(\lambda_1\leftrightarrow\lambda_1,\lambda_3\leftrightarrow\lambda_4)$ and
$S_{7c}'=S_{8c}^*(\lambda_1\leftrightarrow\lambda_1,\lambda_3\leftrightarrow\lambda_4)$. This is not essential since anyway $S_{7b}'+S_{7c}'$ would cancel.

Finally $S_{8a}'$  from \eqref{e87} with $\lambda_1,\lambda_2\leftrightarrow \lambda_3,\lambda_4$,
\beq{92}
S_{8a}'=&&I_72\pi\delta(\omega+\epsilon_{RL}+\lambda_{34})
[\frac{(1-f_R(\epsilon_R'))f_L(\epsilon_L')}{\epsilon_{RL}'-\lambda_{34}+i\eta}
+\frac{(1-f_L(\epsilon_L'))f_R(\epsilon_R')}{-\epsilon_{RL}'+\lambda_{34}+i\eta}]\nonumber\\
&&\qquad [\frac{\eexp{-\beta\lambda_2}}{\lambda_{34}+\lambda_{12}+i\eta}
+\frac{\eexp{-\beta\lambda_1}}{-\lambda_{34}-\lambda_{12}+i\eta}]\eexp{-\beta\lambda_4}
\eeq
so that $S_{8a}'=S_{7a}^*(\lambda_1\leftrightarrow\lambda_2,\lambda_3\leftrightarrow\lambda_4)$,
similarly $S_{8b}'=S_{7b}^*(\lambda_1\leftrightarrow\lambda_2,\lambda_3\leftrightarrow\lambda_4)$
and $S_{8c}'=S_{7c}^*(\lambda_1\leftrightarrow\lambda_2,\lambda_3\leftrightarrow\lambda_4)$, allowing to keep only real parts.\\

{\bf Aharonov-Bohm phase $\chi$ }\\

We comment now on the modification due to adding a Aharonov-Bohm phase $\chi$ in the Hamiltonian, i.e. $\hat u\rightarrow \eexp{i\chi}\hat u$. Note first that for $S_4$ Eqs. (\ref{e57},\ref{e61}) terms with a single P.P. involve $i\eexp{2i\chi}$ that has a real part $\sim\sin 2\chi$ which does not cancel with $S'_4$; yet, this term is $\sim P.P.(\frac{1}{\omega+\lambda_{12}+\lambda_{34}})\ll \delta(\omega+\lambda_{12}+\lambda_{34})$ and a similar term without $\omega$, these are weakly $\omega$ dependent, i.e. they are not a resonance, and are neglected. Same with the P.P. for $S_7$, Eqs. (\ref{e80},\ref{e82},\ref{e84}) and $S_8$, Eqs. (\ref{e87},\ref{e89},\ref{e90}). Therefore,
the noise terms $S_{4a}+h.c.$ and $S_{4b}+h.c.$ are multiplied by $\cos 2\chi$, while all terms of $S_5$ are not changed. All terms $S_{7}$ acquire $\eexp{-2i\chi}$ while for their exchange $R\leftrightarrow L$ the phase $\chi$ cancels, hence $S_{7a}\rightarrow \half S_{7a}(\eexp{-2i\chi}+1)$; together with the h.c. term $S'_{8a}$ we have a factor $\half(\cos 2\chi +1)$. $S_{7b},S_{7c}$ acquire a factor $\half(\eexp{-2i\chi}-1)$, but anyway these two cancel $S_{7b}+S_{7c}=0$, same with $S_{8b},S_{8c}$ and the primed quantities.\\

\subsection{Summary -- two QD}

Consider first the divergent $\sim\delta(\lambda_{12}+\lambda_{34})$ terms, in $S_{5c}$ exchange integration variables $\epsilon_{RL}\leftrightarrow \epsilon'_{RL}$ since $I_4$ is symmetric in these.
\beq{93}
&&S_{4b}+h.c.=4\pi^3I_4\delta(\lambda_{12}+\lambda_{34})\delta(\omega+\epsilon_{RL}'-\lambda_{12})
\delta(\epsilon_{RL}-\lambda_{34})(\eexp{-\beta\lambda_1}+\eexp{-\beta\lambda_2})(\eexp{-\beta\lambda_3}
+\eexp{-\beta\lambda_4})\cos 2\chi\nonumber\\
&&S_{5b}+S_{5c}=8\pi^3I_4\delta(\lambda_{12}+\lambda_{34})\cdot 2\delta(\epsilon_{RL}-\lambda_{12})
\delta(\omega+\epsilon'_{RL}-\lambda_{34})\eexp{-\beta\lambda_2-\beta\lambda_3}\\
&&2(S_{7a}+h.c.)=-4\pi^3I_4\delta(\lambda_{12}+\lambda_{34})\delta(\omega+\epsilon_{RL}-\lambda_{34})
\delta(\epsilon'_{RL}+\lambda_{34})(\eexp{-\beta\lambda_1}+\eexp{-\beta\lambda_2})\eexp{-\beta\lambda_3}
(\cos 2\chi +1)\nonumber\\&&
2(S_{8a}+h.c.)=-4\pi^3I_4\delta(\lambda_{12}+\lambda_{34})\delta(\omega+\epsilon_{RL}-\lambda_{34})
\delta(\epsilon'_{RL}+\lambda_{34})(\eexp{-\beta\lambda_3}+\eexp{-\beta\lambda_4})\eexp{-\beta\lambda_2}
(\cos 2\chi +1)\nonumber
\eeq
where in $S_{7a},S_{8a}$ the factor $(1-f_L(\epsilon'_L))f_R(\epsilon'_R)$ is neglected for $V>0$ and large. In the generic case $g_1\neq g_2$ the factor $\delta(\lambda_{12}+\lambda_{34})$ implies $\lambda_{12}=\lambda_{34}=0$. The divergence cancels exactly (!) (even for $\chi\neq 0$), again using $\epsilon_{RL}\leftrightarrow \epsilon'_{RL}$ in $S_{7a},S_{8a}$.

 We proceed finally to our main result with the resonance terms in $S_{4b},S_{5b},S_{5c}$ which before normalization are
\beq{96}
&&\int_\epsilon S^{res,\,unren}(\omega)=\int_\epsilon(S_{4a}+S_{5a}+S_{5d}) +h.c.\nonumber\\&&
=4\pi^3e^2J_1^2J_2^2(eV)^2N^4(0)\tilde I_4\delta(\omega+\lambda_{12}+\lambda_{34})
(\eexp{-\beta\lambda_2}+\eexp{-\beta\lambda_1})(\eexp{-\beta\lambda_4}+\eexp{-\beta\lambda_3})
\cos 2\chi\nonumber\\
&&\qquad\qquad +(2\pi)^3e^2J_1^2J_2^2(eV)^2N^4(0)\tilde I_4\delta(\omega+\lambda_{12}+\lambda_{34})[\eexp{-\beta\lambda_2-\beta\lambda_3}
+\eexp{-\beta\lambda_1-\beta\lambda_4}]\nonumber\\
\eeq
Recall $\lambda_i=\lambda_0+\half\gamma_i\nu_1\,(i=1,2),\,\lambda_i=\lambda_0+\half\gamma_i\nu_2\,(i=3,4)$. Using the result for $\tilde I_4$ Eq. \ref{e209} and summing on $\gamma_i$ (each line indicates near its end the values of $\gamma_1\gamma_2\gamma_3\gamma_4$), we have before normalization,
\beq{97}
&&\int_\epsilon S^{res,\,unren}(\omega)=4\pi^3e^2J_1^2J_2^2(eV)^2N^4(0)\times\qquad\qquad \qquad \qquad \,\,\, ----,\,++++,\,++--,\,--++\nonumber\\
&&\left\{
4(1+\cos 2\chi)(\eexp{-\half\beta\nu_1-\half\beta\nu_2}+\eexp{-\half\beta\nu_1+\half\beta\nu_2}+
\eexp{\half\beta\nu_1-\half\beta\nu_2}+\eexp{\half\beta\nu_1\half\beta\nu_2})
4\cos^2\half\theta\cos^2\phi\right.\delta(\omega)\,\, \nonumber\\
&&+2(1+\cos 2\chi)\eexp{-\half\beta\nu_1}(\eexp{-\half\beta\nu_2}+\eexp{\half\beta\nu_2})4\sin^2\half\theta
[\delta(\omega+\nu_2)+\delta(\omega-\nu_2)]\qquad \qquad +++-,\,++-+\nonumber\\
&&+2(1+\cos 2\chi)\eexp{\half\beta\nu_1}(\eexp{-\half\beta\nu_2}+\eexp{\half\beta\nu_2})4\sin^2\half\theta
[\delta(\omega+\nu_2)+\delta(\omega-\nu_2)]\qquad\qquad  --+-,\,---+\nonumber\\
&&+2(1+\cos 2\chi)\eexp{-\half\beta\nu_2}(\eexp{-\half\beta\nu_1}+\eexp{\half\beta\nu_1})4\sin^2\half\theta
[\delta(\omega+\nu_1)+\delta(\omega-\nu_1)]\qquad\qquad  +-++,\,-+++\nonumber\\
&&+2(1+\cos 2\chi)\eexp{\half\beta\nu_2}(\eexp{-\half\beta\nu_1}+\eexp{\half\beta\nu_1})4\sin^2\half\theta
[\delta(\omega+\nu_1)+\delta(\omega-\nu_1)]\qquad  \qquad +---,\,-+--\nonumber\\
&&+[\cos 2\chi(\eexp{-\half\beta\nu_1}+\eexp{\half\beta\nu_1})(\eexp{-\half\beta\nu_2}+\eexp{\half\beta\nu_2})
+2(\eexp{\half\beta\nu_1+\half\beta\nu_2}+\eexp{-\half\beta\nu_1-\half\beta\nu_2})]
\times\nonumber\\
&&\qquad\qquad\qquad  \left. 16\cos^2\half\theta[\delta(\omega+\nu_1-\nu_2)
+\delta(\omega-\nu_1+\nu_2)]\right\}\qquad \qquad\qquad \qquad\,\, +--+,\,-++-
\eeq
It is remarkable that the terms $+-+-,\,-+-+$ vanish, i.e. no resonance at $\delta(\omega+\nu_1+\nu_2)$. Finally, normalize by
$(\eexp{-\half\beta\nu_1}+\eexp{\half\beta\nu_1})(\eexp{-\half\beta\nu_2}+\eexp{\half\beta\nu_2})$ to obtain\\
\beq{98}
\boxed{
\begin{aligned}
\int_\epsilon S^{res}(\omega)=&&
64\pi^3e^2J_1^2J_2^2(eV)^2N^4(0)\left\{\,\,(1+\cos 2\chi)\cos^2\half\theta\cos^2\phi\,\delta(\omega)
\right.\nonumber\\
&&+\half(1+\cos 2\chi)\sin^2\half\theta[\delta(\omega+\nu_1)+\delta(\omega-\nu_1)+\delta(\omega+\nu_2)
+\delta(\omega-\nu_2)]\nonumber
\\ &&\left.+\cos^2\half\theta(1+\cos 2\chi+\tanh\half\beta\nu_1\tanh\half\beta\nu_2)
[\delta(\omega+\nu_1-\nu_2)+\delta(\omega-\nu_1+\nu_2]\,\,\right\}\nonumber
\end{aligned}}
 \\
\eeq
\\
\section{Current Noise via Keldysh -- single QD}

\subsection{Tools}

Consider noise for the J-W model, i.e. the effective action and current are
\beq{99}
&&S_{eff}=S_0+\sum_\pm \pm[J\bar\psi_{R\pm}{\bm \sigma}\psi_{L\pm}\bar f_\pm{\bm \tau}f_\pm
+W\eexp{-i\chi}\bar\psi_{R\pm}\hat u^\dagger\psi_{L\pm}+h.c.]\nonumber\\
&&j=ie\sum_\pm [J\bar\psi_{R\pm}{\bm \sigma}\psi_{L\pm}\bar f_\pm{\bm \tau}f_\pm
+W\eexp{-i\chi}\bar\psi_{R\pm}\hat u^\dagger\psi_{L\pm}]+h.c.
\eeq
$\chi$ represents an Aharonov-Bohm phase.

The diagram rules are similar to those of the double QD case except for a new type of vertex, denoted as X in the figures, where an L electron is directly transferred to an R electron, or vice versa, with strength $W$.

We evaluate now a few traces, needed in the following:
\beq{102}
&&\tr[\sigma^i\hat u]=\tr[\sigma^i\eexp{i\sigma_z\phi}\eexp{i\sigma_y\theta/2}]=
\tr[\sigma^i(\cos\phi+i\sigma_z\sin\phi)(\cos\half\theta+i\sigma_y\sin\half\theta)]\nonumber\\&&
\tr[\sigma^z\hat u]=2i\sin\phi\cos\half\theta\nonumber\\&&
\tr[\sigma^x\hat u]=2i\sin\phi\sin\half\theta\nonumber\\&&
\tr[\sigma^y\hat u]=2i\cos\phi\sin\half\theta\nonumber\\
&&\tau^z_{\gamma_1\gamma_2}\tau^z_{\gamma_2\gamma_1}=\delta_{\gamma_1,\gamma_2}\nonumber\\
&&\tau^x_{\gamma_1\gamma_2}\tau^x_{\gamma_2\gamma_1}=\tau^y_{\gamma_1\gamma_2}\tau^y_{\gamma_2\gamma_1}
=\delta_{\gamma_1,-\gamma_2}\nonumber\\
&&\tau^z_{\gamma_1\gamma_2}\tau^x_{\gamma_2\gamma_1}=\tau^z_{\gamma_1\gamma_2}\tau^y_{\gamma_2\gamma_1}
=0\nonumber\\
&&\tau^x_{\gamma_1\gamma_2}\tau^y_{\gamma_2\gamma_1}=
-\tau^y_{\gamma_1\gamma_2}\tau^x_{\gamma_2\gamma_1}=i\delta_{\gamma_1,-\gamma_2}\mbox{sign}\gamma_1
\eeq
Define $I_W$, note that all $\tr$ are imaginary,
\beq{103}
I^W=\sum_{ij}\tr[\sigma^i\hat u]\tr[\sigma^j\hat u^\dagger]\tau^i_{\gamma_1\gamma_2}\tau^j_{\gamma_2\gamma_1}
=4\sin^2\phi\cos^2\half\theta\delta_{\gamma_1\gamma_2}+4\sin^2\half\theta\delta_{\gamma_1,-\gamma_2}
\eeq

\subsection{n=0}

  \begin{figure}  \centering
\includegraphics [width=.5\textwidth]{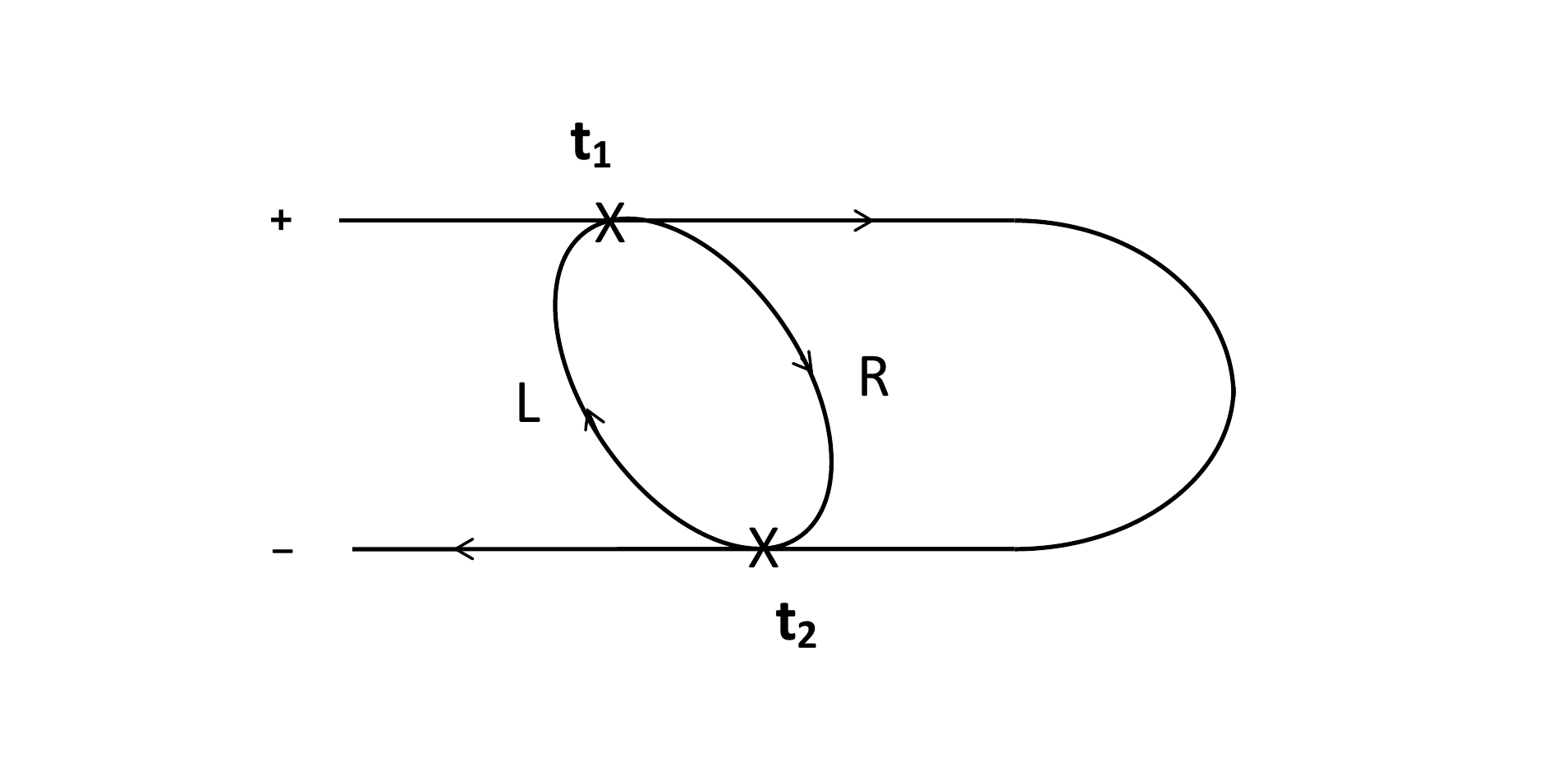}
\caption{Diagram 1W}
\label{diagram1W}
\end{figure}
From Fig. \ref{diagram1W}, for large $eV$ and using $\tr[\hat u^\dagger \hat u]=2$,
\beq{100}
&&S_1^W(t_1,t_2)=e^2W^2\langle \bar\psi_{R+}(t_1)\hat u^\dagger _{\alpha\beta}\psi_{L+}(t_1)
\bar\psi_{L-}(t_2)\hat u_{\alpha'\beta'}\psi_{R-}(t_2)\rangle=\nonumber\\&&
e^2W^2(-i)G^>_R(t_2-t_1)iG^<_L(t_1-t_2)\delta_{\alpha\beta'}\delta_{\beta\alpha'}\hat u^\dagger_{\alpha\beta}\hat u_{\beta\alpha}\nonumber\\&&
S_1^W(\omega)=2e^2W^2\int_{t_1}\eexp{i\omega(t_1-t_2)}(1-f_R(\epsilon_R))\eexp{-i\epsilon_{RL}(t_2-t_1)}
f_L(\epsilon_L)=4\pi e^2W^2N^2(0)eV\nonumber\\&&
S_1^W+S_1^J=2\pi e^2(2W^2+3J^2)eVN^2(0)
\eeq
adding the J term from Eq. \eqref{e36}. The DC current is
\beq{101}
j_{DC}^W=ieW^2\langle \bar\psi_{R+}\hat u^\dagger\psi_{L+}(-i)\bar\psi_{L-}\hat u\psi_{R-}\rangle=S_1^W/e
\eeq

\subsection{Order n=2}
From Fig. \ref{diagram4W}, left side,
\beq{104}
&&S_4^W=e^2J^2W^2\left\langle [\stackrel{1}{\bar\psi}_{R+}\bm\sigma_{\alpha_1\beta_1}\stackrel{2}{\psi}_{L+}\cdot \stackrel{3}{\bar f}_{+}\bm\tau_{\gamma_1\delta_1}\stackrel{4}{f}_{+}]_{t_1}[\stackrel{5}{\bar\psi}_{R+}\bm
\sigma_{\alpha_2\beta_2}\stackrel{6}{\psi}_{L+}\cdot\stackrel{4}{\bar f}_{+}\bm\tau_{\gamma_2\delta_2}\stackrel{3}{f}_{+}]_{t_2}\right.\nonumber\\
&&\left.\qquad [\stackrel{2}{\bar\psi}_{L-} \hat u_{\alpha_3\beta_3}\stackrel{1}{\psi}_{R-}]_{t_3} [\stackrel{6}{\bar\psi}_{L-} \hat u_{\alpha_4\beta_4}\cdot\stackrel{5}{\psi}_{R-}]_{t_4}\right\rangle\eexp{2i\chi}\nonumber\\
&&=e^2J^2W^2(-i)G^>_R(t_3-t_1)iG_L^<(t_1-t_3)(-i)G_R^>(t_4-t_2)iG_L^<(t_2-t_4)
(-i)F^t_{1\gamma_1}(t_2-t_1)iF^t_{1\gamma_2}(t_1-t_2)\times\nonumber\\&&
 \delta_{\alpha_1\beta_3}\delta_{\alpha_3\beta_1}\delta_{\alpha_2\beta_4}\delta_{\alpha_4\beta_2}
\delta_{\gamma_1\delta_2}\delta_{\gamma_2\delta_1}
\sigma^i_{\alpha_1\beta_1}\tau^i_{\gamma_1\delta_1}
\sigma^j_{\alpha_2\beta_2}\tau^j_{\gamma_2\delta_2}
\hat u_{\alpha_3\beta_3}\hat u_{\alpha_4\beta_4}\eexp{2i\chi}
\eeq
This is identical to $S_4$ in \eqref{e54} except traces that are replaced by $-I_W$ and that
$(-i)F^{\bar t}_{2\gamma_3}(t_4-t_3)iF_{2\gamma_4}^{\bar t}(t_3-t_4)$ is missing. The latter is achieved by $\lambda_3,\lambda_4=0$ in the results. Hence from \eqref{e57}
\beq{105}
&&I_4^W=-e^2J^2W^2(1-f_R(\epsilon_R))(1-f_R(\epsilon_R'))f_L(\epsilon_L)f_L(\epsilon_L')I^W\nonumber\\
&&S_{4a}^W=i(2\pi)^2I_4^W\delta(\epsilon_{RL})\delta(\epsilon_{RL}'+\omega)
[\frac{\eexp{-\beta\lambda_2}}{-\omega-\lambda_{12}+i\eta}+
\frac{\eexp{-\beta\lambda_1}}{\omega+\lambda_{12}+i\eta}]\eexp{2i\chi}\nonumber\\
&&\int_\epsilon S_{4a}^W=-4\pi^2J^2W^2(eV)^2N^4(0)I^W[\pi\delta(\omega+\lambda_{12})
(\eexp{-\beta\lambda_2}+\eexp{-\beta\lambda_1})\nonumber\\
&&\qquad\qquad+iP.P.(\frac{1}{\omega+\lambda_{12}})
(\eexp{-\beta\lambda_1}-\eexp{-\beta\lambda_2})]\eexp{2i\chi}
\eeq
 \begin{figure*}  \centering
\includegraphics [width=.45\textwidth]{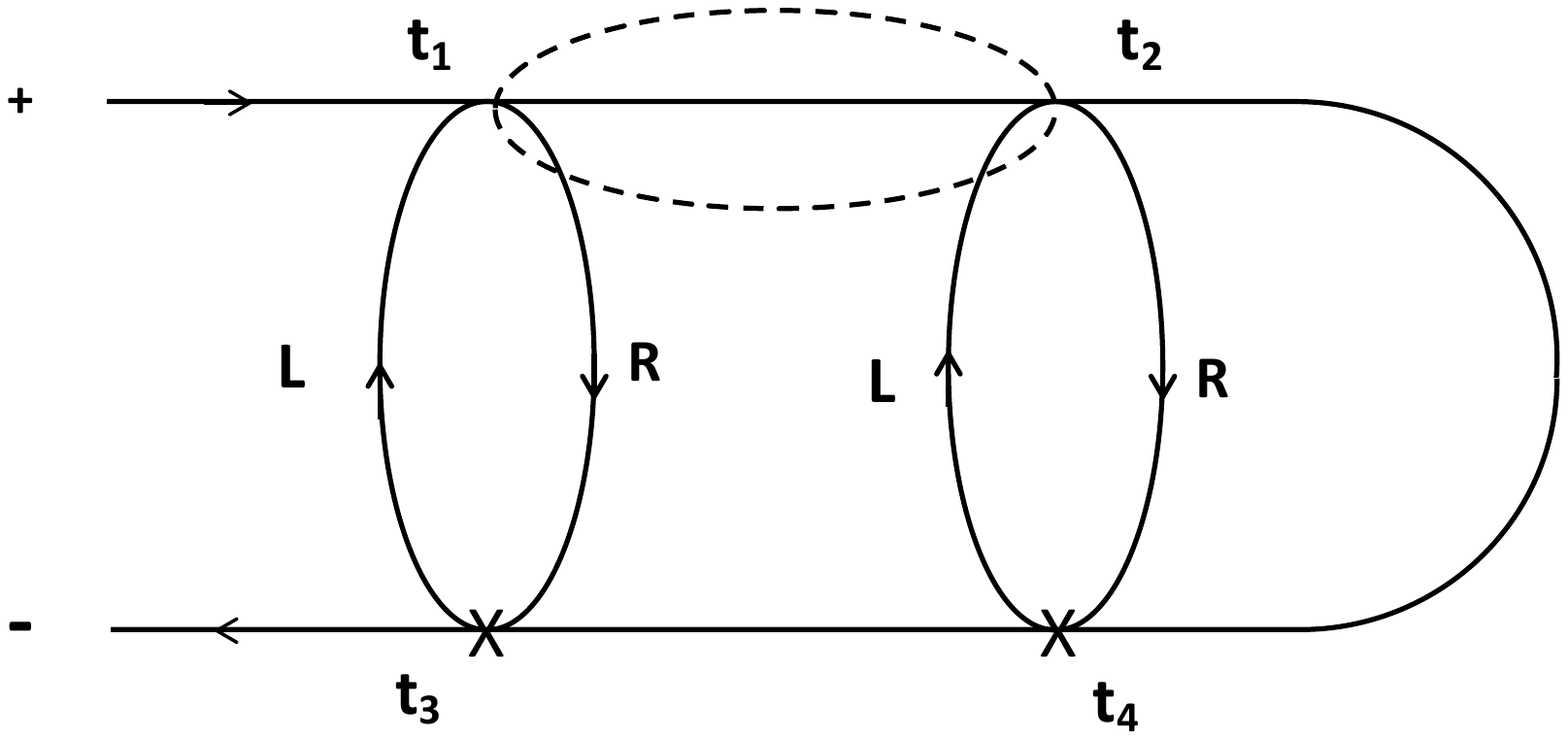}
\includegraphics [width=.45\textwidth]{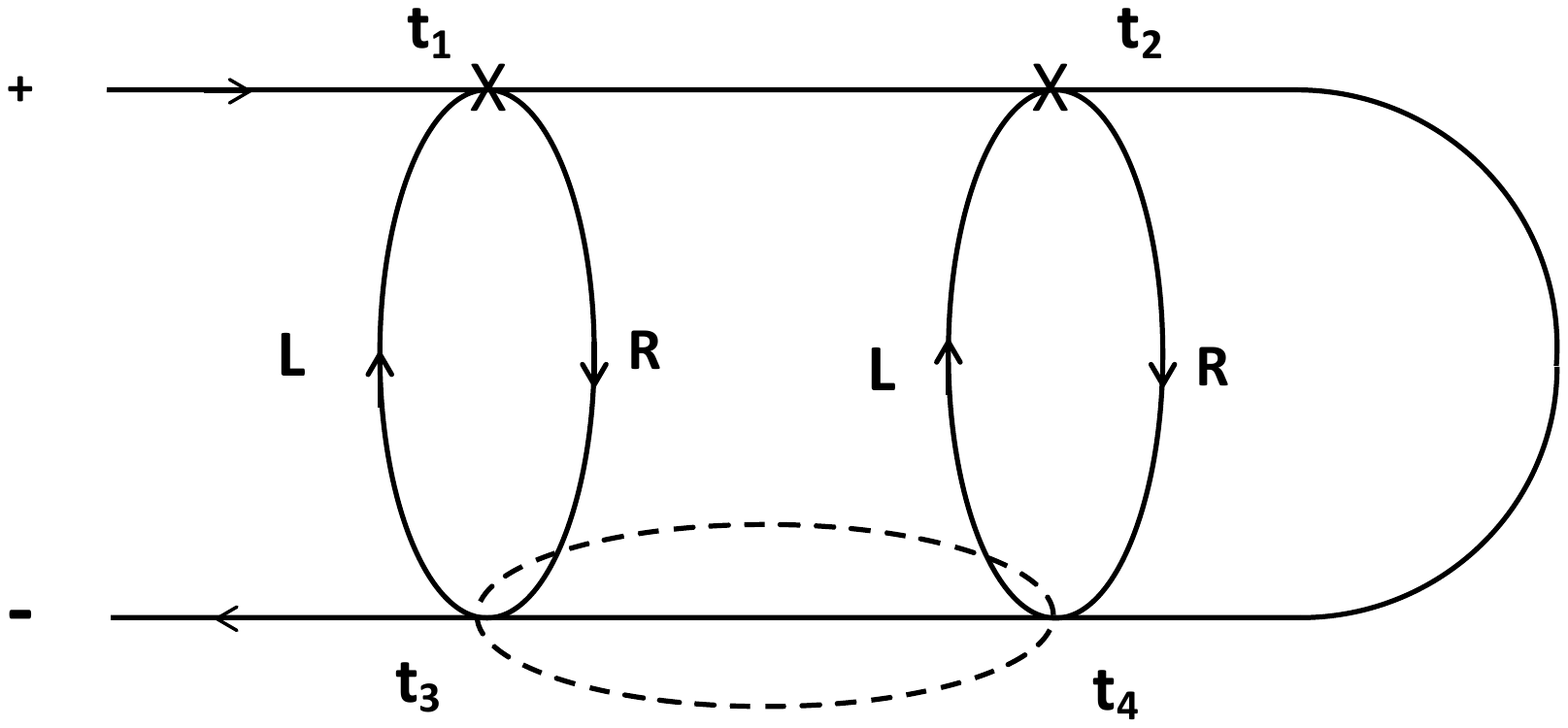}
\caption{Diagram 4W (left) and 4'W (right)}
\label{diagram4W}
\end{figure*}
 From \eqref{e61} with $\lambda_3,\lambda_4=0$
\beq{106}
&&S_{4b}^W=i(2\pi)^2I_4^W\delta(\epsilon_{RL})\delta(\epsilon_{RL}'+\omega)
[\frac{\eexp{-\beta\lambda_2}}{-\lambda_{12}+i\eta}+
\frac{\eexp{-\beta\lambda_1}}{\lambda_{12}+i\eta}]\eexp{2i\chi}\nonumber\\
&&\int_\epsilon S_{4b}^W=-4\pi^2e^2J^2W^2(eV)^2N^4(0)I^W[\pi\delta(\lambda_{12})
(\eexp{-\beta\lambda_2}+\eexp{-\beta\lambda_1})\nonumber\\&&
\qquad\qquad  +iP.P.(\frac{1}{\lambda_{12}})(\eexp{-\beta\lambda_1}-\eexp{-\beta\lambda_2})]\eexp{2i\chi}
\eeq

Consider next Fig. \ref{diagram4W}, right side,
\beq{107}
&&S_4'^W=e^2J^2W^2\left\langle
[\stackrel{2}{\bar\psi}_{R+} \hat u_{\alpha_3\beta_3}^\dagger\stackrel{1}{\psi}_{L+}]_{t_1} [\stackrel{6}{\bar\psi}_{R+} \hat u_{\alpha_4\beta_4}^\dagger\cdot\stackrel{5}{\psi}_{L+}]_{t_2}\right.\nonumber\\
&&\left.[\stackrel{1}{\bar\psi}_{L-}\bm\sigma_{\alpha_1\beta_1}\stackrel{2}{\psi}_{R-}\cdot \stackrel{3}{\bar f}_{-}\bm\tau_{\gamma_1\delta_1}\stackrel{4}{f}_{-}]_{t_3}[\stackrel{5}{\bar\psi}_{L-}\bm
\sigma_{\alpha_2\beta_2}\stackrel{6}{\psi}_{R-}\cdot\stackrel{4}{\bar f}_{-}\bm\tau_{\gamma_2\delta_2}\stackrel{3}{f}_{-}]_{t_4}\right\rangle\eexp{-2i\chi}\nonumber\\
&&=e^2J^2W^2(-i)G^>_R(t_3-t_1)iG_L^<(t_1-t_3)(-i)G_R^>(t_4-t_2)iG_L^<(t_2-t_4)
(-i)F^{\bar t}_{1\gamma_1}(t_4-t_3)iF^{\bar t}_{1\gamma_2}(t_3-t_4)\times\nonumber\\&&
 \delta_{\alpha_1\beta_3}\delta_{\alpha_3\beta_1}\delta_{\alpha_2\beta_4}\delta_{\alpha_4\beta_2}
\delta_{\gamma_1\delta_2}\delta_{\gamma_2\delta_1}
\sigma^i_{\alpha_1\beta_1}\tau^i_{\gamma_1\delta_1}
\sigma^j_{\alpha_2\beta_2}\tau^j_{\gamma_2\delta_2}
\hat u_{\alpha_3\beta_3}^\dagger\hat u_{\alpha_4\beta_4}^\dagger\eexp{-2i\chi}
\eeq
By $t_2\leftrightarrow t_3,\,t_1\leftrightarrow t_4$ it is seen from \eqref{e104} that $S_4'^W=S_4^{W*}$, noting that $I^W$ is invariant under $\hat u\rightarrow \hat u^\dagger$. Furthermore, $\omega$ is defined as incoming on the upper times $t_1,t_2$, hence the change in time variables needs, for both $S'^W_{4a},S'^W_{4b}$ that $S_4'^W(-\omega)=\int_t S_4^{W*}\eexp{-i\omega t}=S_4^{W*}(\omega)$. Hence the $\delta(...)$ terms in \eqref{e105}, \eqref{e106} involve $\cos 2\chi$ while the P.P. terms involve $\sin 2\chi$. The latter are actually neglected since e.g. $P.P.(\frac{1}{\omega+\lambda_{12}})\ll\delta(\omega+\lambda_{12})$, i.e. these are not resonance terms (yet this term is asymmetric around a resonance, so might be of some interest).

 \begin{figure*}  \centering
\includegraphics [width=.45\textwidth]{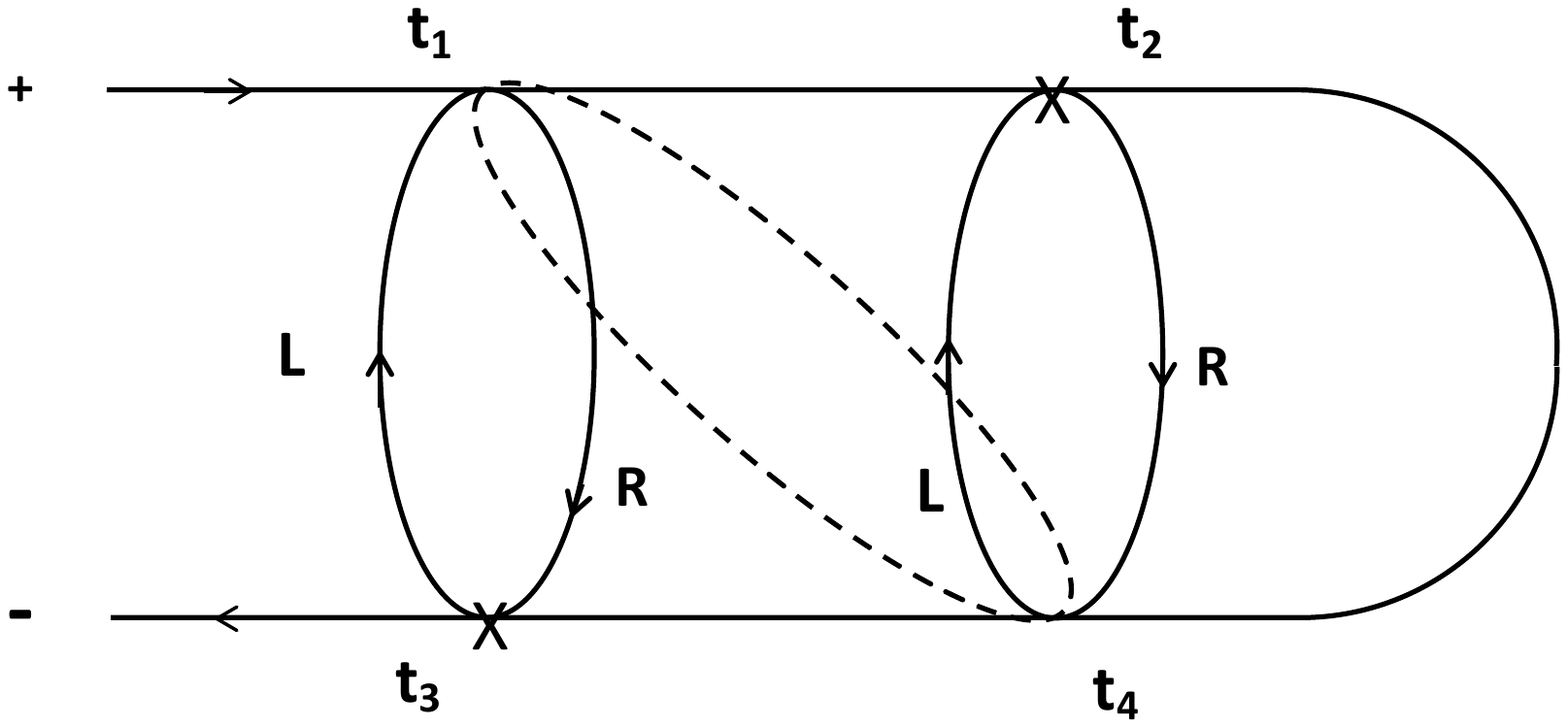}
\caption{Diagram 5W}
\label{diagram5W}
\end{figure*}

Consider next Fig. \ref{diagram5W}, which is $\chi$ independent.
\beq{108}
&&S_5^W=i^4e^2J^2W^2\left\langle [\stackrel{1}{\bar\psi}_{R+}\bm\sigma_{\alpha_1\beta_1}\stackrel{2}{\psi}_{L+}\cdot \stackrel{3}{\bar f}_{+}\bm\tau_{\gamma_1\delta_1}\stackrel{4}{f}_{+}]_{t_1}[\stackrel{5}{\bar\psi}_{R+}\hat u^\dagger _{\alpha_2\beta_2}\stackrel{6}{\psi}_{L+}]_{t_2}\right.\nonumber\\
&&\left.\qquad [\stackrel{2}{\bar\psi}_{L-}\hat u_{\alpha_3\beta_3}\stackrel{1}{\psi}_{R-}]_{t_3} [\stackrel{6}{\bar\psi}_{L-}\bm\sigma_{\alpha_4\beta_4}\cdot\stackrel{5}{\psi}_{R-}\stackrel{4}{\bar f}_{-}\bm\tau_{\gamma_2\delta_2} \stackrel{3}{f}_{-}]_{t_4}\right\rangle\nonumber\\
&&=e^2J^2W^2(-i)G^>_R(t_3-t_1)iG_L^<(t_1-t_3)(-i)G_R^>(t_4-t_2)iG_L^<(t_2-t_4)
 (-i)F^>_{\gamma_1}(t_4-t_1)iF^<_{\gamma_2}(t_1-t_4)\times\nonumber\\&&
\delta_{\alpha_1\beta_3}\delta_{\alpha_3\beta_1}\delta_{\alpha_2\beta_4}\delta_{\alpha_4\beta_2}
\delta_{\gamma_1\delta_2}\delta_{\gamma_2\delta_1}
\sigma^i_{\alpha_1\beta_1}\tau^i_{\gamma_1\delta_1}
\hat u_{\alpha_2\beta_2}\sigma^j_{\alpha_4\beta_4}\tau^j_{\gamma_2\delta_2} \hat u^\dagger_{\alpha_3\beta_3}\\
&&=I^W(1-f_R(\epsilon_R))\eexp{-i\epsilon_{RL}(t_3-t_1)}(-)f_L(\epsilon_L)(1-f_R(\epsilon_R'))
\eexp{-i\epsilon_{RL}'(t_4-t_2)}(-)f_L(\epsilon_L')(-)\eexp{-i\lambda_{12}(t_4-t_1)}
\eexp{-\beta\lambda_2}\nonumber
\eeq
Comparing with \eqref{e64} shows that it needs $\lambda_3,\lambda_4=0$ to obtain $S_5^W$. Hence from
Eqs. (\ref{e65}-\ref{e68})
\beq{109}
&&\int_\epsilon S_{5a}^W=(2\pi)^3I^W e^2J^2W^2(eV)^2N^4(0)\delta(\omega+\lambda_{12})\eexp{-\beta\lambda_2}\nonumber\\&&
\int_\epsilon S_{5b}^W=(2\pi)^3I^W e^2J^2W^2(eV)^2N^4(0)\delta(\lambda_{12})\eexp{-\beta\lambda_2}\nonumber\\&&
\int_\epsilon S_{5c}^W=(2\pi)^3I^W e^2J^2W^2(eV)^2N^4(0)\delta(\lambda_{12})\eexp{-\beta\lambda_2}\nonumber\\&&
\int_\epsilon S_{5d}^W=(2\pi)^3I^W e^2J^2W^2(eV)^2N^4(0)\delta(\omega-\lambda_{12})\eexp{-\beta\lambda_2}\nonumber\\&&
\eeq

 \begin{figure*}  \centering
\includegraphics [width=.45\textwidth]{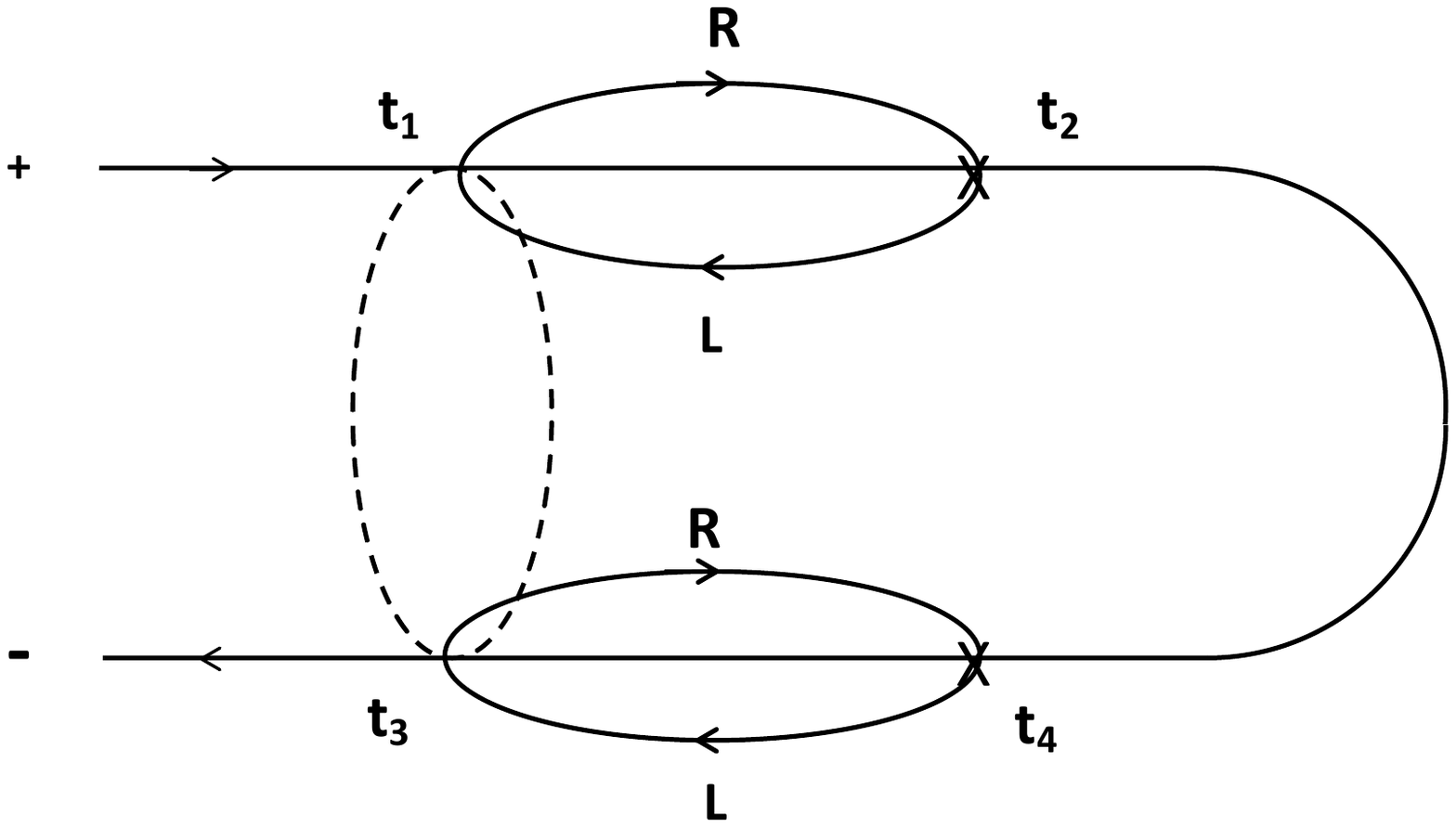}
\caption{Diagram 6W}
\label{diagram6W}
\end{figure*}

Consider next Fig. \ref{diagram6W}.
\beq{110}
&&S_6^W=e^2J^2W^2\left\langle [\stackrel{1}{\bar\psi}_{R+}\bm\sigma_{\alpha_1\beta_1}\stackrel{2}{\psi}_{L+}\cdot \stackrel{3}{\bar f}_{+}\bm\tau_{\gamma_1\delta_1}\stackrel{4}{f}_{+}]_{t_1}[\stackrel{2}{\bar\psi}_{L+}\hat u_{\alpha_2\beta_2}\stackrel{1}{\psi}_{R+}]_{t_2}\right.\nonumber\\
&&\left.\qquad [\stackrel{7}{\bar\psi}_{R-}\bm\sigma _{\alpha_3\beta_3}\stackrel{8}{\psi}_{L-}\cdot\stackrel{4}{\bar f}_{-}\bm\tau_{\gamma_2\delta_2}\stackrel{3}{f}_{-}]_{t_3} [\stackrel{8}{\bar\psi}_{L-}\hat u_{\alpha_4\beta_4}\stackrel{7}{\psi}_{R-}]_{t_4}\right\rangle\eexp{2i\chi}
\eeq
This equals $S_6$ if $-iF_{2\gamma_3}^>(t_4-t_2)iF_{2\gamma_4}^<(t_2-t_4)\rightarrow 1$ and $I_6\rightarrow I^W$. Hence $\lambda_3,\lambda_4=0$, Eq. \eqref{e77}, with an overall factor $\eexp{2i\chi}$, shows then that $S_6^W$ is negligible for $\lambda_{12}\ll eV$. Fig. \ref{diagram6W} with $R\leftrightarrow L$ in either bubble involves either $|\eexp{i\chi}|=1$ or $\eexp{-2i\chi}$, yet summation on the 4 options of external sources leads to cancellation of either term, as in \eqref{e77}.

 \begin{figure*}  \centering
\includegraphics [width=.45\textwidth]{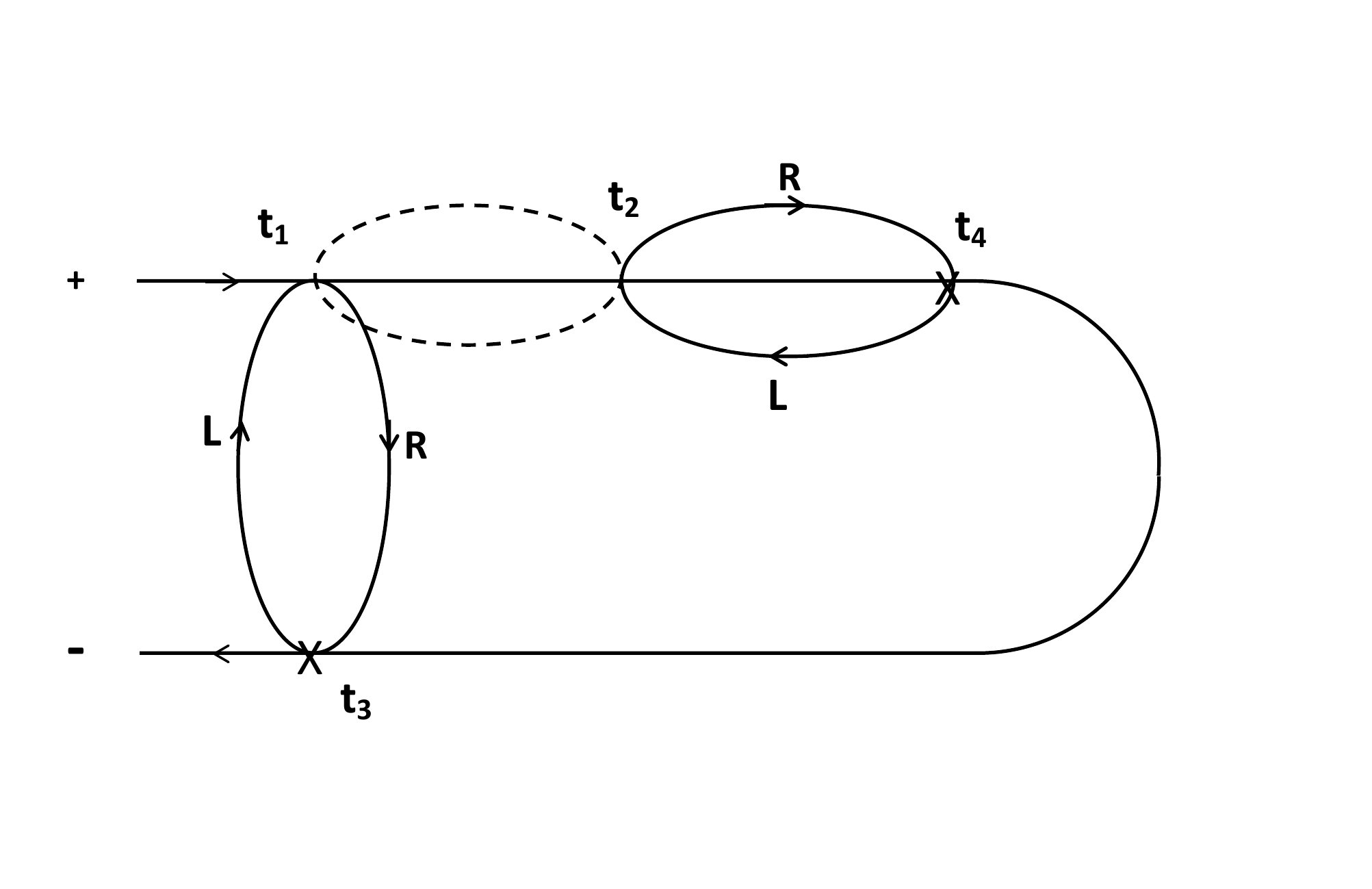}
\includegraphics [width=.45\textwidth]{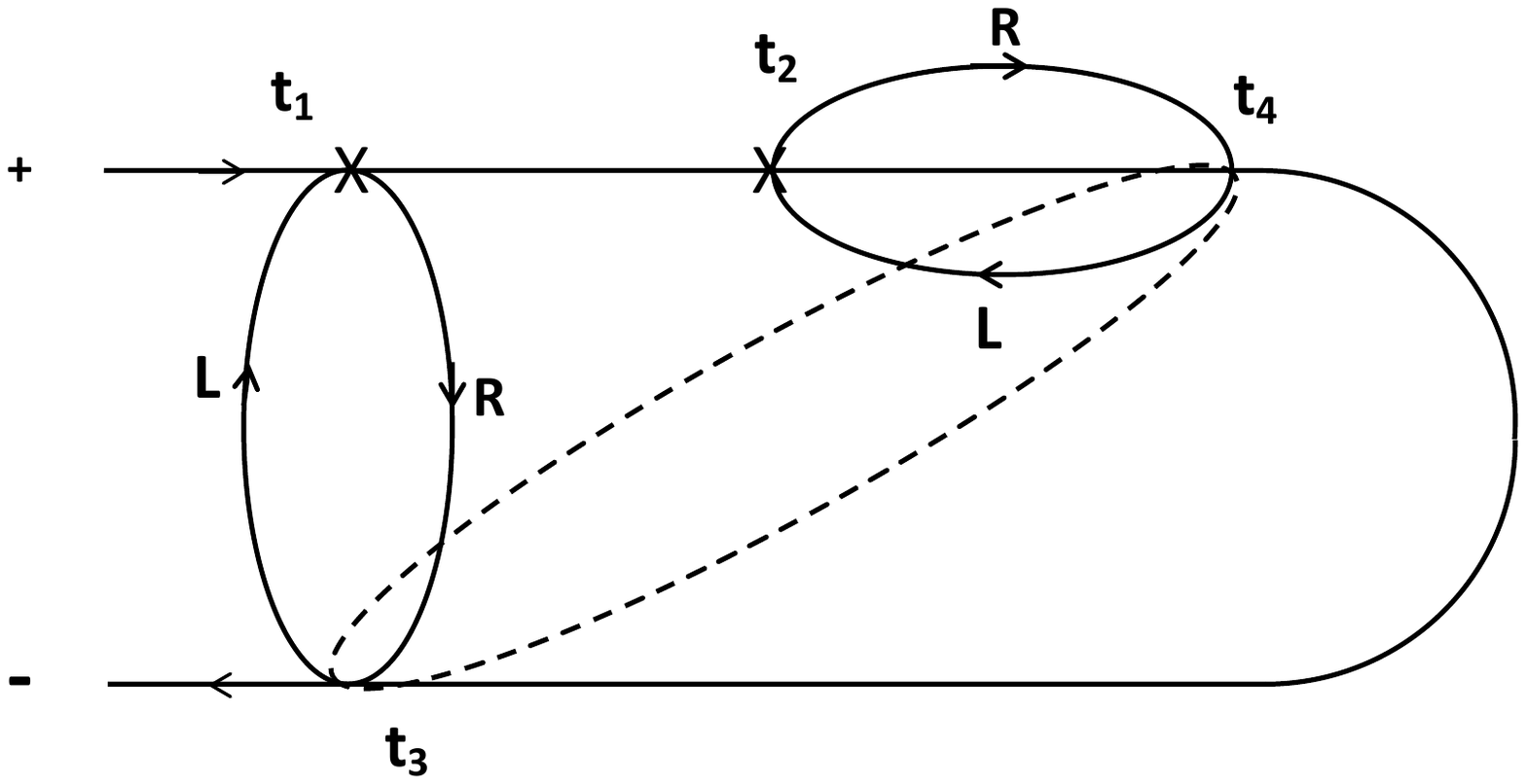}
\caption{Diagram 7W (left) and 7'W (right)}
\label{diagram7W}
\end{figure*}

Consider now Fig. \ref{diagram7W}, sign assumes that external currents are at $t_1,t_3$,
\beq{111}
&&S_7^W=-i^4e^2J_1^2J_2^2\left\langle [\stackrel{1}{\bar\psi}_{R+}\bm\sigma_{\alpha_1\beta_1}\stackrel{2}{\psi}_{L+}\cdot \stackrel{3}{\bar f}_{+}\bm\tau_{\gamma_1\delta_1}\stackrel{4}{f}_{+}]_{t_1}[\stackrel{5}{\bar\psi}_{R+}\bm
\sigma_{\alpha_2\beta_2}\stackrel{6}{\psi}_{L+}\cdot\stackrel{4}{\bar f}_{+}\bm\tau_{\gamma_2\delta_2}\stackrel{3}{f}_{+}]_{t_2}\right.\nonumber\\
&&\left.\qquad [\stackrel{2}{\bar\psi}_{L-}\hat u_{\alpha_3\beta_3}\stackrel{1}{\psi}_{R-}]_{t_3} [\stackrel{6}{\bar\psi}_{L+} \hat u_{\alpha_4\beta_4}\cdot\stackrel{5}{\psi}_{R+}]_{t_4}\right\rangle\eexp{2i\chi}\nonumber\\
&&=-e^2J_1^2J_2^2(-i)G^>_R(t_3-t_1)iG_L^<(t_1-t_3)(-i)G_R^t(t_4-t_2)iG_L^t(t_2-t_4)
 (-i)F^t_{\gamma_1}(t_2-t_1)iF^t_{\gamma_2}(t_1-t_2)\times\nonumber\\&&
 \delta_{\alpha_1\beta_3}\delta_{\alpha_3\beta_1}\delta_{\alpha_2\beta_4}\delta_{\alpha_4\beta_2}
\delta_{\gamma_1\delta_2}\delta_{\gamma_2\delta_1}
\sigma^i_{\alpha_1\beta_1}\tau^i_{\gamma_1\delta_1}
\sigma^j_{\alpha_2\beta_2}\tau^j_{\gamma_2\delta_2} \hat u_{\alpha_3\beta_3}\hat u_{\alpha_4\beta_4}
\eexp{2i\chi}
\eeq
This equals $S_7$ in \eqref{e78} if $\lambda_3,\lambda_4=0$ and $\tilde I_4\rightarrow -I^W$ ($-$ since here $\tr[\sigma^i\hat u]\tr[\sigma^j\hat u]$). Hence from Eqs. (\ref{e80}, \ref{e82}, \ref{e84}), where the 1st P.P. vanishes and the dominant term at $eV\gg k_BT$ is $(1-f_R(\epsilon_R'))f_L(\epsilon_L')$
\beq{112}
&&S_{7a}^W=-I^W2\pi\delta(\omega+\epsilon_{RL})(1-f_R(\epsilon_R))f_L(\epsilon_L)
\cdot (-i)\pi\delta(\epsilon'_{RL})(1-f_R(\epsilon_R'))f_L(\epsilon_L')\nonumber\\&&
\qquad \times[\frac{\eexp{-\beta\lambda_2}}{-\lambda_{12}+i\eta}
+\frac{\eexp{-\beta\lambda_1}}{\lambda_{12}+i\eta}](\eexp{2i\chi}-1)\nonumber\\
&& S_{7b}^W=-I^W2\pi\delta(\omega+\epsilon_{RL})(1-f_R(\epsilon_R))f_L(\epsilon_L)
\cdot (-i)\pi\delta(\epsilon'_{RL})(1-f_R(\epsilon_R'))f_L(\epsilon_L')\nonumber\\
&&\qquad \times[\frac{\eexp{-\beta\lambda_2}}{-\epsilon_{RL}-\lambda_{12}+i\eta}
+\frac{\eexp{-\beta\lambda_1}}{\epsilon_{RL}+\lambda_{12}+i\eta}](\eexp{2i\chi}+1)\nonumber\\
&& S_{7c}^W=I^W2\pi\delta(\omega+\epsilon_{RL})(1-f_R(\epsilon_R))f_L(\epsilon_L)
\cdot (-i)\pi\delta(\epsilon'_{RL}-\omega)(1-f_R(\epsilon_R'))f_L(\epsilon_L')\nonumber\\
&&\qquad [\frac{\eexp{-\beta\lambda_2}}{-\epsilon_{RL}-\lambda_{12}+i\eta}
+\frac{\eexp{-\beta\lambda_1}}{\epsilon_{RL}+\lambda_{12}+i\eta}](\eexp{2i\chi}+1)
\eeq
where the additional $\pm 1$ terms in $(\eexp{2i\chi}\pm 1)$ arise from the
$R\leftrightarrow L$ in the $t$ ordered fermion bubble as obtained from the result above except for $\epsilon_{R}'\leftrightarrow\epsilon_{L}'$; however one trace involves $\tr[\sigma^j\hat u^\dagger]=-\tr[\sigma^j\hat u]$ (leading to $-1$), but also a sign change for a current vertex for $S_{7b}^W,\,S_{7c}^W$ (leading to $+1$). The terms with $P.P.(\frac{1}{\omega+\lambda_{12}})$ or $P.P.\frac{1}{\lambda_{12}}$ are neglected, being much weaker than resonance. Note that for $eV\gg\omega$ the sum $S^W_{7b}+S^W_{7c}=0$.

Consider next Fig. \ref{diagram7W} right side,
\beq{113}
&&S_7'^W=-i^4e^2J^2W^2\left\langle [\stackrel{1}{\bar\psi}_{R+}u^\dagger_{\alpha_1\beta_1}\stackrel{2}{\psi}_{L+}]_{t_1}
[\stackrel{5}{\bar\psi}_{R+}
u_{\alpha_2\beta_2}^\dagger\stackrel{6}{\psi}_{L+}]_{t_2}\right.\nonumber\\
&&\left.\qquad [\stackrel{2}{\bar\psi}_{L-}\bm\sigma_{\alpha_3\beta_3}\stackrel{1}{\psi}_{R-}\cdot\stackrel{7}{\bar f}_{-}\bm\tau_{\gamma_1\delta_1}\stackrel{8}{f}_{-}]_{t_3} [\stackrel{6}{\bar\psi}_{L+}(\bm\sigma_{\alpha_4\beta_4}\stackrel{5}{\psi}_{R+}\cdot\stackrel{8}{\bar f}_{+}\bm\tau_{\gamma_2\delta_2} \stackrel{7}{f}_{2+}]_{t_4}\right\rangle\eexp{-2i\chi}\nonumber\\
&&=-e^2J2W^2(-i)G^>_R(t_3-t_1)iG_L^<(t_1-t_3)(-i)G_R^t(t_4-t_2)iG_L^t(t_2-t_4)
 (-i)F^{<}_{\gamma_1}(t_4-t_3)
iF_{\gamma_2}^{>}(t_3-t_4)\nonumber\\&&
\qquad\times \delta_{\alpha_1\beta_3}\delta_{\alpha_3\beta_1}\delta_{\alpha_2\beta_4}\delta_{\alpha_4\beta_2}
\delta_{\gamma_1\delta_2}\delta_{\gamma_2\delta_1}
\sigma^i_{\alpha_3\beta_3}\tau^i_{\gamma_1\delta_1}
\sigma^j_{\alpha_4\beta_4}\tau^j_{\gamma_2\delta_2}\hat u^\dagger_{\alpha_1\beta_1}
\hat u^\dagger_{\alpha_2\beta_2}\eexp{-2i\chi}
\eeq
This is obtained from \eqref{e78} by $\lambda_1,\lambda_2=0$ and then replacing $\lambda_3,\lambda_4\rightarrow \lambda_1,\lambda_2$, and $\tilde I_4\rightarrow -I^W$. Hence from from Eqs. (\ref{e80}, \ref{e82}, \ref{e84}),  where the 1st P.P. vanishes and the dominant term at $eV\gg k_BT$ is $(1-f_R(\epsilon_R'))f_L(\epsilon_L')$
\beq{114}
&& S_{7a}'^W=-I^W2\pi\delta(\omega+\epsilon_{RL}-\lambda_{12})(1-f_R(\epsilon_R))f_L(\epsilon_L)
\cdot (-i)\pi\delta(\epsilon'_{RL}+\lambda_{12})(1-f_R(\epsilon_R'))f_L(\epsilon_L')\nonumber\\
&&\qquad [\frac{1}{-\lambda_{12}+i\eta}
+\frac{1}{\lambda_{12}+i\eta}]\eexp{-\beta\lambda_1}(\eexp{-2i\chi}-1)\nonumber\\
&& S_{7b}'^W=-I^W2\pi\delta(\omega+\epsilon_{RL}-\lambda_{12})(1-f_R(\epsilon_R))f_L(\epsilon_L)
\cdot (-i)\pi\delta(\epsilon'_{RL}+\lambda_{12})(1-f_R(\epsilon_R'))f_L(\epsilon_L')\nonumber\\
&&\qquad [\frac{1}{-\epsilon_{RL}+i\eta}
+\frac{1}{\epsilon_{RL}+i\eta}]\eexp{-\beta\lambda_1}(\eexp{-2i\chi}+1)\nonumber\\
&& S_{7c}'^W=I^W2\pi\delta(\omega+\epsilon_{RL}-\lambda_{12})(1-f_R(\epsilon_R))f_L(\epsilon_L)
\cdot (-i)\pi\delta(\epsilon'_{RL}+\lambda_{12}-\omega)(1-f_R(\epsilon_R'))f_L(\epsilon_L')\nonumber\\
&&\qquad [\frac{1}{-\epsilon_{RL}+i\eta}
+\frac{1}{\epsilon_{RL}+i\eta}]\eexp{-\beta\lambda_1}(\eexp{-2i\chi}+1)
\eeq
where the additional $\pm 1$ terms in $(\eexp{-2i\chi}\pm 1)$ arise from the
$R\leftrightarrow L$ in the $t$ ordered fermion bubble as obtained from the result above except for $\epsilon_{R}'\leftrightarrow\epsilon_{L}'$; however one trace involves $\tr[\sigma^j\hat u^\dagger]=-\tr[\sigma^j\hat u]$ (leading to $-1$), but also a sign change for a current vertex for $S_{7b}^W,\,S_{7c}^W$ (leading to $+1$). The terms with $P.P.(\frac{1}{\omega+\lambda_{12}})$ or $P.P.\frac{1}{\lambda_{12}}$ are neglected, being much weaker than resonance. Note that for $eV\gg\omega$ the sum $S'^W_{7b}+S'^W_{7c}=0$.

\begin{figure*}  \centering
\includegraphics [width=.45\textwidth]{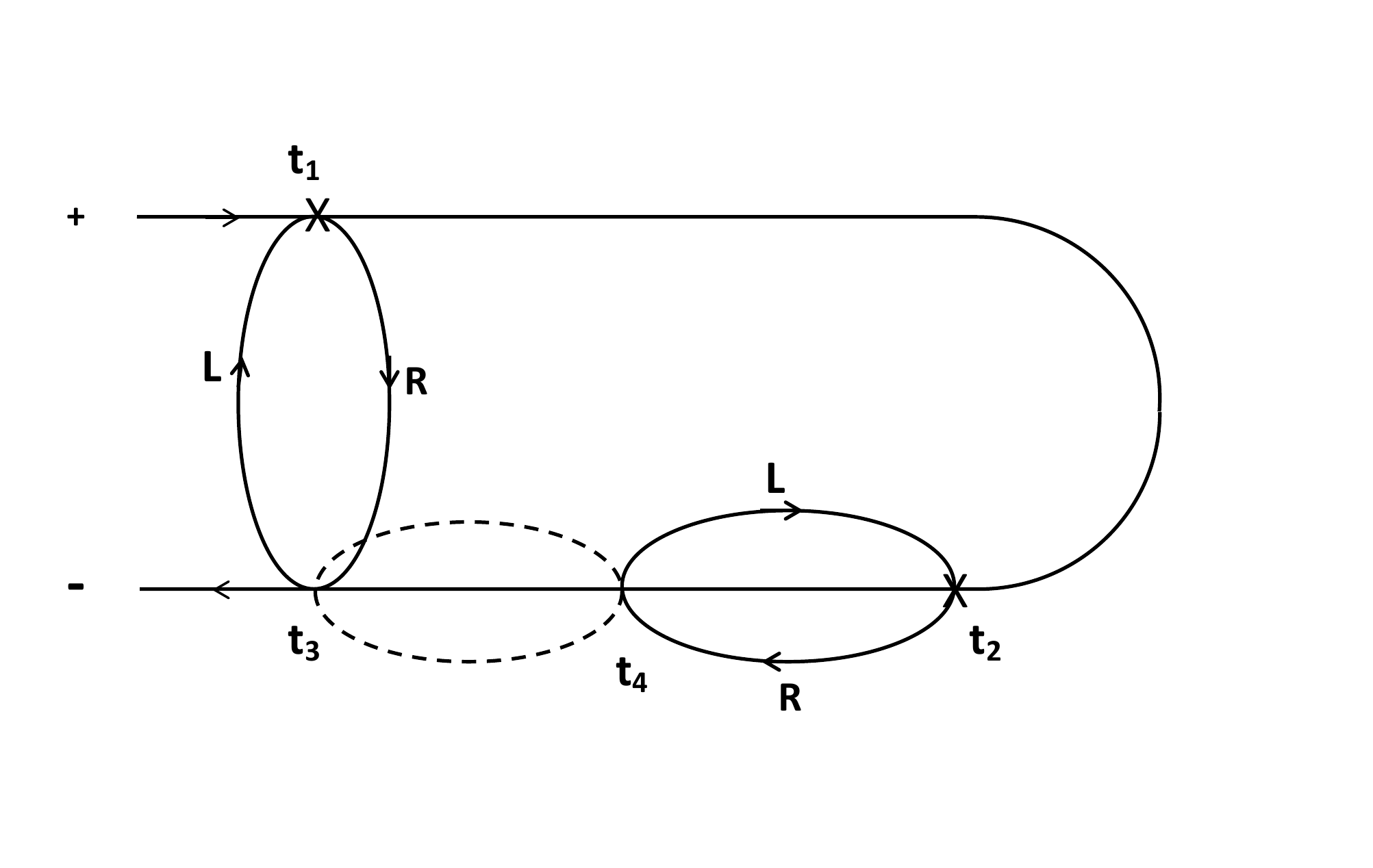}
\includegraphics [width=.45\textwidth]{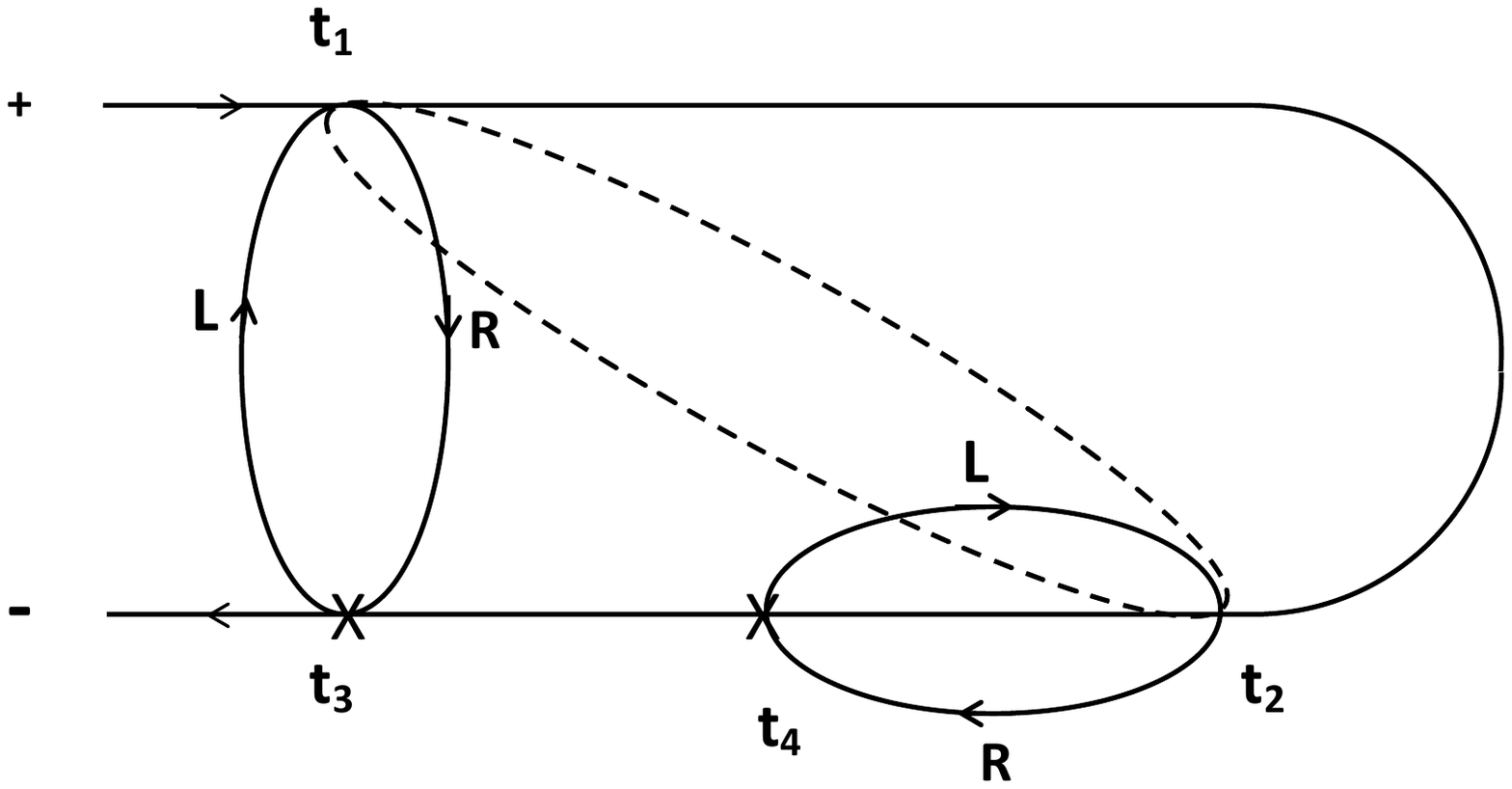}
\caption{Diagram 8W (left) and 8'W (right)}
\label{diagram8W}
\end{figure*}

Consider next Fig. \ref{diagram8W}, left panel.
\beq{115}
&&S_8^W=-i^4e^2J^2W^2\left\langle [\stackrel{1}{\bar\psi}_{R+}u_{\alpha_1\beta_1}^\dagger\stackrel{2}{\psi}_{L+}]_{t_1}
[\stackrel{5}{\bar\psi}_{R-}
u_{\alpha_2\beta_2}^\dagger\stackrel{6}{\psi}_{L-}]_{t_2}\right.\nonumber\\
&&\left.\qquad [\stackrel{2}{\bar\psi}_{L-}\bm\sigma _{\alpha_3\beta_3}\stackrel{1}{\psi}_{R-}\cdot\stackrel{7}{\bar f}_{-}\bm\tau_{\gamma_1\delta_1}\stackrel{8}{f}_{-}]_{t_3} [\stackrel{6}{\bar\psi}_{L-}\bm\sigma  _{\alpha_4\beta_4}\stackrel{5}{\psi}_{R-}\cdot\stackrel{8}{\bar f}_{-}\bm\tau_{\gamma_2\delta_2} \stackrel{7}{f}_{-}]_{t_4}\right\rangle\eexp{-2i\chi}\nonumber\\
&&=-e^2J^2W^2(-i)G^>_R(t_3-t_1)iG_L^<(t_1-t_3)(-i)G_R^{\bar t}(t_4-t_2)iG_L^{\bar t}(t_2-t_4)
(-i)F^{\bar t}_{\gamma_1}(t_4-t_3)iF_{\gamma_2}^{\bar t}(t_3-t_4)\nonumber\\
&&\qquad\times \delta_{\alpha_1\beta_3}\delta_{\alpha_3\beta_1}\delta_{\alpha_2\beta_4}\delta_{\alpha_4\beta_2}
\delta_{\gamma_1\delta_2}\delta_{\gamma_2\delta_1}
\sigma^i_{\alpha_3\beta_3}\tau^i_{\gamma_1\delta_1}
\sigma^j_{\alpha_4\beta_4}\tau^j_{\gamma_2\delta_2} u^\dagger_{\alpha_1\beta_1}
\hat u^\dagger_{\alpha_2\beta_2}\eexp{-2i\chi}
\eeq
This can be obtained from $S_8$ \eqref{e86} by $\lambda_1,\lambda_2=0$ and then replace $\lambda_3,\lambda_4\rightarrow \lambda_1,\lambda_2$ and $\tilde I_4\rightarrow -I^W$.
Hence from Eqs. (\ref{e87},\ref{e89},\ref{e90})
\beq{116}
&& S_{8a}^W=-I^W2\pi\delta(\omega+\epsilon_{RL})(1-f_R(\epsilon_R))f_L(\epsilon_L)
\cdot (-i)\pi\delta(\epsilon'_{RL})(1-f_R(\epsilon_R'))f_L(\epsilon_L')\nonumber\\
&&\qquad [\frac{\eexp{-\beta\lambda_2}}{-\omega-\epsilon_{RL}+\lambda_{12}+i\eta}
+\frac{\eexp{-\beta\lambda_1}}{\omega+\epsilon_{RL}-\lambda_{12}+i\eta}](\eexp{-2i\chi}-1)
=S_{7a}^{W*}\nonumber\\
&& S_{8b}^W=-I^W2\pi\delta(\omega+\epsilon_{RL})(1-f_R(\epsilon_R))f_L(\epsilon_L)
\cdot (-i)\pi\delta(\epsilon'_{RL})(1-f_R(\epsilon_R'))f_L(\epsilon_L')\nonumber\\
&&\qquad [\frac{\eexp{-\beta\lambda_2}}{-\epsilon_{RL}+\lambda_{12}+i\eta}
+\frac{\eexp{-\beta\lambda_1}}{\epsilon_{RL}-\lambda_{12}+i\eta}](\eexp{-2i\chi}+1)=S_{7b}^{W*}\nonumber\\
&& S_{8c}^W=I^W2\pi\delta(\omega+\epsilon_{RL})(1-f_R(\epsilon_R))f_L(\epsilon_L)
\cdot (-i)\pi\delta(\epsilon'_{RL}-\omega)(1-f_R(\epsilon_R'))f_L(\epsilon_L')\nonumber\\
&&\qquad [\frac{\eexp{-\beta\lambda_2}}{-\epsilon_{RL}+\lambda_{12}+i\eta}
+\frac{\eexp{-\beta\lambda_1}}{\epsilon_{RL}-\lambda_{12}+i\eta}](\eexp{-2i\chi}+1)=S_{7c}^{W*}\nonumber\\
\eeq
where the additional $\pm 1$ arise from $R\leftrightarrow L$ as above, and
 for $S_{8b}^W,S_{8c}^W$ an exchange $\lambda_1\leftrightarrow\lambda_2$ is needed for the final relation to $S_7^{W*}$.

Finally, consider Fig. \ref{diagram8W} right panel,
\beq{117}
&&S_8'^W=-i^4e^2J^2W^2\left\langle [\stackrel{1}{\bar\psi}_{R+}\bm\sigma_{\alpha_1\beta_1}\stackrel{2}{\psi}_{L+}\cdot \stackrel{3}{\bar f}_{+}\bm\tau_{\gamma_1\delta_1}\stackrel{4}{f}_{+}]_{t_1}[\stackrel{5}{\bar\psi}_{R-}\bm
\sigma_{\alpha_2\beta_2}\stackrel{6}{\psi}_{L-}\cdot\stackrel{4}{\bar f}_{-}\bm\tau_{\gamma_2\delta_2}\stackrel{3}{f}_{-}]_{t_2}\right.\nonumber\\
&&\left.\qquad [\stackrel{2}{\bar\psi}_{L-}\hat u_{\alpha_3\beta_3}\stackrel{1}{\psi}_{R-}]_{t_3} [\stackrel{6}{\bar\psi}_{L-}\hat u_{\alpha_4\beta_4}\stackrel{5}{\psi}_{R-}]_{t_4}\right\rangle\eexp{2i\chi}\nonumber\\
&&=-e^2J_1^2J_2^2(-i)G^>_R(t_3-t_1)iG_L^<(t_1-t_3)(-i)G_R^{\bar t}(t_4-t_2)iG_L^{\bar t}(t_2-t_4)
 (-i)F^>_{\gamma_1}(t_2-t_1)iF^<_{\gamma_2}(t_1-t_2)\nonumber\\
&&\qquad\times \delta_{\alpha_1\beta_3}\delta_{\alpha_3\beta_1}\delta_{\alpha_2\beta_4}\delta_{\alpha_4\beta_2}
\delta_{\gamma_1\delta_2}\delta_{\gamma_2\delta_1}
\sigma^i_{\alpha_1\beta_1}\tau^i_{\gamma_1\delta_1}
\sigma^j_{\alpha_2\beta_2}\tau^j_{\gamma_2\delta_2}\hat u_{\alpha_3\beta_3}\hat u_{\alpha_4\beta_4}\eexp{2i\chi}\nonumber\\&&
\eeq
This is obtained from $S_8$ \eqref{e86} by $(-i)F^{\bar t}_{2\gamma_3}(t_4-t_3)
iF_{2\gamma_4}^{\bar t}(t_3-t_4)\rightarrow 1$, i.e. $\lambda_3,\lambda_4=0$ and $\tilde I_4\rightarrow -I^W$.
Hence from Eqs. (\ref{e87},\ref{e89},\ref{e90})
\beq{118}
&& S_{8a}'^W=-I^W2\pi\delta(\omega+\epsilon_{RL}+\lambda_{12})(1-f_R(\epsilon_R))f_L(\epsilon_L)
\cdot (-i)\pi\delta(\epsilon'_{RL}-\lambda_{12})(1-f_R(\epsilon_R'))f_L(\epsilon_L')\nonumber\\
&&\qquad [\frac{1}{-\omega-\epsilon_{RL}+i\eta}
+\frac{1}{\omega+\epsilon_{RL}+i\eta}]\eexp{-\beta\lambda_2}(\eexp{2i\chi}-1)=S_{7a}'^{W*}\nonumber\\
&& S_{8b}'^W=-I^W2\pi\delta(\omega+\epsilon_{RL}+\lambda_{12})(1-f_R(\epsilon_R))fL(\epsilon_L)
\cdot (-i)\pi\delta(\epsilon'_{RL}-\lambda_{12})(1-f_R(\epsilon_R'))f_L(\epsilon_L')\nonumber\\
&&\qquad [\frac{1}{-\epsilon_{RL}+i\eta}
+\frac{1}{\epsilon_{RL}+i\eta}]\eexp{-\beta\lambda_2}(\eexp{2i\chi}+1)=S_{7b}'^{W*}\nonumber\\
&& S_{8c}'^W=I^W2\pi\delta(\omega+\epsilon_{RL}+\lambda_{12})(1-f_R(\epsilon_R))fL(\epsilon_L)
\cdot (-i)\pi\delta(\epsilon'_{RL}-\lambda_{12}-\omega)(1-f_R(\epsilon_R'))f_L(\epsilon_L')\nonumber\\
&&\qquad [\frac{1}{-\epsilon_{RL}+i\eta}
+\frac{1}{\epsilon_{RL}+i\eta}]\eexp{-\beta\lambda_2}(\eexp{2i\chi}+1)=S_{7c}'^{W*}
\eeq
where the additional $\pm 1$ arise from $R\leftrightarrow L$ as above,
 in all 3 cases exchange $\lambda_1\leftrightarrow \lambda_2$ is needed for the final relation with $S'^{W*}_7$.

 For $eV\gg\omega,\nu$ we have $S^W_{7b}+S^W_{7c}=S'^W_{7b}+S'^W_{7c}=S^W_{8b}+S^W_{8c}=
 S'^W_{8b}+S'^W_{8c}=0$, hence we need only
 \beq{119}
&& \int_\epsilon [S^W_{7a}+S^W_{8a}]=e^2J^2W^2I^W 4\pi^3(eV)^2\delta(\lambda_{12})
(\eexp{-\beta\lambda_1}+\eexp{-\beta\lambda_2})(\cos 2\chi-1))\nonumber\\&&
\int_\epsilon [S'^W_{7a}+S'^W_{8a}]=e^2J^2W^2I^W 8\pi^3(eV)^2\delta(\lambda_{12})\eexp{-\beta\lambda_1}(\cos 2\chi -1)
\eeq
\\

\subsection{Summary -- single QD}

With a prefactor $e^2J^2W^2(eV)^2N^4(0)$ we have
\beq{120}
&&4a+4'a\qquad \qquad -8\pi^3I^W\delta(\omega+\lambda_{12})(\eexp{-\beta\lambda_1}+\eexp{-\beta\lambda_2})\cos 2\chi\nonumber\\
&&4b+4'b\qquad\qquad  -8\pi^3 I^W\delta(\lambda_{12})(\eexp{-\beta\lambda_1}+\eexp{-\beta\lambda_2})\cos 2\chi\nonumber\\
&&5a \qquad\qquad\qquad\qquad 8\pi^3I^W\delta(\omega+\lambda_{12})\eexp{-\beta\lambda_2}\nonumber\\
&& 5b+5c\qquad \qquad\qquad 16\pi^3I^W\delta(\lambda_{12})\eexp{-\beta\lambda_{1}}\nonumber\\
&& 5d \,\,\qquad\qquad \qquad\qquad 8\pi^3I^W\delta(\omega-\lambda_{12})\eexp{-\beta\lambda_2}\nonumber\\
&& 7a+7'a+8a+8'a\qquad 16\pi^3I^W\delta(\lambda_{12})\eexp{-\beta\lambda_1}(\cos 2\chi -1)
\eeq
The terms $\delta(\lambda_{12})$ precisely cancel, while the resonance terms give (with $\lambda_1\leftrightarrow\lambda_2$ in the 5d term)
\beq{121}
&&16\pi^3I^W\delta(\omega+\lambda_{12})(\eexp{-\beta\lambda_1}+\eexp{-\beta\lambda_2})\sin^2\chi\nonumber\\
&& \delta_{\gamma_1\gamma_2}\rightarrow\qquad\qquad\qquad =16\pi^3\cdot 4\sin^2\phi\cos^2\half\theta
(2\eexp{-\half\beta\nu}+2\eexp{\half\beta\nu})\sin^2\chi\delta(\omega) \nonumber\\
&& \gamma_1=\pm,\gamma_2=\mp\rightarrow\qquad =16\pi^3\cdot 4\sin^2\half\theta(\eexp{-\half\beta\nu}+\eexp{\half\beta\nu})\sin^2\chi\delta(\omega\pm\nu)
\eeq
After normalization, we finally have
\beq{122}
\boxed{
\begin{aligned}
\int_\epsilon S^{res}(\omega)=64\pi^3e^2J^2W^2(eV)^2N^4(0)\sin^2\chi\{2\sin^2\phi\cos^2\half\theta
\delta(\omega)+\sin^2\half\theta[\delta(\omega+\nu)+\delta(\omega-\nu)]\}\nonumber
\end{aligned} }\\
\eeq
Remarkably, this vanishes for $\chi=0$.\\

\section{ Spin relaxation}

We evaluate here the spin relaxation rates using a Lindblad equation, following Refs. \onlinecite{shnirman1},\onlinecite{schlosshauer}. The derivation is within 2nd order perturbation in the system-environment couplings, i.e. the exchange couplings.
This method is equivalent to extending the previous diagrammatic expansion, yet it is more straightforward.

\subsection{General formulation}
For completeness we outline the derivation of Lindblad type equations.
Consider a Hamiltonian, possibly time dependent,
\beq{301}
\tilde{\cal H}(t)=\tilde{\cal H}_S(t)+\tilde{\cal H}_E+\tilde{\cal H}_{SE}
\eeq
for the system, the environment and the coupling between them, respectively. Define $U_e(t)$ as the evolution operator for $\tilde{\cal H}_S$, i.e. $\frac{d}{dt}U_e(t)=-i[\tilde{\cal H}_S(t)+{\cal H}_E]U_e(t)$, where $U_e(t)$ depends implicitly on an initial time $t=0$. The interaction picture is denoted as operators without tilde, e.g.
${\cal H}_{SE}(t)=U_e^\dagger(t)\tilde{\cal H}_{SE} U_e(t)$, and the density matrix satisfies
\beq{302}
\frac{d}{dt}\tilde\rho&=&-i[\tilde{\cal H}(t),\tilde\rho], \qquad\rho(t)=U_e^\dagger(t)\tilde\rho U_e(t)\,,
\qquad\Rightarrow \frac{d}{dt}\rho(t)=-i[{\cal H}_{SE}(t),\rho(t)]
\eeq
We wish to derive the reduced density matrix $\rho_S=\tr_E\rho$. Consider
\beq{303}
\rho(t)&=&\rho(0) -i\int_0^t ds [{\cal H}_{SE}(s),\rho(s)]\nonumber\\
\frac{d}{dt}\rho_S(t)&=&-i \tr_E[{\cal H}_{SE}(t),\rho(0)]-\tr_E\left\{\int_0^tds [{\cal H}_{SE}(t),[{\cal H}_{SE}(s),\rho(s)]]\right\}
\eeq
The first term vanishes, e.g. when ${\cal H}_{SE}$ is linear in environment coordinates as is the case in what follows. The key assumption is that $\rho_{SE}$ factorizes
\beq{304}
\rho(s)=\rho_S(s)\otimes\rho_E
\eeq
This assumption is made after this 2nd order form is obtained, hence entanglement within 2nd order perturbation of $\rho_S$ and $\rho_E$ is already contained in this form.
Changing $t-s$ to $s$,
\beq{305}
\frac{d}{dt}\rho_S(t)=-\int_0^tds \tr_E[{\cal H}_{SE}(t),[{\cal H}_{SE}(t-s),\rho_S(t-s)\otimes\rho_E]]
\eeq
 Consider interactions with products of $A_j(t), B_j(t)$ that are operators in the system and environment space, respectively, $j=-J,...,J$ and $A_j^\dagger=A_{-j},\,B^\dagger_j=B_{-j}$.  $\tr_E$ can be done by using the environment correlation $\Gamma_{jk}(s)$
\beq{306}
{\cal H}_{SE}(t)=\sum_jA_j(t)\otimes B_j(t)\qquad
\Gamma_{jk}(s)=\tr_E[B_j(s)B_k(0)\rho_E]\,, \qquad \Gamma_{jk}^*(s)=\Gamma_{-k,-j}(-s)\nonumber
\eeq
since the environment is stationary. Furthermore,  defining $\gamma_j(\omega)$ as the Fourier of $\Gamma_{j,-j}(s)$, it satisfies $\gamma_j(\omega)=\gamma_j^*(\omega)$, and {\it if} the environment is at equilibrium with temperature $1/\beta$ then the Fourier of $\Gamma_{j,-j}^*(s)$ is $\gamma_j(-\omega)=\eexp{-\beta\omega}\gamma_j(\omega)$ from detailed balance.

Opening the commutators and using cyclic property of the trace
\beq{307}
\frac{d}{dt}\rho_S(t)= \int_0^tds\sum_{jk}\Gamma_{jk}(s)[A_k(t-s)\rho_S(t-s)A_j(t)-A_j(t)A_k(t-s)\rho_S(t-s)]+h.c.\nonumber
\eeq
We assumes now that $\Gamma(s)$ is short ranged (to be checked below), i.e. the Markoff assumption, so that
\beq{308}
\frac{d}{dt}\rho_S(t)= \int_0^tds\sum_{jk}\Gamma_{jk}(s)[A_k(t-s)\rho_S(t)A_j(t)-A_j(t)A_k(t-s)\rho_S(t)]+h.c.
\eeq
which is a local equation.

Each term in the interaction is chosen to have an eigenfrequency $\nu_j$ with $\nu_{-j}=-\nu_j$  i.e.
\beq{309}
{\cal H}_{SE}=\sum_{j=-J,...,J}B_j(t)\otimes A_j\eexp{-i\nu_j t}
\eeq
where $A_j^\dagger=A_{-j}$; the sum may contain a $j=0$ term with $\nu_0=0$. Hence
\beq{310}
\frac{d}{dt}\rho_S(t)&=&\sum_{j,k}\int_0^\infty ds \Gamma_{jk}(s)\eexp{-i\nu_k(t-s)-i\nu_j t}
[A_k\rho_S(t)A_j-A_j A_k\rho_S(t)]+h.c.\nonumber\\
&=&\sum_{j,k}\tilde\Gamma_{jk}(-\nu_k)\eexp{-i(\nu_k+\nu_j)t}[A_k\rho_S(t)A_j-A_j A_k\rho_S(t)]+h.c.
\eeq
where $\tilde\Gamma_{jk}(\omega)=\int_0^\infty ds\Gamma_{jk}(s)\eexp{i\omega s}$.

A standard assumption is a secular one, i.e. keeping only terms $j,-j$ in the sum so that the RHS is time independent. Terms $j,k$ ($j\neq -k$) can be neglected if the resulting linewidth is $\ll \nu_j+\nu_k$. In the following some of the non-secular correlations indeed satisfy this inequality, though not all.
Assuming for now that the only relevant environment correlations are
$\Gamma_{jk}(s)=\delta_{j,-k}\Gamma_j(s)$ the Lindblad form is obtained
\beq{312}
&&\frac{d}{dt}\rho_S(t)=\sum_{j}\tilde\Gamma_j(\nu_j)[A_j^\dagger\rho_S(t)A_j-A_j A_j^\dagger\rho_S(t)]+h.c.\nonumber\\
&&\tilde\Gamma_{j}(\omega)=\int_0^\infty ds\Gamma_{j}(s)\eexp{i\omega s}=\half \gamma_{j}(\omega)+i\im \tilde\Gamma_{j}(\omega)
\eeq
 Noting that
$\im\tilde\Gamma(\nu_j)$ cancels in the first term of $\frac{d}{dt}\rho_S(t)$, it has the form of shifting the Hamiltonian, i.e. a Lamb shift \cite{shnirman1} with ${\cal H}_{LS}=\sum_j\im\tilde\Gamma(\nu_j)A_j^\dagger A_j$, hence
\beq{313}
\frac{d}{dt}\rho_S(t)=-i[{\cal H}_{LS},\rho_S]+\sum_{j}\gamma_j(\nu_j)[A_j^\dagger\rho_S(t)A_j-\half A_j A_j^\dagger\rho_S(t)-\half\rho_SA_j A_j^\dagger]
\eeq

In the following we encounter cases of degeneracy, i.e. operators $j,-j$ and $j',-j'$ having the same eigenfrequency. We therefore use the form \eqref{e310} keeping the off-diagonal terms involving degenerate terms. The secular approximation can be applied only to products with (sufficiently) different eigenfrequencies. The resulting equation is time independent, similar to \eqref{e312}, even with these degenerate off-diagonal $j,j'$ terms.

\subsection{Single QD}

Consider the Hamiltonian Eq. (3) in the main text
\beq{314}
{\cal H}=\half\nu_L\tau_z+[Jc_R^\dagger{\bm \sigma}c_L\cdot{\bm\tau}+W\eexp{-i\chi}c_R^\dagger\hat u^\dagger c_L+h.c.]+{\cal H}_{R,L}
\eeq
$\tau_i$ are Pauli matrices in the localized spin space, $\nu_L$ is the Larmor frequency, $\hat u=\eexp{i\sigma_z\phi}\eexp{\half i\sigma_y\theta}$ represents spin-orbit and
${\cal H}_{R,L}=\sum_k\epsilon_{R,L}c^\dagger_{R,L}c_{R,L}$ are the lead Hamiltonians with $\epsilon, c^\dagger, c$ implicitly momentum $k$ dependent.

The interaction picture is obtained by
\beq{315}
&&U_e=\eexp{-i(\half\nu_L\tau_z+\epsilon_Lc^\dagger_L c_L+\epsilon_Rc^\dagger_R c_R)t}\nonumber\\
&& U_e^\dagger c_R^\dagger c_L U_e=c_R^\dagger c_L \eexp{i\epsilon_{RL}t}\qquad \epsilon_{RL}=\epsilon_R-\epsilon_L,\qquad U_e^\dagger \tau_- U_e=\tau_-\eexp{-i\nu_Lt}
\eeq
Using ${\bm\sigma}\cdot{\bm\tau}=2\sigma_+\tau_-+2\sigma_-\tau_++\sigma_z\tau_z$ the interactions in the interaction picture become
\beq{316}
{\cal H}_{SE}=2Jc_R^\dagger \sigma_+c_L\tau_-\eexp{i\epsilon_{RL}t-i\nu_Lt}+
2Jc_R^\dagger \sigma_-c_L\tau_+\eexp{i\epsilon_{RL}t+i\nu_Lt}+
Jc_R^\dagger \sigma_zc_L\tau_z\eexp{i\epsilon_{RL}t}+Wc_R^\dagger\hat u^\dagger c_L\eexp{i\epsilon_{RL}t}+h.c.\nonumber\\
\eeq
Hence the form of Eq. \eqref{e309} with
\beq{317}
&&A_1=\tau_-\qquad\nu_1=\nu_L\qquad B_1=2J(c_R^\dagger\sigma_+c_L\eexp{i\epsilon_{RL}t}+
c_L^\dagger\sigma_+c_R\eexp{-i\epsilon_{RL}t})\nonumber\\
&&A_{-1}=\tau_+\qquad \nu_{-1}=-\nu_L\qquad B_{-1}=B_1^\dagger\nonumber\\
&&A_z=\tau_z\qquad\nu_z=0\qquad B_z=Jc_R^\dagger\sigma_z c_L\eexp{i\epsilon_{RL}t}+h.c.\nonumber\\
&&A_{0}=1\qquad \nu_{0}=0\qquad B_{0}=W\eexp{-i\chi}c_R^\dagger\hat u^\dagger c_L\eexp{i\epsilon_{RL}t}+h.c.
\eeq
Consider the correlation (integration on $\epsilon_R,\epsilon_L$ is implicit)
\beq{318}
&&\Gamma_1(s)=4J^2\tr[(c_R^\dagger\sigma_+c_L\eexp{i\epsilon_{RL}s}+
c_L^\dagger\sigma_+c_R\eexp{-i\epsilon_{RL}s})(c_L^\dagger\sigma_-c_R+
c_R^\dagger\sigma_-c_L)\rho_E]\nonumber\\&&
=4J^2\{f_R(\epsilon_R)(1-f_L(\epsilon_L))\tr[\sigma_+\sigma_-]
\eexp{-i\epsilon_{RL}s}+f_L(\epsilon_L)(1-f_R(\epsilon_R))\tr[\sigma_-\sigma_+]
\eexp{i\epsilon_{RL}s}\}
\eeq
and $\Gamma_{-1}(s)=\Gamma_{1}(s)$ since $\tr[\sigma_+\sigma_-]=\tr[\sigma_-\sigma_+]=1$. Hence FDT is expected for $\gamma_1(\omega)$ in equilibrium.
With a convergence factor \cite{shnirman1} $\eta$,
$\int_0^\infty \eexp{i\epsilon_{RL}s+i\omega s-\eta s}ds=\frac{-1}{i(\epsilon_{RL}+\omega+i\eta)}$,
\beq{319}
&&\tilde\Gamma_1(\omega)=4J^2N^2(0)\int_{\epsilon_R,\epsilon_L}\{\frac{-f_L(\epsilon_L)(1-f_R(\epsilon_R))}
{i(-\epsilon_{RL}+\omega+i\eta)}+\frac{-f_R(\epsilon_R)(1-f_L(\epsilon_L))}
{i(\epsilon_{RL}+\omega+i\eta)}\}\Rightarrow\nonumber\\&&
\gamma_1(\omega)=8\pi J^2N^2(0)\int_{\epsilon_L}[f_L(\epsilon_L)(1-f_R(\epsilon_L+\omega))+
f_R(\epsilon_L-\omega)(1-f_L(\epsilon_L))]\nonumber\\&&
=8\pi J^2N^2(0)\{(eV+\omega)\frac{\eexp{\beta(eV+\omega)}}{\eexp{\beta(eV+\omega)}-1}+
(-eV+\omega)\frac{\eexp{\beta(-eV+\omega)}}{\eexp{\beta(-eV+\omega)}-1}\}
\eeq
For $V=0$ FDT is obeyed with the bath temperature, however, not for $V\neq 0$,
\beq{320}
&&V=0:\qquad \gamma_1(\omega)=16\pi J^2N^2(0)\omega\frac{\eexp{\beta\omega}}{\eexp{\beta\omega}-1}\Rightarrow \qquad
\gamma_1(-\omega)=\eexp{-\beta\omega}\gamma_1(\omega)\nonumber\\&&
eV\gg 1/\beta :\qquad \gamma_1(\omega)=8\pi J^2N^2(0)(eV+\omega)\Rightarrow\qquad \eexp{-\beta^*\omega}\equiv\frac{eV-\omega}{eV+\omega}
\eeq
$1/\beta^*$ defines an effective temperature for the spin population, which in general depends on the frequency $\omega=\nu_L$, if $eV\gg 1/\beta,\omega$ then $\beta^*\rightarrow 2/eV$.

For the imaginary part (that shifts $\nu_L$) we have
\beq{321}
\im\tilde\Gamma_1(\omega)=
-8J^2N^2(0)\omega P.P.\int_{\epsilon_R,\epsilon_L}\frac{f_L(\epsilon_L)(1-f_R(\epsilon_R))
+f_R(\epsilon_R)(1-f_L(\epsilon_L))}{\epsilon_{RL}^2-\omega^2}\approx -8J^2N^2(0)\omega\ln\frac{eV}{\epsilon_F}
\eeq
where a cutoff $\epsilon_F$ is needed assuming $\epsilon_F\gtrsim eV$. For $\,eV\gg\omega$ we have $\im\tilde\Gamma_1(\omega)\ll\gamma_1(\omega)$ and $\im\tilde\Gamma_1(\omega)$ can be neglected. We note that $\tilde\Gamma_1(\omega)$, and the following $\tilde\Gamma_{ij}(\omega)$ are weakly $\omega$ dependent, hence $\Gamma_1(s)$ has short range, as needed for the Markoff assumption.

The subspace $z,0$ is degenerate so that off diagonal terms are needed. The P.P. terms are same as for $\Gamma_1(s)$ (with $\omega\rightarrow 0$), for the real parts consider $eV>0$ and $eV\gg \omega,1/\beta$ so that the term $f_L(\epsilon_L)(1-f_R(\epsilon_R)$ dominates.
Consider first
\beq{322}
\Gamma_{zz}(s)&&=J^2\tr[(c_R^\dagger\sigma_z c_L\eexp{i\epsilon_{RL}s}+h.c.)
(c_R^\dagger\sigma_z c_L+h.c.)\rho_E]=J^2f_L(\epsilon_L)(1-f_R(\epsilon_R))\tr[\sigma_z^2]
\eexp{-i\epsilon_{RL}s}\nonumber\\
\half\gamma_z(\omega=0)&&=2J^2\re\int_{\epsilon_R,\epsilon_L}\frac{-f_L(\epsilon_L)(1-f_R(\epsilon_R))}
{i(-\epsilon_{RL}+i\eta)}=2\pi J^2eVN^2(0)
\eeq
We also find $\Gamma_{00}(s)=\frac{W^2}{J^2}\Gamma_{zz}(s)$, but actually it is of no interest since in Eq. \eqref{e310} it multiplies $\rho_S-\rho_S=0$.
 $\Gamma_{0z}$ multiplies in \eqref{e310} $\tau_z\rho_S(t)\cdot 1-\tau_z\cdot 1\rho_S(t)=0$ and  therefore is also of no interest. Consider then
\beq{323}
&&\Gamma_{z0}(s)=JW\tr[(c_R^\dagger\sigma_z c_L\eexp{i\epsilon_{RL}t}+h.c.)(\eexp{-i\chi}c_R^\dagger\hat u^\dagger c_L+h.c.)\rho_E]\\&&
=\{f_R(\epsilon_R)(1-f_L(\epsilon_L))\eexp{i\chi}\tr[\sigma_z\hat u]
\eexp{i\epsilon_{RL}s}+f_L(\epsilon_L)(1-f_R(\epsilon_R))\eexp{-i\chi}\tr[\sigma_z\hat u^\dagger]
\eexp{-i\epsilon_{RL}s}\}\nonumber\\&&
\tilde\Gamma_{z0}(\omega)=JWi\sin\phi\cos\half\theta\int_{\epsilon_R,\epsilon_L}
\{\frac{f_L(\epsilon_L)(1-f_R(\epsilon_R))}
{i(-\epsilon_{RL}+\omega+i\eta)}\eexp{-i\chi}-\frac{f_R(\epsilon_R)(1-f_L(\epsilon_L))}
{i(\epsilon_{RL}+\omega+i\eta)}\eexp{i\chi}\}\nonumber\\&&
=JW\sin\phi\cos\half\theta[-\pi i eVN^2(0)\eexp{-i\chi}+P.P.\int_{\epsilon_R,\epsilon_L}
(\frac{f_L(\epsilon_L)f_R(\epsilon_R)}{\epsilon_{RL}-\omega}\eexp{-i\chi}+
\frac{f_L(\epsilon_L)f_R(\epsilon_R)}{\epsilon_{RL}+\omega}\eexp{i\chi})]\nonumber
\eeq
using $\tr[\sigma_z\hat u]=2i\sin\phi\cos\half\theta=(\tr[\sigma_z\hat u^\dagger])^*$,
neglecting $f_R(\epsilon_R)(1-f_L(\epsilon_R+\omega))$ for large $V>0$, and using $P.P.\int_{\epsilon_R} 1/(\epsilon_{RL}\pm\omega)=0$. Note that the h.c. term in \eqref{e310} cancels the $P.P.$ term since
\[\tilde\Gamma_{z0}(1\rho_S\tau_z-\tau_z\rho_S1)+\tilde\Gamma_{z0}^*(\tau_z\rho_S1-1\rho_S\tau_z)
=2i\im\tilde\Gamma_{z0}(\rho_S\tau_z-\tau_z\rho_S)\]

Denoting $\gamma_1=\gamma_1(\nu_L),\,\gamma_{-1}=\gamma_{-1}(-\nu_L)=\gamma_1(-\nu_L)$
Lindblad's Eq. \eqref{e310} for $\rho_S=\half\cdot 1+\rho_z(t)\tau_z+\rho_{+}(t)\tau_++\rho_{-}(t)\tau_-$ becomes
\beq{324}
&&\frac{d}{dt}[\rho_z(t)\tau_z+\rho_{+}(t)\tau_++\rho_{-}\tau_-]=
\gamma_z[\tau_z\rho_S\tau_z-\rho_S]+
\half\gamma_1[\tau_+\rho_S\tau_--\tau_-\tau_+\rho_S+h.c.]
\nonumber\\&& +\half\gamma_{-1}[\tau_-\rho_S\tau_+-\tau_+\tau_-\rho_S+h.c.]
+2i\im\tilde\Gamma_{z0}(\rho_S\tau_z-\tau_z\rho_S)\nonumber\\&&
=\gamma_z[\rho_{+}(\tau_z\tau_+\tau_z-\tau_+)+h.c.]+\half\gamma_1
[\tau_z+\rho_z(\tau_+\tau_z\tau_--\tau_-\tau_+\tau_z+h.c.)-\rho_{-}\tau_--\rho_{+}\tau_+)]\nonumber\\&&
+\half\gamma_{-1}
[-\tau_z+\rho_z(\tau_-\tau_z\tau_+-\tau_+\tau_-\tau_z+h.c.)-\rho_{+}\tau_+-\rho_{-}\tau_-)]
+4i\im\tilde\Gamma_{z0}(-\rho_{+}\tau_++\rho_{-}\tau_-)\nonumber\\&&
=-2\gamma_z(\rho_{+}\tau_++\rho_{-}\tau_-)-\half\gamma_1(-\tau_z+2\rho_z\tau_z+
\rho_{-}\tau_-+\rho_{+}\tau_+)\nonumber\\&&-\half\gamma_{-1}(\tau_z+2\rho_z\tau_z+
\rho_{+}\tau_++\rho_{-}\tau_-)+4i\im\tilde\Gamma_{z0}
(-\rho_{+}\tau_++\rho_{-}\tau_-)
\eeq
using $\tau_z\tau_\pm=\pm\tau_\pm,\, \tau_+\tau_z\tau_--\tau_-\tau_+\tau_z=-\tau_z,\,\tau_+\tau_-\tau_+=\tau_+,\,[\tau_+,\tau_-]=\tau_z$. Comparing coefficients
\beq{325}
&&\frac{d\rho_z}{dt}=-(\gamma_1+\gamma_{-1})\rho_z
+\half(\gamma_1-\gamma_{-1})=-\frac{1}{T_1}(\rho_z-\rho_z^0)\nonumber\\&&
\frac{d\rho_{+}}{dt}=-2\gamma_z\rho_{+}-\half(\gamma_1+\gamma_{-1})\rho_+
-4i\im\tilde\Gamma_{z0}\rho_{+}=-\frac{1}{T_2}\rho_{+}-4i\im\tilde\Gamma_{z0}\rho_{+}\nonumber\\
&&\frac{1}{T_1}=16\pi J^2eVN^2(0),\qquad\frac{1}{T_2}=\frac{1}{2T_1}+2\gamma_0=16\pi J^2eVN^2(0)
\eeq
The effective temperature $1/\beta^*$ relates $\gamma_1(-\nu_L)=\eexp{-\beta^*\nu_L}\gamma_1(\nu_L)$, hence
$\rho_z^0=\half\langle \sigma_z\rangle_0=\half\tanh\half\beta^*\nu_L$, hence from \eqref{e319}, with full dependence on parameters,
\[\rho_z^0=\frac{\nu_L}{\frac{eV+\nu_L}{\tanh\half\beta(eV+\nu_L)}+\frac{eV-\nu_L}
{\tanh\half\beta(eV-\nu_L)}}=\frac{\nu_L}{2eV}[1+O(\frac{\nu_L}{eV},\frac{T}{eV})]\]
a result known from studies of the Kondo problem \cite{parcolet}. For large $eV$ the spin population tends to be equal in the two states.

The $-4i\im\tilde\Gamma_{00'}\rho_{01}$ term can be absorbed into a redefinition of $\tilde\rho_{01}=
\rho_{01}\eexp{-i\delta\nu_Lt}$ with
\beq{326}
\delta\nu_L=4\im\tilde\Gamma_{z0}=-4\pi JWeVN^2(0)\sin\phi\cos\half\theta\cos\chi
\eeq
since the imaginary of the P.P. term in \eqref{e323} at $\omega=0$ vanishes.
For $W\gg J$ this shift is larger than those neglected, from $\im\Gamma_{\pm 1}$ and in particular it is larger than the linewidth found above.

It is of interest to compare with a study \cite{paaske,rosch} of the $W=0$ case, using a diagrammatic expansion. They find that indeed the  longitudinal and  transverse  rates  are  identical, i.e. $T_1=T_2$, with results that are consistent with Eq. \eqref{e325}.

\subsection{Double QD}

Consider the Hamiltonian Eq. (6) in the main text, in terms of the product space of Pauli matrices $\tau_i\otimes\tau_j$ of spin 1 and 2, respectively, and $\sigma_i$ for the electron spin,
\beq{328}
{\cal H}=\half\nu_{L1}\tau_z\otimes 1+\half\nu_{L2} 1\otimes\tau_z+
[J_1c_R^\dagger{\bm \sigma}c_L\cdot{\bm\tau}\otimes 1+J_2\eexp{i\chi}c_R^\dagger{\bm \sigma}\hat u c_L\cdot 1\otimes{\bm\tau}+h.c.]+{\cal H}_{R,L}
\eeq
The interaction picture yields
\beq{329}
&&U_e=\eexp{-i[\half\nu_{L1}\tau_z\otimes 1+\half\nu_{L2} 1\otimes\tau_z+\epsilon_Lc^\dagger_L c_L+\epsilon_Rc^\dagger_R c_R]t}\\
&& {\cal H}_{SE}=2J_1c_R^\dagger \sigma_+c_L\tau_-\otimes 1\eexp{i\epsilon_{RL}t-i\nu_{L1}t}+
2J_1c_R^\dagger \sigma_-c_L\tau_+\otimes 1\eexp{i\epsilon_{RL}t+i\nu_{L1}t}+
J_1c_R^\dagger \sigma_zc_L\tau_z\otimes 1\eexp{i\epsilon_{RL}t}+\nonumber\\&&
J_2\eexp{i\chi}[2c_R^\dagger \sigma_+\hat uc_L1\otimes\tau_-\eexp{i\epsilon_{RL}t-i\nu_{L2}t}+
2c_R^\dagger \sigma_-\hat uc_L1\otimes\tau_+\eexp{i\epsilon_{RL}t+i\nu_{L2}t}+
c_R^\dagger \sigma_z\hat uc_L1\otimes\tau_z\eexp{i\epsilon_{RL}t}]+h.c.\nonumber
\eeq
Hence ${\cal H}_{SE}$ has the form \eqref{e309} with
\beq{330}
&&A_1=\tau_-\otimes 1\qquad \nu_1=\nu_{L1}\qquad \qquad B_1=2J_1(c_R^\dagger\sigma_+c_L\eexp{i\epsilon_{RL}t}+
c_L^\dagger\sigma_+c_R\eexp{-i\epsilon_{RL}t})\nonumber\\&&
A_{-1}=\tau_+\otimes 1\qquad \nu_{-1}=-\nu_{L1}\qquad B_{-1}=B_1^\dagger\nonumber\\&&
A_z=\tau_z\otimes 1\qquad \nu_z=0\qquad \qquad\,\,\, B_z=J_1c_R^\dagger\sigma_zc_L\eexp{i\epsilon_{RL}t}+h.c.\nonumber\\&&
A_2=1\otimes\tau_-\qquad \nu_2=\nu_{L2}\qquad \qquad B_2=2J_2(\eexp{i\chi}c_R^\dagger\sigma_+\hat u c_L\eexp{i\epsilon_{RL}t}+
\eexp{-i\chi}c_L^\dagger\hat u^\dagger\sigma_+ c_R\eexp{-i\epsilon_{RL}t})\nonumber\\&&
A_{-2}=1\otimes\tau_+\qquad \nu_{-2}=-\nu_{L2}\qquad B_{-2}=B_2^\dagger\nonumber\\&&
A_{z'}=1\otimes\tau_z\qquad  \nu_{z'}=0\qquad \qquad B_{z'}=J_2\eexp{i\chi}c_R^\dagger\sigma_z\hat u c_L\eexp{i\epsilon_{RL}t}+h.c.
\eeq
The terms $B_z,B_{z'}$ are degenerate and their off diagonal terms are kept. All other terms are treated within the secular scheme keeping only their diagonal form, denoted by $\gamma_{\pm 1},\gamma_{\pm 2}$ (i.e. Eq. \ref{e319} with $J_1,\,J_2$, respectively); note that $\Gamma_2(s)=\tr[B_2(s)B_{-2}(0)\rho_E]$ has $\hat u^\dagger \hat u=1$ so that $\Gamma_2(s)=(J_2^2/J_1^2)\Gamma_1(s)$,
 similarly $\Gamma_{z'z'}(s)=(J_2^2/J_1^2)\Gamma_{zz}(s)$. The corresponding imaginary terms are neglected as in \eqref{e321}. The off diagonal term is
\beq{331}
&&\Gamma_{zz'}=\tr[B_z(s)B_{z'}(0)\rho_E]=J_1J_2\langle(c_R^\dagger\sigma_z c_L\eexp{i\epsilon_{RL}s}
+c_L^\dagger\sigma_zc_R\eexp{-i\epsilon_{RL}s})(\eexp{i\chi}c_R^\dagger\sigma_z\hat u c_L+\eexp{-i\chi}c_L^\dagger\hat u^\dagger \sigma_z c_R)\rangle\nonumber\\&&
=J_1J_2\{ f_R(\epsilon_R)(1-f_L(\epsilon_L))\eexp{i\epsilon_{RL}t}\tr[\sigma_z^2\hat u^\dagger]\eexp{-i\chi}+
f_L(\epsilon_L)(1-f_R(\epsilon_R))\eexp{-i\epsilon_{RL}t}\tr[\sigma_z^2\hat u]\eexp{i\chi}\}\nonumber\\&&
\tilde\Gamma_{zz'}(\omega)=2J_1J_2N^2(0)\cos\half\theta\cos\phi\int_{\epsilon_R,\epsilon_L}
\{\frac{-f_L(\epsilon_L)(1-f_R(\epsilon_R))}{i(\omega-\epsilon_{RL}+i\eta)}\eexp{i\chi}+
\frac{-f_R(\epsilon_R)(1-f_L(\epsilon_L))}{i(\omega+\epsilon_{RL}+i\eta)}\eexp{-i\chi}\}\nonumber\\&&
\Rightarrow\qquad\tilde\Gamma_{zz'}(0)=\gamma_{z}\eexp{i\chi},\qquad \gamma_{z}=2\pi J_1J_2 N^2(0)eV\cos\half\theta\cos\phi
\eeq
using $\tr[\hat u]=\tr[\hat u^\dagger]=2\cos\half\theta\cos\phi$,
$ P.P.\int_{\epsilon_R}\frac{1}{\epsilon_R-\epsilon_L}=0$ and neglecting in the last line the $f_R(\epsilon_L))(1-f_L(\epsilon_L))$ term when $eV>0$, yields the result $\gamma_{z}$. Note also $\Gamma_{z'z}(s)=\Gamma_{zz'}(s,\chi\rightarrow -\chi)$ so that $\tilde\Gamma_{z'z}(0)=\gamma_{z}\eexp{-i\chi}$.

Consider a general form
$\rho_S(t)=\sum_{\alpha,\beta}\rho_{\alpha,\beta}(t)\tau_\alpha\otimes\tau_\beta$, with $\alpha,\beta=0,z,+,-$ so that $\tau_\alpha=1,\tau_z,\tau_+,\tau_-$. The diagonal terms produce the same terms as in \eqref{e325}, while the off diagonals change $\rho_{\pm,\pm}$ as well as adding imaginary terms when $\chi\neq 0$ (see below),
\beq{332}
&&\frac{d\rho_{z0}}{dt}=-(\gamma_1+\gamma_{-1})(\rho_{z0}-\rho_{z0}^0),\qquad
\frac{d\rho_{0z}}{dt}=-(\gamma_2+\gamma_{-2})(\rho_{0z}-\rho_{0z}^0),\qquad \rho_{z0}^0=\half\frac{\gamma_1-\gamma_{-1}}{\gamma_1+\gamma_{-1}}\nonumber\\&&
\frac{d\rho_{+0}}{dt}=-\frac{1}{T_2^{(1)}}\rho_{+0},\qquad
\frac{d\rho_{0+}}{dt}=-\frac{1}{T_2^{(2)}}\rho_{0+},\qquad\qquad\qquad\qquad\qquad\qquad
\rho_{0z}^0=\half\frac{\gamma_2-\gamma_{-2}}{\gamma_2+\gamma_{-2}}\nonumber\\&&
\frac{d\rho_{zz}}{dt}=-(\gamma_1+\gamma_{-1})\rho_{zz}+(\gamma_1-\gamma_{-1})\rho_{0z}
-(\gamma_2+\gamma_{-2})\rho_{zz}+(\gamma_2-\gamma_{-2})\rho_{z0}\nonumber\\&&
\frac{d\rho_{z+}}{dt}=-(\gamma_1+\gamma_{-1})\rho_{z+}+(\gamma_1-\gamma_{-1})\rho_{0+}
-\frac{1}{T_2^{(2)}}\rho_{z+}\nonumber\\&&
\frac{d\rho_{+z}}{dt}=-\frac{1}{T_2^{(1)}}\rho_{+z}-(\gamma_2+\gamma_{-2})\rho_{+z}+
(\gamma_2-\gamma_{-2})\rho_{+0}\nonumber\\&&
\frac{d\rho_{\alpha,\pm\alpha}}{dt}=-(\frac{1}{T_2^{(1)}}+\frac{1}{T_2^{(2)}})
\rho_{\alpha,\pm\alpha}+\mbox{off diagonal terms, see \eqref{e35}}\qquad
\alpha=+,-
\eeq
where $T_2^{(1),(2)}$ are the corresponding $T_2$ relaxation times. Equations with $+\rightarrow -$ are related by c.c. ($\rho_S$ is hermitian), e.g. $\rho_{z-}=\rho_{z+}^*$.
In steady state $\rho_{z0}=\rho_{z0}^0=\half\tanh\half\beta^*\nu_{L1}$, $\rho_{0z}=\rho_{0z}^0=\half\tanh\half\beta^*\nu_{L2}$, but also $\rho_{zz}=\half\tanh\half\beta^*\nu_{L1}
\tanh\half\beta^*\nu_{L2}$ is finite.

Consider next $\tilde\Gamma_{zz'}$ (at $\omega=0$) term in \eqref{e310}, allowing now $\chi\neq 0$,
\beq{333}
&&\tilde\Gamma_{zz'}[A_{z'}\rho_S A_z-A_z A_{z'}\rho_S]+h.c.=\tilde\Gamma_{zz'}\sum_{\alpha,\beta}\rho_{\alpha\beta}
[1\otimes\tau_z\cdot\tau_\alpha\otimes\tau_\beta\cdot \tau_z\otimes 1\nonumber\\&&
-\tau_z\otimes 1\cdot 1\otimes \tau_z\cdot\tau_\alpha\otimes\tau_\beta]+h.c.
=\tilde\Gamma_{zz'}\sum_{\alpha,\beta}\rho_{\alpha\beta}
[\tau_\alpha\tau_z\otimes\tau_z\tau_\beta-\tau_z\tau_\alpha\otimes\tau_z\tau_\beta]+h.c.\nonumber\\&&
=\tilde\Gamma_{zz'}\sum_{\pm,\beta}2\rho_{\pm\beta}(\mp)\tau_{\pm}\otimes \tau_z\tau_\beta+h.c.
=\gamma_{z}\sum_{\pm,\beta=0,z'}2[\eexp{i\chi}\rho_{\pm\beta}(\mp)\tau_\pm\otimes\tau_z\tau_\beta
+\eexp{-i\chi}\rho_{\mp\beta}(\mp)\tau_{\mp}\otimes\tau_z\tau_\beta]+\nonumber\\&&
\gamma_{z}\sum_{\pm,\pm}2[\eexp{i\chi}\rho_{\pm,\pm}(\mp)\tau_\pm\otimes(\pm)\tau_\pm
+\eexp{-i\chi}\rho_{\mp,\mp}(\mp)\tau_{\mp}\otimes (\pm)\tau_\mp]\nonumber\\&&=
4i\gamma_z\sin\chi\sum_{\pm,\beta=0,z'}\rho_{\pm\beta}(\mp)\tau_\pm\otimes\tau_z\tau_\beta+
4\gamma_{z}\cos\chi\sum_{\pm,\pm}\rho_{\pm,\pm}(\mp)\tau_\pm\otimes(\pm)\tau_\pm
\eeq
the indices $\pm$ for $\alpha,\beta$ are uncorrelated; on the 3rd line above only
 $\alpha=\pm$ contributes, using $\tau_z\tau_\pm=\pm\tau_\pm,\,\tau_\pm\tau_z=\mp\tau_\pm$  and $\tilde\Gamma_{zz'}$ is real. Similarly, for $\tilde\Gamma_{z'z}$ only $\beta=\pm$ contributes, hence
\beq{334}
&&\tilde\Gamma_{z'z}[A_{z}\rho_S A_{z'}-A_{z'} A_{z}\rho_S]+h.c.=\tilde\Gamma_{z'z}\sum_{\alpha,\beta}\rho_{\alpha\beta}
[\tau_z\otimes 1\cdot\tau_\alpha\otimes\tau_\beta\cdot 1\otimes \tau_z\nonumber\\&&
-1\otimes \tau_z\cdot \tau_z\otimes 1\cdot\tau_\alpha\otimes\tau_\beta]+h.c.
=\tilde\Gamma_{z'z}\sum_{\alpha,\beta}\rho_{\alpha\beta}
[\tau_z\tau_\alpha\otimes\tau_\beta\tau_z-\tau_z\tau_\alpha\otimes\tau_z\tau_\beta]+h.c.\nonumber\\&&
=\tilde\Gamma_{z'z}\sum_{\alpha,\pm}2\rho_{\alpha,\pm}\tau_z\tau_\alpha\otimes(\mp)\tau_\pm+h.c.
=\gamma_{z}\sum_{\alpha=0,z,\beta=\pm}2[\eexp{-i\chi}\rho_{\alpha,\pm}\tau_z\tau_\alpha
\otimes(\mp)\tau_\pm+\eexp{i\chi}\rho_{\alpha,\mp}\tau_z\tau_\alpha\otimes(\mp)\tau_\mp]\nonumber\\&&
+\gamma_{z}\sum_{\pm,\pm}2[\eexp{-i\chi}\rho_{\pm,\pm}(\pm)\tau_\pm\otimes(\mp)\tau_\pm+
\eexp{i\chi}\rho_{\mp,\mp}(\pm)\tau_\mp\otimes
(\mp)\tau_\mp]\nonumber\\&&
=-4i\sin\chi\sum_{\alpha=0,z,\beta=\pm}\rho_{\alpha\pm}\tau_z\tau_\alpha\otimes(\mp)\tau_\pm+
4\gamma_{z}\cos\chi\sum_{\pm,\pm}\rho_{\pm,\pm}(\pm)\tau_\pm\otimes(\mp)\tau_\pm
\eeq
The full equations, including off diagonal terms with $\gamma_z$, become
\beq{335}
&&\frac{d\rho_{z0}}{dt}=-(\gamma_1+\gamma_{-1})(\rho_{z0}-\rho_{z0}^0),\qquad\qquad
\frac{d\rho_{0z}}{dt}=-(\gamma_2+\gamma_{-2})(\rho_{0z}-\rho_{0z}^0),\qquad \rho_{z0}^0=\half\frac{\gamma_1-\gamma_{-1}}{\gamma_1+\gamma_{-1}}\nonumber\\&&
\frac{d\rho_{+0}}{dt}=-\frac{1}{T_2^{(1)}}\rho_{+0}-4i\gamma_z\sin\chi\rho_{+z},\qquad
\frac{d\rho_{0+}}{dt}=-\frac{1}{T_2^{(2)}}\rho_{0+}-4i\gamma_z\sin\chi\rho_{z+},\qquad
\rho_{0z}^0=\half\frac{\gamma_2-\gamma_{-2}}{\gamma_2+\gamma_{-2}}\nonumber\\&&
\frac{d\rho_{zz}}{dt}=-(\gamma_1+\gamma_{-1})\rho_{zz}+(\gamma_1-\gamma_{-1})\rho_{0z}
-(\gamma_2+\gamma_{-2})\rho_{zz}+(\gamma_2-\gamma_{-2})\rho_{z0}\nonumber\\&&
\frac{d\rho_{z+}}{dt}=-(\gamma_1+\gamma_{-1})\rho_{z+}+(\gamma_1-\gamma_{-1})\rho_{0+}
-\frac{1}{T_2^{(2)}}\rho_{z+}-4i\gamma_z\sin\chi\rho_{0+}\nonumber\\&&
\frac{d\rho_{+z}}{dt}=-\frac{1}{T_2^{(1)}}\rho_{+z}-(\gamma_2+\gamma_{-2})\rho_{+z}+
(\gamma_2-\gamma_{-2})\rho_{+0}-4i\gamma_z\sin\chi\rho_{+0}\\
&&\frac{d\rho_{\alpha,\pm\alpha}}{dt}=-(\frac{1}{T_2^{(1)}}+\frac{1}{T_2^{(2)}}\pm 8\gamma_z\cos\chi)
\rho_{\alpha\alpha}
=-16\pi eVN^2(0)[J_1^2+J_2^2\pm J_1J_2\cos\half\theta\cos\phi\cos\chi]\rho_{\alpha\alpha}\nonumber
\eeq

Note that $\rho_{z+}$ and $\rho_{0+}$ are mixed in \eqref{e335}, neglecting the small terms
$\gamma_1-\gamma_{-1},\gamma_2-\gamma_{-2}$, the coupling is via $\gamma=4\gamma_z\sin\chi$. The pairs $\rho_{+0},\rho_{+z}$ and $\rho_{0+},\rho_{z+}$ are coupled, both having the form
\beq{337}
&&\frac{dy}{dt}=-\Gamma_1 y-i\gamma z,\qquad\qquad   \frac{dz}{dt}=-\Gamma_2 z-i\gamma y\nonumber\\
&& \tilde y=y\eexp{\Gamma_1 t},\,\tilde z=z\eexp{\Gamma_1 t}\Rightarrow\qquad \frac{d\tilde y}{dt}=-i\gamma \tilde z,\qquad \frac{d}{dt}(\tilde z\eexp{\Gamma_2 t-\Gamma_1 t})=-i\gamma\tilde y\eexp{\Gamma_2 t-\Gamma_1 t}\nonumber\\&&
\frac{d^2\tilde z}{dt^2}+(\Gamma_2-\Gamma_1)\frac{d\tilde z}{dt}+\gamma^2\tilde z=0
,\qquad \tilde z\sim \eexp{\lambda t}\Rightarrow
\nonumber\\&& y,z\sim \eexp{-\half(\Gamma_1+\Gamma_2)t\pm\half\sqrt{(\Gamma_2-\Gamma_1)^2-4\gamma^2}\,\,t}
\eeq
If $\gamma<|\Gamma_1-\Gamma_2|$ the decay rates are mixed, while if $\gamma>|\Gamma_1-\Gamma_2|$ then the decay rates of $y,z$ become equal while their resonance frequencies shift.

The resonances at $\nu_{L1},\,\nu_{L2}$ that appear in the current noise involve vertices $\tr[\sigma_\pm\sigma_z\hat u]$, hence their linewidth is given by $\rho_{z+}$ (not by $\rho_{0+}$). Hence Eqs. (\ref{e332},\ref{e335}) identify the linewidth of the corresponding combination of spin propagator, which for $\chi=0$ are
\beq{338}
&&\mbox{ Linewidth of}\,\,\, \nu_{L1}=\frac{1}{T_2^{(1)}}+\frac{1}{T_1^{(2)}}=16\pi eVN^2(0)(J_1^2+J_2^2)\nonumber\\
&&\mbox{ Linewidth of}\,\,\, \nu_{L2}=\frac{1}{T_2^{(2)}}+\frac{1}{T_1^{(1)}}=16\pi eVN^2(0)(J_1^2+J_2^2)\nonumber\\
&&\mbox{ Linewidth of}\,\,\, \delta(\omega)=\frac{1}{T_1^{(1)}}+\frac{1}{T_1^{(2)}}=16\pi eVN^2(0)(J_1^2+J_2^2)\\
&&\mbox{ Linewidth of}\,\,\, \nu_{L1}-\nu_{L2}=\frac{1}{T_2^{(1)}}+\frac{1}{T_1^{(2)}}-8\tilde\Gamma_{zz'}
=16\pi eVN^2(0)[J_1^2+J_2^2-J_1J_2\cos\half\theta\cos\phi]\nonumber
\eeq
where $\delta(\omega)$ in the noise involves $\tr[\sigma_z^2\hat u]$ and is therefore related to the decay of $\rho_{zz}$.

\end{widetext}

\end{document}